\documentstyle[12pt]{article}
\newcommand{\nc}{\newcommand}
\nc{\rnc}{\renewcommand}
\rnc{\baselinestretch}{1.25}
\rnc{\arraystretch}{0.8}
\rnc{\thesection}{\arabic{section}\,.}
\rnc{\thesubsection}{\arabic{section}.\arabic{subsection}}
\rnc{\theequation}{\arabic{section}.\arabic{equation}}
\nc{\sect}[1]{\section{#1}\setcounter{equation}{0}}
\nc{\sub}[1]{\subsection{#1}}
\nc{\subsub}[1]{\subsubsection{#1}}
\rnc{\title}[1]{{\Large\bf#1\bigskip\\}}
\rnc{\author}[1]{{\large#1\\}}
\nc{\address}[1]{{\em#1\\}}
\nc{\be}{\begin{equation}}
\nc{\ee}{\end{equation}}
\nc{\bc}{\begin{center}}
\nc{\ec}{\end{center}}
\nc{\bpic}{\begin{picture}}
\nc{\epic}{\end{picture}}
\nc{\ba}[1]{\begin{array}{@{}#1@{}}}
\nc{\ea}{\end{array}}
\nc{\bea}{\begin{eqnarray}}
\nc{\eea}{\end{eqnarray}}
\nc{\el}[1]{\label{#1}}
\nc{\er}[1]{(\ref{#1})}
\newcounter{subeq}
\newcounter{tab}
\nc{\itm}[1]{\\\noindent$\bullet$\ \ #1}
\nc{\ds}{\displaystyle}
\nc{\ts}{\textstyle}
\rnc{\ss}{\scriptstyle}
\nc{\sss}{\scriptscriptstyle}
\nc{\ru}[1]{\rule[-#1ex]{0ex}{#1ex}}
\rnc{\vec}[1]{\mbox{\boldmath$#1$}}
\font\tenmsb=msbm10 scaled \magstep1
\font\sevenmsb=msbm7 scaled \magstep1
\font\fivemsb=msbm5 scaled \magstep1
\newfam\msbfam
\textfont\msbfam=\tenmsb
\scriptfont\msbfam=\sevenmsb
\scriptscriptfont\msbfam=\fivemsb
\def\Bbb#1{{\fam\msbfam\relax#1}}
\nc{\p}[2]{\makebox(0,0)[#1]{$#2$}}
\nc{\pp}[2]{\makebox(0,0)[#1]{$\ss#2$}}
\nc{\ppp}[2]{\makebox(0,0)[#1]{$\sss#2$}}
\nc{\text}[6]{\begin{picture}(#1,#2)\put(#3,#4){\p{#5}{\ds#6}}\end{picture}}
\rnc{\a}{\alpha}
\nc{\ab}{\bar{a}}
\rnc{\b}{\beta}
\nc{\bb}{\bar{b}}
\nc{\C}{\Bbb C}
\rnc{\d}{\delta}
\nc{\e}{\epsilon}
\nc{\ep}{{\cal E}}
\nc{\et}{{\sss1\!/\!8}}
\nc{\E}{E_{\!A}}
\nc{\F}{{\cal F}}
\nc{\g}{\gamma}
\rnc{\gg}{\frac{g}{2}}
\nc{\G}{{\cal G}}
\nc{\Gc}{\tilde{\G}}
\nc{\Gh}{\hat{\G}}
\nc{\hf}{{\sss1\!/\!2}}
\nc{\M}{{\cal M}}
\nc{\MM}{{\sss M}}
\nc{\N}{{\sss N}}
\nc{\NN}{\Bbb N}
\rnc{\l}{\lambda}
\nc{\Y}{Y}
\nc{\Yt}{\tilde{Y}}
\nc{\x}{\xi}
\nc{\mi}{\!-\!}
\nc{\mipl}{\!\mp\!}
\nc{\n}{g}
\nc{\pl}{\!+\!}
\nc{\plmi}{\!\pm\!}
\nc{\qt}{{\sss1\!/\!4}}
\nc{\r}{F}
\nc{\slt}{\hat{s\ell}(2)}
\nc{\T}{{\cal T}}
\nc{\B}[7]{B^{#1}\!\!\left(\left.\!#4\ba{@{\;\,}c@{\;\,}c@{\:}}#5&#6\\
#2&#3\ea\right|#7\right)}
\nc{\W}[5]{W\!\left(\,\begin{array}{@{}cc|@{\:}}#4&#3\\
#1&#2\end{array}\;#5\right)}
\nc{\Wf}[3]{W^{#1}\!\!\left(\,#2\;#3\right)}
\nc{\Wi}[8]{\begin{array}{@{}c@{\:}c@{\:}c|@{\:}}#7&#6&#5\\#8&&#4\\#1&#2&#3
\end{array}}
\nc{\Q}[7]{Q^{#1}\!\!\left(\,\begin{array}{@{}c@{\;\,}c@{}}
#6&#5\\#7&#4\\#2&#3\end{array}\,\right)}
\nc{\Qf}[8]{Q^{#1}\!\!\left(\,\begin{array}{@{}c@{\:}c@{\:}c@{}}
#7&#6&#5\\#8&&#4\\#2&1&#3\end{array}\,\right)}
\nc{\Z}{\Bbb Z}

\begin{document}
\begin{center}
\vspace*{1mm}
\title{Integrable and Conformal Boundary Conditions\\
for $\slt$ $A$--$D$--$E$ Lattice Models\\
and Unitary Minimal Conformal Field Theories}
\author{Roger E. Behrend}
\address{C. N. Yang Institute for Theoretical Physics\\
State University of New York\\Stony Brook, NY 11794-3840, USA\\
\footnotesize\tt Roger.Behrend@sunysb.edu}
\medskip
\author{Paul A. Pearce}
\address{Department of Mathematics and Statistics\\University of Melbourne\\
Parkville, Victoria 3052, Australia\\
\footnotesize\tt P.Pearce@ms.unimelb.edu.au}
\begin{abstract}
\noindent Integrable boundary conditions are studied for 
critical $A$--$D$--$E$ and general graph-based
lattice models of statistical mechanics.  In
particular, using techniques associated with the Temperley-Lieb
algebra and fusion, a set of boundary Boltzmann weights which
satisfies the boundary Yang-Baxter equation is obtained for each
boundary condition.  When appropriately specialized, these boundary
weights, each of which depends on three spins, decompose into more
natural two-spin edge weights.  The specialized boundary conditions
for the $A$--$D$--$E$ cases
are naturally in one-to-one correspondence with the conformal
boundary conditions of $\slt$ unitary minimal conformal field
theories.  Supported by this and further evidence, we conclude that,
in the continuum scaling limit, the integrable boundary conditions
provide realizations of the complete set of conformal boundary
conditions in the corresponding field theories.
\end{abstract}
\end{center}
\newpage
\sect{Introduction and Overview} 
The notion of conformal boundary conditions in conformal field theories,
in the sense introduced in \cite{Car89}, and the notion of integrable
boundary conditions in integrable lattice models, in the sense
introduced in \cite{Skl88}, are both well developed.  It is also well
known that conformal field theories can be identified with the
continuum scaling limit of certain critical integrable lattice models
of statistical mechanics.  A natural question to address, therefore,
is whether similar associations can be made between the corresponding
boundary conditions.  We shall demonstrate here that indeed they can.
In so doing, not only is the existence of deep connections between
conformal and integrable boundary conditions established, but a means
is also provided for gaining insights, generally not available within
conformal field theory alone, into the physical nature of conformal
boundary conditions.

That there should be a relationship between such integrable and
conformal boundary conditions is not immediately apparent and,
accordingly, the correspondence is somewhat subtle.  It is known that
nonintegrable boundary conditions can be identified with certain
conformal boundary conditions and, conversely, we expect that
integrable boundary conditions will only lead to conformal boundary
conditions upon suitable specialization.  Nevertheless, we shall
explicitly show for all of the $\slt$ cases considered here that an
integrable boundary condition can, after such specialization, be
naturally associated with every conformal boundary condition.
Furthermore, the generality of the approach used here suggests that it
is possible to obtain an integrable boundary condition
corresponding to each allowable conformal boundary condition in any
rational conformal field theory which is realizable as the continuum
scaling limit of a Yang-Baxter-integrable lattice model.

The basic context for this work is provided by certain general results
on boundary conditions.  For statistical mechanical lattice models, it
is well known from~\cite{Bax82b} that a model is integrable on a torus
by commuting transfer matrix techniques if its Boltzmann weights
satisfy the Yang-Baxter equation.  A result of~\cite{Skl88} then
states that such a model is similarly integrable on a cylinder with
particular boundary conditions if corresponding boundary Boltzmann
weights are used which satisfy a boundary Yang-Baxter equation.  A
further result, obtained in \cite{BehPea96}, is that such boundary
weights can be constructed using a general procedure involving the
process of fusion.  For conformal field theories, there exists a
fundamental consistency equation of \cite{Car89} associated with the
conformal boundary conditions of any rational theory on a
cylinder. Furthermore, as shown in
\cite{PraSagSta96,SagSta96,BehPeaZub98,BehPeaPetZub98,BehPeaPetZub00},
the general task of solving this equation and completely classifying
the conformal boundary conditions is essentially equivalent to that of
classifying the representations, using matrices with nonnegative
integer entries, of the fusion algebra.

In this paper, we consider the critical unitary $A$--$D$--$E$
integrable lattice models, as introduced in \cite{Pas87a}, and the
critical series of $\slt$ unitary minimal conformal field theories
with central charge $c<1$, as first identified in
\cite{BelPolZam84,FriQiuShe84} and classified on a torus using an
$A$--$D$--$E$ scheme in \cite{CapItzZub87a,CapItzZub87b,Kat87}.  As
shown in \cite{Hus84,Pas87d,Pas87e}, the continuum scaling limit of
these lattice models is described by these field theories.
Furthermore, the task of classifying the conformal boundary conditions
of these theories on a cylinder was carried out in full in
\cite{BehPeaPetZub00}, leading to an $A$--$D$--$E$ scheme matching
that for the classification of the theories on a torus.  Lattice
realizations of some of these conformal boundary conditions have been
identified and studied in
\cite{Car89,BehPeaZub98,Car86b,SalBau89,DifZub90b,PasSal90,Dif92,Dor93,
Chi96,ObrPeaWar96,ObrPeaWar97,Bat98,AffOshSal98}, but in most of these
cases the lattice boundary conditions are not integrable in the sense
of being implemented by boundary weights which satisfy the boundary
Yang-Baxter equation.  Here, we use the fusion procedure
of~\cite{BehPea96} to obtain systematically such integrable boundary
weights and we show that, when appropriately specialized, these can be
interpreted as providing realizations of the complete sets of
conformal boundary conditions, as classified in \cite{BehPeaPetZub00},
of the corresponding field theories on a cylinder.

In obtaining the integrable boundary conditions for the critical unitary
$A$--$D$--$E$ models, it is convenient to consider a somewhat larger class 
of integrable lattice models.  These models, essentially introduced in
\cite{OwcBax87}, are restricted-solid-on-solid
interaction-round-a-face models and each is based on a graph $\G$.
In such a model, the spin states are the
nodes of $\G$, there is an adjacency condition on the states of
neighboring spins given by the edges of $\G$ and the bulk Boltzmann
weights are defined in terms of a particular eigenvalue and
eigenvector of the adjacency matrix of $\G$.  The critical unitary 
$A$--$D$--$E$ models are obtained by taking $\G$ as an $A$, $D$ or $E$
Dynkin diagram and using the Perron-Frobenius eigenvalue and eigenvector
of its adjacency matrix.  These general graph-based models are also
closely related to the Temperley-Lieb algebra, it being possible to
express their bulk weights in terms of matrices of a certain
representation of this algebra involving $\G$.  The fact that these
weights satisfy the Yang-Baxter equation is then a simple consequence
of the defining relations of the algebra.

In Section~2, we consider the abstract Temperley-Lieb algebra. The
results are thus independent of its representation and apply to the
class of lattice models associated with the Temperley-Lieb algebra,
this including the graph-based models of interest as well as certain
vertex models.  The main emphasis of Section~2 is on the process of
fusion and its use in the construction of boundary operators which
correspond to boundary weights.  In particular, we list various
important properties satisfied by the operators which implement
fusion, including projection and push-through properties, and we
obtain several results on the properties of the constructed boundary
operators, including the facts that they satisfy an operator form of
the boundary Yang-Baxter equation and that they can be expressed as a
linear combination of fusion operators.

In Section~3, we specialize to representations of the Temperley-Lieb
algebra involving graphs and study the corresponding graph-based integrable
lattice models.  For any such model, we then obtain an integrable boundary
condition and a related set of boundary weights for each pair $(r,a)$,
where $r$ is a fusion level and $a$ is a spin state or node of $\G$.
In terms of the procedure of \cite{BehPea96}, the $(r,a)$ boundary
weights are constructed from a fused double block of bulk weights of
width $r-1$, with the spin states on the corners of one end of the
block fixed to $a$.  These boundary weights each depend on three
spins, but we find using results from Section~2 that, upon appropriate
specialization, all of the weights in a set can be simultaneously
decomposed into physically more natural two-spin edge weights.
We refer to the point at which this occurs as the conformal point,
since it is here that we expect correspondence with conformal boundary
conditions.  While certain integrable boundary weights for some of the
models considered here have been obtained in previous studies,
specifically
\cite{BehPea96,BehPeaZub98,FanHouShi95,AhnKoo96a,BehPeaObr96,BehPea97,
AhnYou98}, this crucial decomposition had not previously been
observed.  We conclude Section~3 by considering in detail the symmetry
properties of a lattice on a cylinder with particular left and right
boundary conditions.

In Section~4, we specialize to the critical unitary $A$--$D$--$E$
models.  We obtain completely explicit expressions for the boundary
edge weights of the $A$ and $D$ cases and study $A_3$, $A_4$, $D_4$
and $E_6$ as important examples.  We also obtain an important relation
through which any $A$--$D$--$E$ partition function can be expressed as
a sum of certain $A$ partition functions.  For all of these
$A$--$D$--$E$ models, we find that, at the conformal point, the
$(r,a)$ and $(g\mi r\mi1,\ab)$ boundary conditions are equivalent,
where $g$ is the Coxeter number of the $A$--$D$--$E$ Dynkin diagram
$\G$ and $a\mapsto\ab$ is a
particular involution of the nodes of $\G$.  This implies that these
boundary conditions are in one-to-one correspondence with the
conformal boundary conditions of the unitary minimal theories
$\M(A_{g-2},\G)$, as classified in \cite{BehPeaPetZub00}.  We then use
this and further evidence to argue that the integrable boundary
conditions obtained here provide realizations of the
$\M(A_{g-2},\G)$ conformal boundary conditions.

In Section~5, we briefly discuss ways in which the formalism of this
paper could be applied to other models.

\sect{\protect Relevant Results on the Temperley-Lieb\\ Algebra} 
In this section, we list and obtain various results on the
Temperley-Lieb algebra.  The defining relations of this algebra were
first identified in \cite{TemLie71} and the formalism used here is
largely based on that of \cite{DegAkuWad88,DegWadAku88,WadDegAku89}.

Our primary objective in this section is to study operators in the
abstract algebra, which, in certain representations of the algebra,
correspond to bulk weights and boundary weights of a lattice model.
In particular, motivated by the procedure of \cite{BehPea96}, we shall
construct certain boundary operators.  This procedure involves fusion,
which can be regarded as a means whereby new fused operators
satisfying the Yang-Baxter equation are formed by applying certain
fusion operators to blocks of unfused operators.  The process of
fusion was introduced in \cite{KulResSkl81} and first used in the
context of the Temperley-Lieb algebra in 
\cite{DegAkuWad88,DegWadAku88}.

\sub{The Temperley-Lieb Algebra and General Notation}
The Temperley-Lieb algebra $\T(L,\l)$, with $L\in\Z_{\ge0}$ and $\l\in\C$,  
is generated by the identity $I$ together
with operators $e_1,\ldots,e_L$ which satisfy
\be\el{TLA}\ba{l}\ru{2}e_j^2\;=\;2\cos\!\l\:e_j\\
\ru{2}e_j\:e_k\:e_j\;=\;e_j,\quad|j\mi k|=1\\
\ru{0}e_j\:e_k\;=\;e_k\:e_j,\quad|j\mi k|>1\,.\ea\ee

The various operators to be studied in Section 2 will all be
elements of $\T(L,\l)$ for some fixed $L$ and $\l$.

Throughout this paper, we shall use the notation
\be\el{sD}s_r(u)\;=\;\left\{\ba{l}\ru{3.5}\ds\frac{\sin(r\l\pl u)}{\sin\!\l}\,,
\qquad\l/\pi\not\in\Z\\
(-1)^{(r+1)\l/\pi}\:\bigl(r\pl u\bigr),\qquad\l/\pi\in\Z\,,\ea\right.\ee
for any $r\in\Z$ and $u\in\C$, with $\l$ being the same 
constant as in $\T(L,\l)$.  

When $\T(L,\l)$ is applied to a lattice model,
$\l$ is the crossing parameter.  For the $A$--$D$--$E$ models, this parameter
is always a rational but noninteger multiple of $\pi$ so that the first case
of \er{sD} applies. We also note that the second case of \er{sD} is
simply a limiting case of the first,
\be s_r(u)\Big|_{\l/\pi\,\in\,\Z}\;=\;\lim_{\l'\rightarrow\l}
\frac{\sin(r\l'\pl(\l'\mi\l)u)}{\sin\!\l'}\,,\ee
so that in proving identities satisfied by these functions it is often
sufficient to consider only the first case.

We shall also denote, for any $r\in\Z$,
\be\el{SD}S_r=s_r(0)\,.\ee
We see, as examples, that for any $\l$,
\be S_0=0\,,\qquad S_1=1\,,\qquad S_2=2\cos\!\l\,.\ee

\sub{Face Operators}
We now introduce face operators $X_j(u)$, for 
$j\in\{1,\ldots,L\}$ and $u\in\C$, as
\be\el{X}X_j(u)\;=\;s_1(-u)\:I\;+\;s_0(u)\:e_j\,.\ee
These operators correspond in a lattice model to the bulk Boltzmann
weights of faces of the lattice and in this context $u$ is
the spectral parameter.

We shall represent the face operators diagrammatically as
\setlength{\unitlength}{5mm}
\be\raisebox{-1.75\unitlength}[1.75\unitlength][1.75\unitlength]{
\text{1.8}{3.5}{0.9}{2}{}{X_j(u)}
\text{1.8}{3.5}{0.9}{2}{}{=}
\bpic(2,3.5)\multiput(0,2)(1,-1){2}{\line(1,1){1}}
\multiput(0,2)(1,1){2}{\line(1,-1){1}}
\put(1,2){\pp{}{u}}
\multiput(0,0.5)(0,0.25){13}{\pp{}{.}}
\multiput(1,0.5)(0,0.25){2}{\pp{}{.}}
\multiput(1,3.25)(0,0.25){2}{\pp{}{.}}
\multiput(2,0.5)(0,0.25){13}{\pp{}{.}}
\put(-0.05,0){\pp{b}{j-1}}\put(1.1,0){\pp{b}{j}}\put(2.15,0){\pp{b}{j+1}}
\put(2.4,1.7){\p{}{.}}\epic}\ee

From the Temperley-Lieb relations~\er{TLA} and properties
of the functions \er{sD}, it follows that the
face operators satisfy the operator form of the Yang-Baxter equation,
\setlength{\unitlength}{6mm}
\be\el{TLYBE}\ba{r@{}c@{}l}
X_j(u)\;X_{j+1}(u\pl v)\;X_j(v)&=&X_{j+1}(v)\;X_j(u\pl v)\;X_{j+1}(u)\\
\bpic(3,6.3)\multiput(0,2)(1,-1){2}{\line(1,1){2}}
\multiput(0,4)(1,1){2}{\line(1,-1){2}}
\put(0,2){\line(1,-1){1}}\put(0,4){\line(1,1){1}}
\put(1,2){\pp{}{u}}\put(2,3){\pp{}{u+v}}\put(1,4){\pp{}{v}}
\multiput(0,0.5)(0,0.25){21}{\pp{}{.}}
\multiput(1,0.5)(0,0.25){2}{\pp{}{.}}
\multiput(1,5.25)(0,0.25){2}{\pp{}{.}}
\multiput(2,0.5)(0,0.25){6}{\pp{}{.}}
\multiput(2,4.25)(0,0.25){6}{\pp{}{.}}
\multiput(3,0.5)(0,0.25){21}{\pp{}{.}}
\put(-0.05,0){\pp{b}{j-1}}\put(1.1,0){\pp{b}{j}}
\put(2,0){\pp{b}{j+1}}\put(3.15,0){\pp{b}{j+2}}\epic
&\text{2.6}{5.5}{1.3}{3}{}{=}&
\bpic(3,5.5)\multiput(0,3)(1,1){2}{\line(1,-1){2}}
\multiput(0,3)(1,-1){2}{\line(1,1){2}}
\put(2,1){\line(1,1){1}}\put(2,5){\line(1,-1){1}}
\put(2,2){\pp{}{v}}\put(1,3){\pp{}{u+v}}
\put(2,4){\pp{}{u}}
\multiput(0,0.5)(0,0.25){21}{\pp{}{.}}
\multiput(1,0.5)(0,0.25){6}{\pp{}{.}}
\multiput(1,4.25)(0,0.25){6}{\pp{}{.}}
\multiput(2,0.5)(0,0.25){2}{\pp{}{.}}
\multiput(2,5.25)(0,0.25){2}{\pp{}{.}}
\multiput(3,0.5)(0,0.25){21}{\pp{}{.}}
\put(-0.05,0){\pp{b}{j-1}}\put(1.1,0){\pp{b}{j}}
\put(2,0){\pp{b}{j+1}}\put(3.15,0){\pp{b}{j+2}}
\put(3.4,2.7){\p{}{.}}\epic\ea\ee

We see that the face operators also satisfy the commutation relation
\be\el{XC}X_j(u)\:X_j(v)\;=\;X_j(v)\:X_j(u)\ee
and the operator form of the inversion relation
\be\el{TLIR}X_j(-u)\:X_j(u)\;=\;s_1(-u)\:s_1(u)\:I\,.\ee

\sub{Fusion Operators}\label{SFO}
We now proceed to a consideration of some aspects of the process
of fusion.  In particular, we shall state some important properties
of fusion operators.  The proofs of these or similar properties can
be found in \cite{DegAkuWad88,DegWadAku88}.

In general, fusion levels correspond to representations of
a Lie algebra.  In the case considered here, this algebra is
$\slt$ and the fusion levels are labeled by a single integer
$r\in\{1,2,\ldots,g\}$, where the maximum fusion level $g$ depends on
$\l$ according to
\be\el{g}\n\;=\;\left\{\ba{l}\ru{2}\;h,\qquad \l=k\pi/h,\;\;\mbox{$k$
and $h$ coprime integers},\;\;h>1\\
\;\infty,\qquad\mbox{otherwise}\,.\ea\right.\ee
We note that for the $A$--$D$--$E$ lattice models, the first case of
\er{g} always applies.

We now introduce fusion operators $P^r_j$, for
$r\in\{1,\ldots,\min(\n,L\pl2)\}$ and $j\in\{1,\ldots,L\pl3\mi r\}$,
these being defined recursively by
\be\el{FO}\ba{l}\ru{3}P^1_j\;=\;P^2_j\;=\;I\\ \ds\ru{3}P^r_j\;=\;
\frac{1}{S_{r-1}}\;P^{r-1}_{j+1}\;X_j(-(r\mi2)\l)\:P^{r-1}_{j+1}\,,
\quad r\ge3\,.\ea\ee
We note that, for finite $\n$, the restriction of fusion levels
to $r\le\n$ is necessary
in order to avoid $S_{r-1}=0$ in the denominator in~\er{FO}.

We shall represent the fusion operators diagrammatically as
\setlength{\unitlength}{5mm}
\be\raisebox{-1.75\unitlength}[1.75\unitlength][1.75\unitlength]{
\text{1}{3.5}{0.5}{2}{}{P^r_j}
\text{1.8}{3.5}{0.9}{2}{}{=}
\bpic(6,3.5)\multiput(0,2)(5,-1){2}{\line(1,1){1}}
\multiput(0,2)(5,1){2}{\line(1,-1){1}}
\multiput(1,1)(0,2){2}{\line(1,0){4}}
\multiput(0,0.5)(0,0.25){13}{\pp{}{.}}
\multiput(1,0.5)(0,0.25){2}{\pp{}{.}}
\multiput(1,3.25)(0,0.25){2}{\pp{}{.}}
\multiput(5,0.5)(0,0.25){2}{\pp{}{.}}
\multiput(5,3.25)(0,0.25){2}{\pp{}{.}}
\multiput(6,0.5)(0,0.25){13}{\pp{}{.}}
\put(-0.05,0){\pp{b}{j-1}}\put(1.1,0){\pp{b}{j}}
\put(5.3,0){\pp{br}{j+r-3}}\put(5.7,0){\pp{bl}{j+r-2}}
\put(6.4,1.7){\p{}{.}}\epic}\ee

In general, $P^r_j$ can be expressed in terms of 
$I$ and $e_j,\ldots,e_{j+r-3}$, the cases of the 
first few fusion levels being
\be\ba{c}\ru{4}\ds P^1_j\;=\;P^2_j\;=\;I,\qquad P^3_j\;=\;I-\frac{1}{S_2}e_j\\
\ds P^4_j\;=\;I-\frac{S_2}{S_3}(e_j+e_{j+1})+
\frac{1}{S_3}(e_je_{j+1}+e_{j+1}e_j)\,.\ea\ee

The fusion operators can also be expressed as
\setlength{\unitlength}{6mm}
\be\el{PE}\raisebox{-5.75\unitlength}[5.75\unitlength][5.75\unitlength]{
\text{4.1}{11.5}{4.08}{6}{r}{P^r_j\:=\:\prod_{k=2}^{r-1}\!(S_k)^{k-r}}\,
\bpic(6,11.5)\multiput(0,6)(1,1){2}{\line(1,-1){5}}
\put(2,8){\line(1,-1){4}}\put(3,9){\line(1,-1){3}}
\put(4,10){\line(1,-1){2}}\put(5,11){\line(1,-1){1}}
\multiput(0,6)(1,-1){2}{\line(1,1){5}}
\put(2,4){\line(1,1){4}}\put(3,3){\line(1,1){3}}
\put(4,2){\line(1,1){2}}\put(5,1){\line(1,1){1}}
\put(1,6){\pp{}{\!\mi(r\mi2)\l}}
\put(4,3){\pp{}{-2\l}}\put(4,9){\pp{}{-2\l}}
\put(5,2){\pp{}{-\l}}\put(5,4){\pp{}{-\l}}
\put(5,8){\pp{}{-\l}}\put(5,10){\pp{}{-\l}}
\multiput(0,0.5)(0,0.25){45}{\pp{}{.}}
\multiput(1,0.5)(0,0.25){18}{\pp{}{.}}
\multiput(1,7.25)(0,0.25){18}{\pp{}{.}}
\multiput(5,0.5)(0,0.25){2}{\pp{}{.}}
\multiput(5,11.25)(0,0.25){2}{\pp{}{.}}
\multiput(6,0.5)(0,0.25){45}{\pp{}{.}}
\put(-0.1,0){\pp{b}{j-1}}\put(1,0){\pp{b}{j}}
\put(5.3,0){\pp{br}{j+r-3}}\put(5.7,0){\pp{bl}{j+r-2}}\epic
\text{4.38}{11.5}{4.37}{6}{r}{=\;\prod_{k=2}^{r-1}\!(S_k)^{k-r}}\,
\bpic(6,11.5)\multiput(6,6)(-1,1){2}{\line(-1,-1){5}}
\put(4,8){\line(-1,-1){4}}\put(3,9){\line(-1,-1){3}}
\put(2,10){\line(-1,-1){2}}\put(1,11){\line(-1,-1){1}}
\multiput(6,6)(-1,-1){2}{\line(-1,1){5}}
\put(4,4){\line(-1,1){4}}\put(3,3){\line(-1,1){3}}
\put(2,2){\line(-1,1){2}}\put(1,1){\line(-1,1){1}}
\put(5,6){\pp{}{\!\mi(r\mi2)\l}}
\put(2,3){\pp{}{-2\l}}\put(2,9){\pp{}{-2\l}}
\put(1,2){\pp{}{-\l}}\put(1,4){\pp{}{-\l}}
\put(1,8){\pp{}{-\l}}\put(1,10){\pp{}{-\l}}
\multiput(0,0.5)(0,0.25){45}{\pp{}{.}}
\multiput(1,0.5)(0,0.25){2}{\pp{}{.}}
\multiput(1,11.25)(0,0.25){2}{\pp{}{.}}
\multiput(5,0.5)(0,0.25){18}{\pp{}{.}}
\multiput(5,7.25)(0,0.25){18}{\pp{}{.}}
\multiput(6,0.5)(0,0.25){45}{\pp{}{.}}
\put(-0.1,0){\pp{b}{j-1}}\put(1,0){\pp{b}{j}}
\put(5.3,0){\pp{br}{j+r-3}}\put(5.7,0){\pp{bl}{j+r-2}}
\put(6.4,5.7){\p{}{.}}\epic}\ee

A key property of the fusion operators is that they are projectors,
\be\el{PP}(P^r_j)^2\;=\;P^r_j\,.\ee
In fact, more generally, we have the property
\be\el{PPG}P^{r'}_{j'}\:P^r_j\;=\;P^r_j\:P^{r'}_{j'}\;=\;P^r_j\,,\quad
\mbox{if $0\le j'\mi j\le r\mi r'$}\,.\ee
A useful case of this is $r'=3$, from which we have
\be e_{j'}\:P^r_j\;=\;P^r_j\:e_{j'}\;=\;0,
\quad\mbox{if $j\le j'\le j\pl r\mi3$}\,.\ee

The fusion operators also satisfy the Hecke-like identity
\be S_{r-1}^2\:P^r_{j+1}\:P^r_j\:P^r_{j+1}\:-\:P^r_{j+1}\;=\;
S_{r-1}^2\:P^r_j\:P^r_{j+1}\:P^r_j\:-\:P^r_j\;=\;S_{r-2}\:S_r\:P^{r+1}_j \,.
\refstepcounter{equation}\el{Y}\addtocounter{equation}{-1}\ee

\sub{Fused Row Operators}
We now introduce operators $\Y_j^r(u)$ and $\Yt^r_j(u)$, for
$r\in\{1,\ldots,\min(\n\mi1,L\pl1)\}$ and $j\in\{1,\ldots,L\pl2\mi r\}$,
which correspond to fused rows of faces. These operators are related
to products of $r\mi1$ face operators, for $r\ge2$, by
\rnc{\theequation}{\arabic{section}.\arabic{equation}\alph{subeq}}
\setcounter{subeq}{1}\setlength{\unitlength}{6mm}
\be\ba{r@{}c@{}l}
\ru{2.5}\ds\prod_{k=1}^{r-2}\!s_k(-u)\;\Y^r_j(u)&=&
P^r_{j+1}\:X_j(u\mi(r\mi2)\l)\,\ldots\,X_{j+r-3}(u\mi\l)\:X_{j+r-2}(u)\\
&=&X_j(u)\:X_{j+1}(u\mi\l)\,\ldots\,X_{j+r-2}(u\mi(r\mi2)\l)\:P^r_j\\
&\text{1.8}{8.3}{0.9}{4}{}{=}&\;\;\;
\bpic(5,7.5)
\multiput(0,3)(1,-1){2}{\line(1,1){4}}\put(0,3){\line(1,-1){2}}
\multiput(1,4)(2,2){2}{\line(1,-1){1}}\multiput(4,3)(0,4){2}{\line(1,-1){1}}
\multiput(2,1)(0,2){2}{\line(1,0){2}}\put(4,1){\line(1,1){1}}
\put(0.8,3.32){\pp{}{u-}}\put(1.04,2.9){\pp{}{(r\mi2)\l}}
\put(4,6){\pp{}{u}}
\multiput(0,0.5)(0,0.25){29}{\pp{}{.}}
\multiput(1,0.5)(0,0.25){6}{\pp{}{.}}
\multiput(1,4.25)(0,0.25){14}{\pp{}{.}}
\multiput(4,0.5)(0,0.25){2}{\pp{}{.}}
\multiput(4,3.25)(0,0.25){7}{\pp{}{.}}
\multiput(4,7.25)(0,0.25){2}{\pp{}{.}}
\multiput(5,0.5)(0,0.25){29}{\pp{}{.}}
\put(-0.1,0){\pp{b}{j-1}}\put(1,0){\pp{b}{j}}
\put(4.3,0){\pp{br}{j+r-2}}\put(4.7,0){\pp{bl}{j+r-1}}\epic
\text{2.6}{7.5}{1.3}{4}{}{=}
\bpic(5,7.5)
\multiput(0,2)(1,-1){2}{\line(1,1){4}}
\put(3,7){\line(1,-1){2}}
\multiput(1,3)(2,2){2}{\line(1,-1){1}}\multiput(0,2)(0,4){2}{\line(1,-1){1}}
\multiput(1,5)(0,2){2}{\line(1,0){2}}\put(0,6){\line(1,1){1}}
\put(3.8,5.32){\pp{}{u-}}\put(4.04,4.9){\pp{}{(r\mi2)\l}}
\put(1,2){\pp{}{u}}
\multiput(0,0.5)(0,0.25){29}{\pp{}{.}}
\multiput(1,0.5)(0,0.25){2}{\pp{}{.}}
\multiput(1,3.25)(0,0.25){7}{\pp{}{.}}
\multiput(1,7.25)(0,0.25){2}{\pp{}{.}}
\multiput(4,0.5)(0,0.25){14}{\pp{}{.}}
\multiput(4,6.25)(0,0.25){6}{\pp{}{.}}
\multiput(5,0.5)(0,0.25){29}{\pp{}{.}}
\put(-0.1,0){\pp{b}{j-1}}\put(1,0){\pp{b}{j}}
\put(4.3,0){\pp{br}{j+r-2}}\put(4.7,0){\pp{bl}{j+r-1}}\epic\ea\ee
and
\be\stepcounter{subeq}\addtocounter{equation}{-1}\ba{r@{}c@{}l}
\ru{2.5}\ds\prod_{k=0}^{r-3}\!s_{-k}(-u)\;\Yt^r_j(u)&=&
P^r_j\:X_{j+r-2}(u)\:X_{j+r-3}(u\pl\l)\,\ldots\,X_j(u\pl(r\mi2)\l)\\
&=&X_{j+r-2}(u\pl(r\mi2)\l)\,\ldots\,X_{j+1}(u\pl\l)\:X_j(u)\:P^r_{j+1}\\
&\text{1.8}{8.3}{0.9}{4}{}{=}&\;\;\;
\bpic(5,7.5)
\multiput(5,3)(-1,-1){2}{\line(-1,1){4}}\put(5,3){\line(-1,-1){2}}
\multiput(3,3)(-2,2){2}{\line(1,1){1}}\multiput(0,2)(0,4){2}{\line(1,1){1}}
\multiput(1,1)(0,2){2}{\line(1,0){2}}\put(0,2){\line(1,-1){1}}
\put(0.8,6.32){\pp{}{u+}}\put(1.04,5.9){\pp{}{(r\mi2)\l}}
\put(4,3){\pp{}{u}}
\multiput(0,0.5)(0,0.25){29}{\pp{}{.}}
\multiput(1,0.5)(0,0.25){2}{\pp{}{.}}
\multiput(1,3.25)(0,0.25){7}{\pp{}{.}}
\multiput(1,7.25)(0,0.25){2}{\pp{}{.}}
\multiput(4,0.5)(0,0.25){6}{\pp{}{.}}
\multiput(4,4.25)(0,0.25){14}{\pp{}{.}}
\multiput(5,0.5)(0,0.25){29}{\pp{}{.}}
\put(-0.1,0){\pp{b}{j-1}}\put(1,0){\pp{b}{j}}
\put(4.3,0){\pp{br}{j+r-2}}\put(4.7,0){\pp{bl}{j+r-1}}\epic
\text{2.6}{7.5}{1.3}{4}{}{=}
\bpic(5,7.5)
\multiput(5,2)(-1,-1){2}{\line(-1,1){4}}\put(0,5){\line(1,1){2}}
\multiput(3,2)(-2,2){2}{\line(1,1){1}}\multiput(4,1)(0,4){2}{\line(1,1){1}}
\multiput(2,5)(0,2){2}{\line(1,0){2}}\put(4,7){\line(1,-1){1}}
\put(3.8,2.32){\pp{}{u+}}\put(4.04,1.9){\pp{}{(r\mi2)\l}}
\put(1,5){\pp{}{u}}
\multiput(0,0.5)(0,0.25){29}{\pp{}{.}}
\multiput(1,0.5)(0,0.25){14}{\pp{}{.}}
\multiput(1,6.25)(0,0.25){6}{\pp{}{.}}
\multiput(4,0.5)(0,0.25){2}{\pp{}{.}}
\multiput(4,3.25)(0,0.25){7}{\pp{}{.}}
\multiput(4,7.25)(0,0.25){2}{\pp{}{.}}
\multiput(5,0.5)(0,0.25){29}{\pp{}{.}}
\put(-0.1,0){\pp{b}{j-1}}\put(1,0){\pp{b}{j}}
\put(4.3,0){\pp{br}{j+r-2}}\put(4.7,0){\pp{bl}{j+r-1}}
\put(5.9,3.7){\p{}{.}}\epic\ea\ee
\rnc{\theequation}{\arabic{section}.\arabic{equation}}
The second (or fourth) equalities in \er{Y} imply
the push-through relations 
\be\el{PTR}\ba{l}\ru{2.5}
\Y^r_j(u)\;=\;P^r_{j+1}\:\Y^r_j(u)\;=\;\Y^r_j(u)\:P^r_j\\
\Yt^r_j(u)\;=\;P^r_j\:\Yt^r_j(u)\;=\;\Yt^r_j(u)\:P^r_{j+1}\,,\ea\ee
and can be derived using~\er{PE} and 
repeated application of~\er{TLYBE}.

The fused row operators can be written in terms of fusion operators as 
\be\el{YA}\ba{l}\ru{2.5}
\Y^r_j(u)\;=\;S_{r-1}\:s_1(u)\:P^r_{j+1}\:P^r_j\;-\;S_r\:s_0(u)\:P^{r+1}_j\\
\Yt^r_j(u)\;=\;
S_{r-1}\:s_{r-1}(u)\:P^r_j\:P^r_{j+1}\;-\;S_r\:s_{r-2}(u)\:P^{r+1}_j
\,.\ea\ee
The equivalence of \er{Y} and \er{YA} follows from several of the
properties of fusion operators listed in Section~\ref{SFO}.

We note, as examples, that for the first two fusion levels,
\be\Y^1_j(u)\;=\;-s_0(u)\:I,\quad\Yt^1_j(u)\;=\;s_1(-u)\:I,\quad
\Y^2_j(u)\;=\;\Yt^2_j(u)\;=\;X_j(u)\,.\ee

An important property of the fused row operators is that they satisfy
the mixed Yang-Baxter equations,
\bea\el{YBE1}\ru{2.5}X_j(u\mi v)\;\Y^r_{j+1}(u)\;\Y^r_j(v)&=&
\Y^r_{j+1}(v)\;\Y^r_j(u)\;X_{j+r-1}(u\mi v)\\
\el{YBE2}\ru{2.5}X_{j+r-1}(u\mi v)\;\Yt^r_j(u)\;\Yt^r_{j+1}(v)&=&
\Yt^r_j(v)\;\Yt^r_{j+1}(u)\;X_j(u\mi v)\\
\el{YBE3}\Y^r_j(u)\;X_{j+r-1}(u\pl v)\;\Yt^r_j(v)&=&
\Yt^r_{j+1}(v)\;X_j(u\pl v)\;\Y^r_{j+1}(u)\,.\eea These equations can
be obtained using \er{Y} and repeated application of~\er{TLYBE}.

The fused row operators also satisfy the product identities
\be\el{PI}\ba{l}\ru{2.5}\Y^r_j(u)\:\Yt^r_j(v)\;=\;
s_1(u)\:s_{r-1}(v)\:P^r_{j+1}\;-\;S_r\:s_0(u\pl v)\:P^{r+1}_j\\
\Yt^r_j(u)\:\Y^r_j(v)\;=\;s_{r-1}(u)\:s_1(v)\:P^r_j\;-\;S_r\:s_0(u\pl v)\:
P^{r+1}_j\,,\ea\ee
these being most easily derived using \er{YA} and
properties of the fusion operators.

\sub{Boundary Operators}
We now introduce boundary operators $K^r_j(u,\x)$, with $\x\in\C$,
as products of two fused row operators,
\be\el{K}K^r_j(u,\x)\;=\;-\,\Y^r_j(u\mi\l\mi\x)\;\Yt^r_j(u\pl\l\pl\xi)\,.\ee
These operators correspond in a lattice model to boundary Boltzmann
weights and in this context $\x$ is a boundary field parameter.

{}From~\er{PI}, we see that the boundary operators can
be written in terms of fusion operators as
\be\el{K1}K^r_j(u,\x)\;=\;
s_0(\x\mi u)\:s_r(\x\pl u)\:P^r_{j+1}\;+\;S_r\:s_0(2u)\:P^{r+1}_j\,.\ee
An alternative expression, which follows using \er{FO} on the
second term on the right side of \er{K1}, is
\be\el{K2}K^r_j(u,\x)\;=\;
s_0(\x\pl u)\:s_r(\x\mi u)\:P^{r+1}_j\;+\;\frac{S_{r-1}}{S_r}\:
s_0(\xi\mi u)\:s_r(\x\pl u)\:P^r_{j+1}\:e_j\:P^r_{j+1}\,.\ee
We see, as examples, that for the first two fusion levels,
\be K^1_j(u,\x)\:=\:s_0(\x\pl u)\,s_1(\x\mi u)\,I,\quad
K^2_j(u,\x)\:=\:s_0(\x\pl u)\,s_2(\x\mi u)\,I\,-\,s_0(2u)\,e_j.\ee

The key property of the face and boundary operators is that they 
satisfy the operator form of the boundary Yang-Baxter equation,
\be\el{TLBYBE}\ba{l}\ru{2.5}
X_j(u\mi v)\;K^r_{j+1}(u,\x)\;X_j(u\pl v)\;K^r_{j+1}(v,\x)\;=\\
\qquad\qquad\qquad\qquad\qquad\qquad
K^r_{j+1}(v,\x)\;X_j(u\pl v)\;K^r_{j+1}(u,\x)\;X_j(u\mi v)\,.\ea\ee
This can be proved by substituting~\er{K} into the left side of \er{TLBYBE},
using \er{YBE3} followed by \er{YBE1} to bring $X_{j+r-1}(u\mi v)$ 
adjacent to
$X_{j+r-1}(u\pl v)$, interchanging the order of these face operators 
using~\er{XC}, and then using
\er{YBE2} followed by \er{YBE3} to give the right side of \er{TLBYBE}. 

In terms of the construction procedure of \cite{BehPea96}, the boundary 
operators $K^r_{j}(u,\x)$ can be considered as a family of solutions,
one solution for each value of $r$, of~\er{TLBYBE} with given $X_j(u)$, these
solutions having been constructed by starting with the identity
solution of~\er{TLBYBE} and adding a fused double row of faces of
width $r\mi1$.
  
The boundary operators also satisfy the operator form of the 
boundary inversion relation,
\be\el{TLBIR}K^r_j(u,\x)\;K^r_j(-u,\x)\;=\;s_0(\x\mi u)\:s_0(\x\pl u)\:
s_r(\x\mi u)\:s_r(\x\pl u)\:P^r_{j+1}\,,\ee
this being most easily obtained using~\er{K1} and properties of the fusion
operators.

\newpage
\sect{Lattice Models Based on Graphs} 
In this section, we consider representations of the Temperley-Lieb
algebra involving a graph and the associated graph-based lattice
models.  Our general treatment is motivated by that introduced in
\cite{Pas87a,OwcBax87} and, when considering fusion for these models,
our approach is also motivated by certain results of
\cite{DatJimMiwOka86,DatJimMiwOka87a,DatJimKunMiwOka88,ZhoPea94a}.

A graph-based lattice model of the type to be considered here can be
associated with any pair $\G$ and $\psi$, where $\G$ is a connected
graph containing only unoriented, single edges and $\psi$ is an
eigenvector of the adjacency matrix of $\G$ with all nonzero entries.
All of the formalism and results of Section~3 are applicable to any
choice of $\G$ and $\psi$, and only in Section~4 will we eventually
specialize $\G$ to be an $A$, $D$ or $E$ Dynkin diagram and $\psi$ to
be the Perron-Frobenius eigenvector of its adjacency matrix, thus
giving the critical unitary $A$--$D$--$E$ models.

In a lattice model based on $\G$, a spin is attached to each site of a
two-dimensional square lattice, with the possible states of each spin being
the nodes of $\G$ and there being a lattice adjacency condition stipulating
that, in any assignment of spin states to the lattice, the states on each pair
of nearest-neighbor sites must correspond to an edge of $\G$. These
models are interaction-round-a-face models, so that a bulk Boltzmann weight is
associated with each set of four spin states adjacent around a square face. 
Here, we shall also obtain and use sets of boundary weights, each of these
weights being associated with three adjacent spin states.  The partition
function of the model on a cylinder
is then the sum, over all possible spin assignments, of
products of Boltzmann weights, with each square face in the bulk of the
lattice contributing a bulk weight and each alternate triplet of
neighboring sites on the boundaries contributing a boundary weight.

The key property of the boundary weights obtained in this section is that 
they satisfy the boundary Yang-Baxter equation for interaction-round-a-face
models.  This equation was first used in
\cite{BehPeaObr96,Kul96} and is based on the reflection equation
introduced in \cite{Che84}.  However, we note that 
the boundary weights obtained here depend on certain
additional fusion indices, which gives them and the 
boundary Yang-Baxter equation they satisfy a somewhat
more general form than the form of those used in all previous studies.
We shall show that a further important property of the three-spin
boundary weights used here is that, at a certain point, they 
simultaneously decompose into more natural two-spin boundary edge weights.  

Finally, using these and other local properties of the bulk and boundary
weights, we shall identify various symmetry properties of the partition
function and transfer matrices, including the invariance of the partition
function under interchanging parts or all of the left and right boundaries,
and the facts that the transfer matrices form a commuting family
and, in certain cases, have all nonnegative eigenvalues.

\sub{Graphs and Paths} 
Throughout Section 3, we shall be considering a finite graph $\G$ with
an associated adjacency matrix $G$. We require that $\G$ contain only
unoriented, single edges, implying that $G$ is symmetric and that each
of its nonzero entries is~$1$.  In fact, the formalism of Section~3
can be generalized straightforwardly to encompass graphs with multiple
edges, but since this is not needed for the $A$--$D$--$E$ cases of
primary interest, we shall for simplicity restrict our attention to
graphs with only single edges.

We also require that $\G$ be connected, implying that $G$ is
indecomposable.  Perron-Frobenius theory then implies that $G$ has a
unique maximum eigenvalue with an associated eigenvector whose entries
are all positive.

We shall denote the set of all $r$-point paths on $\G$, for $r\in\Z_{\ge1}$,
by $\G^r$; that is,
\be\G^r\;=\;\Bigl\{(a_0,\ldots,a_{r-1})\,\Big|\,a_j\in\G,\;
\prod_{j=0}^{r-2}G_{a_ja_{j+1}}=1\Bigr\}\,.\ee
We note that $\G^1$ corresponds to the set of nodes of $\G$
and that $\G^2$ is the set of edges of $\G$. 

We shall also denote the set of all $r$-point paths between 
$a$ and $b$ in $\G$ by $\G^r_{ab}$; that is,
\be\G^r_{ab}\;=\;\Bigl\{(a_0,\ldots,a_{r-1})\in\G^r\,\Big|\,a_0=a,
\:a_{r-1}=b\Bigr\}\,.\ee
It follows that
\be|\G^r_{ab}|\;=\;(G^{r-1})_{ab}\,.\ee

\sub{Graph Representations of the Temperley-Lieb Algebra}
A graph representation involving $\G$ of the Temperley-Lieb algebra $\T(L,\l)$ 
exists for each $\l$ for which $2\cos\!\l$ is an eigenvalue of $G$ with an 
associated eigenvector $\psi$ whose entries are all nonzero; 
that is, for each $\l$ for which there exists a vector $\psi$ satisfying
\be\el{GE}\sum_{b\:\in\:\G}G_{ab}\,\psi_b\;=\;
2\cos\!\l\;\psi_a\mbox{ \ and \ } 
\psi_a\ne0\,,\quad\mbox{for each $a\in\G$.}\ee
There is always at least one such case, namely that provided by the
Perron-Frobenius eigenvector and eigenvalue.  We shall assume,
throughout the rest of Section~3, that a fixed choice of $\l$ and
$\psi$ has been made.  We note that, since $G$ is symmetric,
$\cos\!\l$ must be real so that each $\psi_a$ can, and will, be
assumed to be real.  In the rest of this paper, we shall often use
$\psi_a^\hf$ and $\psi_a^\qt$, which we take to be positive for
positive $\psi_a$ or to have arguments $\pi/2$ and $\pi/4$
respectively for negative $\psi_a$.

The elements of the representation of $\T(L,\l)$ associated with $\G$ and 
$\psi$ are matrices 
with rows and columns labeled by the paths of $\G^{L+2}$, with 
the generators $e_j$ being defined by
\be\el{e}
e_{j\:\raisebox{-0.2ex}{$\ss(a_0,\ldots,a_{L+1}),(b_0,\ldots,b_{L+1})$}}
\;=\;\frac{\psi_{a_j}^\hf\;\psi_{b_j}^\hf}{\psi_{a_{j-1}}}\;\,
\delta_{a_{j-1}a_{j+1}}\,\prod_{k=0\atop k\ne j}^{L+1}\delta_{a_k b_k}\,.\ee
It follows straightforwardly that the defining relations~\er{TLA} of
$\T(L,\l)$ are satisfied, with the first relation depending on~\er{GE}.

\sub{Bulk Weights} 
We now proceed to a consideration of the lattice model based on the
graph $\G$ and associated with the adjacency matrix eigenvector
$\psi$.

The bulk weights for this model are given explicitly, for each
$(a,b,c,d,a)\in\G^5$, by
\be\el{W}\W{a}{b}{c}{d}{u}\;=\;s_1(-u)\:\d_{ac}\;+\;
\frac{s_0(u)\:\psi_a^\hf\:\psi_c^\hf}{\psi_b}\:\d_{bd}\,,\ee
where $u$ is the spectral parameter and $\l$, on which $s$ depends
through \er{sD}, is the crossing parameter.  The spectral parameter
can be considered as a measure of anisotropy, with $u=\l/2$ being an
isotropic point and $u=0$ and $u=\l$ being completely anisotropic
points.  We note that the number of bulk weights is $\mbox{tr}(G^4)$.

We shall represent the bulk weights diagrammatically as
\setlength{\unitlength}{8mm}
\be\raisebox{-1\unitlength}[1\unitlength][1\unitlength]
{\text{3.2}{2}{1.6}{1}{}{\W{a}{b}{c}{d}{u}}
\text{1.4}{2}{0.7}{1}{}{=}
\bpic(2,2)\multiput(0.5,0.5)(1,0){2}{\line(0,1){1}}
\multiput(0.5,0.5)(0,1){2}{\line(1,0){1}}
\put(0.48,0.63){\pp{bl}{\searrow}}
\put(0.45,0.45){\pp{tr}{a}}\put(1.55,0.45){\pp{tl}{b}}
\put(1.55,1.55){\pp{bl}{c}}\put(0.45,1.55){\pp{br}{d}}
\put(1,1){\pp{}{u}}\put(2.1,0.8){\p{}{.}}\epic}\ee

These bulk weights are related to the face operators~\er{X} by
\be\el{WX}\W{a}{b}{c}{d}{u}\;=\;
X_j(u)_{(e_0,\ldots,e_{j-2},d,a,b,e_{j+2},\ldots,e_{L+1}),
(e_0,\ldots,e_{j-2},d,c,b,e_{j+2},\ldots,e_{L+1})}\,,\ee
where $X_j(u)$ is taken in the graph representation of
$\T(L,\l)$.

We see that the bulk weights satisfy reflection symmetry,
\be\el{RS}\W{a}{b}{c}{d}{u}\;=\;\W{c}{b}{a}{d}{u}\;=\;\W{a}{d}{c}{b}{u},\ee
crossing symmetry,
\be\el{CS}\W{a}{b}{c}{d}{u}\;=\;
\frac{\psi_a^\hf\;\psi_c^\hf}{\psi_b^\hf\;\psi_d^\hf}\;
\W{b}{a}{d}{c}{\l\mi u},\ee
and the anisotropy property
\be\el{CAP}\W{a}{b}{c}{d}{0}\;=\;\delta_{ac}\,.\ee
It follows from \er{TLYBE}, \er{TLIR} and \er{WX} that the bulk weights
also satisfy the Yang-Baxter equation,
\setlength{\unitlength}{8mm}
\be\el{YBE}\ba{c}\ru{3.5}\ds
\sum_{\stackrel{\ru{0.6}\ss g\:\in\:\G}{\sss 
(G_{bg}G_{dg}G_{fg}=1)}}\!\!\!\!
\W{a}{b}{g}{f}{u\mi v}\,\W{b}{c}{d}{g}{u}\,\W{g}{d}{e}{f}{v}\;=\\
\ds\qquad\qquad\qquad\qquad
\sum_{\stackrel{\ru{0.6}\ss g\:\in\:\G}{\sss
(G_{ag}G_{cg}G_{eg}=1)}}\!\!\!\!
\W{b}{c}{g}{a}{v}\,\W{a}{g}{e}{f}{u}\,\W{g}{c}{d}{e}{u\mi v}\\
\bpic(4,3.6)\multiput(0.5,1.5)(1,-1){2}{\line(1,1){1}}
\multiput(0.5,1.5)(1,1){2}{\line(1,-1){1}}
\multiput(2.5,0.5)(0,1){3}{\line(1,0){1}}
\multiput(2.5,0.5)(1,0){2}{\line(0,1){2}}
\put(0.62,1.56){\pp{l}{\downarrow}}
\multiput(2.48,0.63)(0,1){2}{\pp{bl}{\searrow}}
\put(1.5,1.5){\pp{}{u-v}}\put(3,1){\pp{}{u}}\put(3,2){\pp{}{v}}
\put(0.4,1.5){\pp{r}{a}}\put(1.5,0.4){\pp{t}{b}}
\put(2.5,0.4){\pp{t}{b}}\put(3.55,0.45){\pp{tl}{c}}
\put(3.6,1.5){\pp{l}{d}}\put(3.55,2.55){\pp{bl}{e}}
\put(1.5,2.6){\pp{b}{f}}\put(2.5,2.6){\pp{b}{f}}
\put(2.5,1.5){\pp{}{\bullet}}
\multiput(1.75,0.5)(0.25,0){3}{\pp{}{.}}
\multiput(1.75,2.5)(0.25,0){3}{\pp{}{.}}\epic
\text{1}{3}{0.5}{1.5}{}{=}
\bpic(4,3)\multiput(1.5,1.5)(1,-1){2}{\line(1,1){1}}
\multiput(1.5,1.5)(1,1){2}{\line(1,-1){1}}
\multiput(0.5,0.5)(0,1){3}{\line(1,0){1}}
\multiput(0.5,0.5)(1,0){2}{\line(0,1){2}}
\put(1.62,1.56){\pp{l}{\downarrow}}
\multiput(0.48,0.63)(0,1){2}{\pp{bl}{\searrow}}
\put(2.5,1.5){\pp{}{u-v}}\put(1,1){\pp{}{v}}\put(1,2){\pp{}{u}}
\put(0.4,1.5){\pp{r}{a}}\put(0.45,0.45){\pp{tr}{b}}
\put(1.5,0.4){\pp{t}{c}}\put(2.5,0.4){\pp{t}{c}}
\put(3.6,1.5){\pp{l}{d}}\put(1.5,2.6){\pp{b}{e}}
\put(2.5,2.6){\pp{b}{e}}\put(0.45,2.55){\pp{br}{f}}
\put(1.5,1.5){\pp{}{\bullet}}
\multiput(1.75,0.5)(0.25,0){3}{\pp{}{.}}
\multiput(1.75,2.5)(0.25,0){3}{\pp{}{.}}
\put(4.2,1.3){\p{}{,}}\epic\ea\ee
for each $(a,b,c,d,e,f,a)\in\G^7$, and the inversion relation
\setlength{\unitlength}{7mm}
\be\el{IR}\ba{c}\ds\text{10.8}{3}{5.4}{1.2}{}{
\sum_{\stackrel{\ru{0.5}\ss e\:\in\:\G}{\sss(G_{be}G_{de}=1)}}\!\!
\W{a}{b}{e}{d}{\!\mi u}\;\W{e}{b}{c}{d}{u}\;=}
\bpic(5,3)
\multiput(0.5,1.5)(3,-1){2}{\line(1,1){1}}
\multiput(0.5,1.5)(3,1){2}{\line(1,-1){1}}
\put(1.5,2.5){\line(1,-1){2}}\put(1.5,0.5){\line(1,1){2}}
\multiput(1.72,0.5)(0.26,0){7}{\pp{}{.}}
\multiput(1.72,2.5)(0.26,0){7}{\pp{}{.}}
\multiput(0.62,1.54)(2,0){2}{\pp{l}{\downarrow}}
\put(1.35,1.5){\pp{}{-u}}\put(3.5,1.5){\pp{}{u}}
\put(0.4,1.5){\pp{r}{a}}\multiput(1.5,0.4)(2,0){2}{\pp{t}{b}}
\put(4.6,1.5){\pp{l}{c}}\multiput(1.5,2.6)(2,0){2}{\pp{b}{d}}
\put(2.5,1.5){\pp{}{\bullet}}\epic\\
\qquad\qquad=\;s_1(u)\,s_1(-u)\:\delta_{ac}\,,\ea\ee
for each $(a,b,c,d,a)\in\G^5$.
In these and all subsequent diagrams, solid circles are used to indicate 
spins whose states are summed over and dotted lines are used 
to connect identical spins.

\sub{Fusion}
We now consider various objects related to the process of fusion
in graph-based lattice models.

\subsub{Maximum Fusion Level}\label{MFL}
Having chosen a graph $\G$ and an adjacency matrix eigenvector with
eigenvalue $2\cos\!\l$, the maximum fusion level $g$ is then
determined by $\l$ according to \er{g}.  We note that although $\l$
itself is only determined by the eigenvector up to sign and shifts of
$2\pi$, such changes in $\l$ do not affect $g$ or any other properties
of interest.

As we shall see in more detail in Section~4.1, if $\G$ is an $A$, $D$
or $E$ Dynkin diagram, then $g$ is the Coxeter number of $\G$, since
any eigenvector of $G$ with all nonzero entries has an eigenvalue
$2\cos(k\pi/h)$, where $h\in\Z_{\ge2}$ is the Coxeter number of $\G$
and $k\in\{1,\ldots,h\mi1\}$ is a Coxeter exponent coprime to $h$.

Although the $A$--$D$--$E$ cases are of primary interest, simply as a
further example, we briefly discuss the $A^{(1)}$, $D^{(1)}$ and
$E^{(1)}$ Dynkin diagrams of the affine Lie algebras.  If $\G$ is one
of these graphs, then any eigenvalue of $G$ can be written as
$2\cos(k\pi/h)$, where $h$ is the Coxeter number of $\G$ and
$k\in\{0,\ldots,h\}$ is a Coxeter exponent.  For all of these graphs,
$k=0$ is a Coxeter exponent, giving a maximal eigenvalue of $2$, and
in some cases $k=h$ is also a Coxeter exponent, giving a minimal
eigenvalue of $-2$, so if either of these eigenvalues is chosen, then
$g=\infty$ (and, incidentally, the second case of \er{sD} applies).
However, if an eigenvalue corresponding to a Coxeter exponent $0<k<h$
is chosen, then $g$ is finite with $g\le h$.

\subsub{Fusion Matrices}
We now introduce, for each $r\in\{1,\ldots,\n\}$ and
$a,b\in\G$ satisfying $(G^{r\mi1})_{ab}>0$,
a fusion matrix $P^r(a,b)$ with
rows and columns labeled by the paths of $\G^r_{ab}$ and entries given by
\be\el{FM}\ba{l}\ru{2}
P^r(a,b)_{(a,c_1,\ldots,c_{r\mi2},b),(a,d_1,\ldots,d_{r\mi2},b)}\;=\\
\qquad\qquad\qquad\qquad\qquad\qquad
\left\{\ba{l}\ru{2}\,1\,,\quad r=1\\
\,P^r_{1\:\raisebox{-0.2ex}{$\ss(a,c_1,\ldots,c_{r\mi2},b),
(a,d_1,\ldots,d_{r\mi2},b)$}}\,,\quad r=2,\ldots,g\,,\ea\right.\ea\ee
where the fusion operator $P^r_1$ is defined by \er{FO} and taken in the
graph representation of $\T(r\mi2,\l)$.  

It follows from~\er{PP} that each fusion matrix is a projector, 
\be\el{FMP}P^r(a,b)^2\;=\;P^r(a,b)\,.\ee 
It also follows, from \er{FO}, \er{WX} and the first equality of \er{RS}, 
that each fusion matrix is symmetric,
\be\el{FMT}P^r(a,b)^T\;=\;P^r(a,b)\,\ee
and, from \er{PE} and \er{WX}, that $a$ and $b$ can interchanged according to
\be\el{FMI}
P^r(a,b)_{(a,c_1,\ldots,c_{r\mi2},b),(a,d_1,\ldots,d_{r\mi2},b)}\;=\;
P^r(b,a)_{(b,c_{r\mi2},\ldots,c_1,a),(b,d_{r\mi2},\ldots,d_1,a)}\,.\ee
We note that the fusion matrices also satisfy more general 
projection-type relations corresponding to those, \er{PPG}, satisfied by
the fusion operators.

\subsub{Fused Adjacency Matrices}
We next  introduce fused adjacency matrices, $\r^1,\ldots,\r^\n$, 
which are defined by the $\slt$ recursion,
\be\el{RR}\ba{c}\r^1\;=\,I\,;\qquad\r^2\,=\,G\,;
\qquad\r^r\;=\;G\,\r^{r-1}\,-\,\r^{r-2}\,,
\quad r=3,\ldots,\n\,.\ea\ee
We see that each fused adjacency matrix is a polynomial in $G$ 
and hence that the set of these matrices is mutually commuting.   
In fact, the polynomial form is
given by the Type II Chebyshev polynomials ${\cal U}_r$,
\be\el{CP}\r^r\;=\;{\cal U}_{r-1}(G/2)\,.\ee
We note that at various places in Section~4, we shall be considering the
fused adjacency matrices of two different graphs and in these cases we
shall explicitly indicate the dependence on the graph as $\r(\G)^r$.

The main relevance of the fused adjacency 
matrices at this point is that their entries give
the ranks of the fusion matrices, according to
\be\el{r}\r^r_{ab}\;=\;\left\{\ba{l}\ru{2}0\,,\quad(G^{r-1})_{ab}=0\\
\mbox{rank}P^r(a,b)\,,\quad(G^{r-1})_{ab}>0\,.\ea\right.\ee
This result can be proved by defining matrices $\tilde{\r}^r$ 
with entries $\tilde{\r}^r_{ab}$ given by the right side of~\er{r}
and showing that these matrices satisfy relations~\er{RR}, so that
$\tilde{\r}^r=\r^r$.  Showing that $\tilde{\r}^r$ satisfy the
initial conditions of~\er{RR} is straightforward.
Meanwhile, showing that $\tilde{\r}^r$ satisfy
the recursion relation of~\er{RR} can be done by
using the fact that the rank of any projector is given by its trace
(since by idempotence each eigenvalue is either $0$ or $1$, 
the rank is the number
of eigenvalues which are $1$ and the trace is the sum of eigenvalues)
and then using definitions~\er{X}, \er{e} and \er{FM}, the recursion 
relation of~\er{FO}, which is needed twice, relations~\er{PP} and~\er{GE},
and general properties of the trace and of the numbers $S_r$.

We note that \er{r} also implies that $\r^r_{ab}$ are nonnegative
integers.  In fact, the property that the entries of $\r^r$ are
integers follows immediately from the properties that the entries of
$G$ are integers and that each $\r^r$ is a polynomial, with integer
coefficients, in $G$, but the property that the entries of $\r^r$ are
also nonnegative is less trivial and depends on the restriction $r\le
g$ and on some of the additionally-assumed properties of $G$, such as
its being symmetric. Since the result of nonnegativity is of some
interest in its own right, we note that \er{r} specifically implies
that if $G$ is any symmetric indecomposable matrix with each entry in
$\{0,1\}$ and if $F^r$ is defined in terms of $G$ by \er{CP}, then
each entry of $F^r$ is a nonnegative integer for any
$r\in\{1,\ldots,g_{\mbox{\scriptsize max}}\}$, where
$g_{\mbox{\scriptsize max}}$ is the largest of the values $g$
corresponding to the eigenvectors of $G$ with all nonzero entries.

\subsub{Fusion Vectors} 
It can be shown using standard results of linear algebra that any
symmetric idempotent matrix $P=P^T=P^2$ with complex entries can be
orthonormally diagonalized; that is, written as $P=UD\,U^{-1}$, where
the matrix of eigenvectors $U$ satisfies $U^{-1}=U^T$ and $D$ has
$\mbox{rank}P$ $1$'s on the diagonal and all other entries $0$.  We
note that if $P=P^*$ then this simply amounts to an orthonormal
diagonalization of a real symmetric matrix, but that if $P\ne P^*$
then it differs from a unitary diagonalization and that, in fact, $P$
is then not normal and a unitary diagonalization is not possible.

{}From this result, it follows that each fusion matrix $P^r(a,b)$ can be
decomposed using $\r^r_{ab}$ orthonormal eigenvectors with eigenvalue
$1$.  We shall denote, for $r\in\{1,\ldots,\n\}$ and $a,b\in\G$
satisfying $\r^r_{ab}>0$, such eigenvectors as $U^r(a,b)_\a$, where
$\a=1,\ldots,\r^r_{ab}$, and refer to these as fusion vectors.  

The decomposition and orthonormality are then
\be\el{FMD}\sum_{\a=1}^{\r^r_{ab}}\;U^r(a,b)_\a\:\:U^r(a,b)_\a^{\;\:T}
\;=\;P^r(a,b)\ee and
\be\el{FMON}U^r(a,b)_\a^{\;\:T}\:\:U^r(a,b)_{\a'}\;=\;\d_{\a\a'}\,,\ee
or, more explicitly,
\[\sum_{\a=1}^{\r^r_{ab}}\,
U^r(a,b)_{\a,(a,c_1,\ldots,c_{r\mi2},b)}\;
U^r(a,b)_{\a,(a,d_1,\ldots,d_{r\mi2},b)}\;=\;
P^r(a,b)_{(a,c_1,\ldots,c_{r\mi2},b),(a,d_1,\ldots,d_{r\mi2},b)}\]
and
\[\sum_{(a,c_1,\ldots,c_{r\mi2},b)\:\in\:\G^r_{ab}}
\!\!\!\!U^r(a,b)_{\a,(a,c_1,\ldots,c_{r\mi2},b)}\;
U^{r}(a,b)_{\a',(a,c_1,\ldots,c_{r\mi2},b)}\;=\;\d_{\a\a'}\,.\]

We note, as examples, that for the first two fusion levels,
\be P^1(a,a)_{(a),(a)}\;=\;\pm\,U^1(a,a)_{1,(a)}\;=\;1\,,\quad\mbox{for
each }a\in\G\,,\ee
and
\be P^2(a,b)_{(a,b),(a,b)}\;=\;\pm\,U^2(a,b)_{1,(a,b)}\;=\;1\,,
\quad\mbox{for each }(a,b)\in\G^2\,.\ee

We shall assume that a specific choice of $U^r(a,b)_\a$ has been made.
All other possible choices are then given by transformations,
\be\el{UGT}U^r(a,b)_\a\;\mapsto\;\sum_{\a'=1}^{\r^r_{ab}}\;R^r(a,b)_{\a\a'}\;
U^r(a,b)_{\a'}\,,\ee
where $R^r(a,b)$ is an orthonormal matrix,
\be\el{RGT}\sum_{\a''=1}^{\r^r_{ab}}\:
R^r(a,b)_{\a\a''}\,R^r(a,b)_{\a'\a''}\;=\;
\d_{\a\a'}\,.\ee
We shall see that all of the lattice model properties of interest will
be invariant under such transformations and thus independent of the
choice of fusion vectors.

\sub{Boundary Weights}\label{BWS}
We now use the boundary operators \er{K} to obtain a set of boundary
Boltzmann weights for the lattice model for each pair $(r,a)$, where
$r\in\{1,\ldots,\n\mi1\}$ is a fusion level and $a$ is a node of $\G$.
It is thus natural to regard these pairs as labeling the boundary
conditions,
\be\Bigl\{\mbox{boundary conditions}\Bigr\}\;\longleftrightarrow\;
\Bigl\{(r,a)\mid r\in\{1,\ldots,\n\mi1\},\;a\in\G\Bigr\}\,.\ee
The $(r,a)$ boundary weights are given, for each $(b,c,d)\in\G^3$ with
$\r^r_{ba}\,\r^r_{da}>0$ and $\b\in\{1,\ldots,\r^r_{ba}\}$, 
$\d\in\{1,\ldots,\r^r_{da}\}$, by
\be\el{B}\ba{l}\ru{5}\B{ra}{b}{\b}{c}{d}{\d}{u,\x}\;=\\
\ds\frac{\psi_c^\hf}{s_0(2\x)\:\psi_{b}^\qt\:\psi_{d}^\qt}
\sum_{\stackrel{\ru{0.8}\ss(b,e_1,\ldots,e_{r\mi2},a)\:\in\:\G^r_{ba}}
{\ss(d,f_1,\ldots,f_{r\mi2},a)\:\in\:\G^r_{da}}}
\!\!\!\!\!U^r(b,a)_{\b,(b,e_1,\ldots,e_{r\mi2},a)}\,
U^r(d,a)_{\d,(d,f_1,\ldots,f_{r\mi2},a)}\\
\qquad\qquad\qquad\qquad\qquad\qquad\qquad\qquad\times\;
K^r_1(u,\x)_{(c,b,e_1,\ldots,e_{r\mi2},a),
(c,d,f_1,\ldots,f_{r\mi2},a)}\,,\ea\ee
where $K^r_1(u,\x)$ is taken in the graph representation of
$\T(r\mi1,\l)$ and $\x$ corresponds to a boundary field.
We shall represent the boundary weights diagrammatically as
\setlength{\unitlength}{6mm}
\be\raisebox{-1.5\unitlength}[1.5\unitlength][1.5\unitlength]{
\text{4.4}{3}{2.2}{1.5}{}{\B{ra}{b}{\b}{c}{d}{\d}{u,\x}}
\text{1.5}{3}{0.9}{1.5}{}{=}
\bpic(2.5,3)
\multiput(2,0.5)(0,0.3){7}{\line(0,1){0.2}}
\multiput(1.5,0.5)(0,2){2}{\line(1,0){0.5}}
\put(0.5,1.5){\line(1,-1){1}}\put(0.5,1.5){\line(1,1){1}}
\put(1.5,1.7){\pp{}{r,\,a}}\put(1.5,1.3){\pp{}{u,\,\x}}
\put(1.45,0.47){\pp{tr}{b}}\put(0.4,1.5){\pp{r}{c}}
\put(1.45,2.53){\pp{br}{d}}
\put(2.1,0.5){\pp{tl}{\b}}\put(2.1,2.5){\pp{bl}{\d}}
\put(2.7,1.3){\p{}{.}}\epic}\ee

We see that
\be\mbox{number of $(r,a)$ boundary weights}\;=\;
((\r^r)^2\,G^2)_{aa}\,.\ee

We shall refer to a boundary weight \er{B} as being of diagonal type
if $b=d$ and $\b=\d$ and as being of nondiagonal type otherwise, and
we note that in the majority of previous studies involving such
boundary weights, only diagonal weights were considered.

We also note, as an example, that the $(1,a)$ boundary weights are
all diagonal and given by
\be\el{B1a}\B{1a}{a}{1}{c}{a}{1}{u,\x}\;=\;
\frac{s_0(\x\pl u)\:s_1(\x\mi u)\;\psi_c^\hf}{s_0(2\x)\;\psi_a^\hf}\,.\ee
A $(1,a)$ boundary condition is thus one in which the state of every
alternate boundary spin is fixed to be $a$, while each other boundary
spin, whose state can be any $c$ adjacent to $a$ on $\G$, is associated
with a weight proportional to $\psi_c^\hf$.  
We shall refer to such a boundary condition, in which the state of
every alternate boundary spin is fixed, as semi-fixed.

We now find that the boundary weights can also be expressed as
\setlength{\unitlength}{12mm}
\be\el{CE}\ba{l}\B{ra}{b}{\b}{c}{d}{\d}{u,\x}\;=\;-\:\Bigl(\ds s_0(2\x)\,
\prod_{k=1}^{r-2}s_{-k-1}(u\mi\x)\:s_k(u\pl\x)\Bigr)^{-1}\;\times\\
\text{4.5}{3.9}{4.6}{1.8}{r}{
\frac{\psi_d^\qt\:\psi_g^\hf}{\psi_{b}^\qt\:\psi_{a}^\hf}}
\bpic(6,3.6)\put(0,0.27){
\bpic(6,3)\multiput(0.5,0.5)(0,1){3}{\line(1,0){5}}
\multiput(0.5,0.5)(1,0){3}{\line(0,1){2}}
\multiput(4.5,0.5)(1,0){2}{\line(0,1){2}}
\multiput(0.48,0.6)(0,1){2}{\pp{bl}{\searrow}}
\multiput(1.48,0.6)(0,1){2}{\pp{bl}{\searrow}}
\multiput(4.48,0.6)(0,1){2}{\pp{bl}{\searrow}}
\put(0.6,1.17){\pp{l}{u-\x}}\put(1.45,0.9){\pp{r}{-(r\mi1)\l}}
\put(0.55,2.17){\pp{l}{-u-\x}}\put(1.45,1.9){\pp{r}{-(r\mi2)\l}}
\put(1.6,1.17){\pp{l}{ u-\x}}\put(2.45,0.9){\pp{r}{-(r\mi2)\l}}
\put(1.55,2.17){\pp{l}{-u-\x}}\put(2.45,1.9){\pp{r}{-(r\mi3)\l}}
\put(5,1){\pp{}{u-\x-\l}}\put(5,2){\pp{}{-u-\x}}
\put(0.48,0.48){\pp{tr}{b}}\put(0.44,1.5){\pp{r}{c}}
\put(0.48,2.52){\pp{br}{d}}
\put(5.52,0.48){\pp{tl}{a}}\put(5.61,1.5){\pp{l}{g}}
\put(5.52,2.52){\pp{bl}{a}}
\put(1.5,0.39){\pp{t}{e_1}}\put(2.5,0.39){\pp{t}{e_2}}
\put(4.6,0.39){\pp{t}{e_{r-2}}}
\put(1.5,2.61){\pp{b}{f_1}}\put(2.5,2.61){\pp{b}{f_2}}
\put(4.6,2.61){\pp{b}{f_{r-2}}}
\multiput(1.5,0.5)(0,1){3}{\pp{}{\bullet}}
\multiput(2.5,0.5)(0,1){3}{\pp{}{\bullet}}
\multiput(4.5,0.5)(0,1){3}{\pp{}{\bullet}}
\put(5.5,1.5){\pp{}{\bullet}}
\multiput(6,0.75)(0,0.15){11}{\pp{}{.}}
\multiput(5.75,0.55)(0,1.9){2}{\pp{}{.}}
\multiput(5.91,0.63)(0,1.74){2}{\pp{}{.}}
\put(6.25,1.4){\p{}{,}}\epic}
\put(3.1,0){\p{b}{U^r(b,a)_{\b,(b,e_1,e_2,\ldots,e_{r\mi2},a)}}}
\put(3.1,3.6){\p{t}{U^r(d,a)_{\d,(d,f_1,f_2,\ldots,f_{r\mi2},a)}}}
\epic\ea\ee
this following straightforwardly from \er{K}, \er{Y}, \er{FM}, \er{W}, 
\er{CS}, \er{FMD} and \er{FMON}. 
In this form, we see that the $(r,a)$ boundary weights can be
considered as having been constructed by starting on the right at node
$a$, essentially with the level $1$ boundary weights \er{B1a}, and
adding a fused double row of bulk weights of width $r\mi1$.

The boundary weights, together with the bulk weights \er{W},
satisfy the boundary Yang-Baxter equation,
\setlength{\unitlength}{7mm}
\be\el{BYBE}\ba{c}
\ds\sum_{\stackrel{\ru{0.7}\ss(g,h)\:\in\:\G^2}
{\sss(G_{bg}G_{dg}G_{eh}\r^r_{ha}>0)}}
\sum_{\g=1}^{\r^r_{ha}}\:
\W{c}{b}{g}{d}{u\mi v}\B{ra}{b}{\b}{g}{h}{\g}{u,\x}\;\times\\
\ru{5}
\qquad\qquad\qquad\qquad\qquad\qquad\qquad
\W{d}{g}{h}{e}{\l\mi u\mi v}\B{ra}{h}{\g}{e}{f}{\d}{v,\x}\;=\\
\ds\sum_{\stackrel{\ru{0.7}\ss(g,h)\:\in\:\G^2}
{\sss(G_{ch}G_{dg}G_{fg}\r^r_{ha}>0)}}
\sum_{\g=1}^{\r^r_{ha}}\:
\B{ra}{b}{\b}{c}{h}{\g}{v,\x}\W{d}{c}{h}{g}{\l\mi u\mi v}\;\times\\
\qquad\qquad\qquad\qquad\qquad\qquad\qquad
\B{ra}{h}{\g}{g}{f}{\d}{u,\x}\W{e}{d}{g}{f}{u\mi v}\\
\bpic(4.5,5.7)
\multiput(0.5,1.5)(2,2){2}{\line(1,-1){1}}
\put(0.5,1.5){\line(1,1){3}}
\put(1.5,2.5){\line(1,-1){2}}
\put(1.5,0.5){\line(1,1){2}}
\multiput(3.5,0.5)(0,2){3}{\line(1,0){0.5}}
\multiput(4,0.5)(0,0.3){6}{\line(0,1){0.2}}\put(4,2.3){\line(0,1){0.4}}
\multiput(4,2.8)(0,0.3){6}{\line(0,1){0.2}}
\multiput(0.62,1.56)(1,1){2}{\pp{l}{\downarrow}}
\put(1.5,1.5){\pp{}{u-v}}\put(2.56,2.5){\pp{}{\l-u-v}}
\put(3.5,1.7){\pp{}{r,\,a}}\put(3.5,1.3){\pp{}{u,\,\x}}
\put(3.5,3.7){\pp{}{r,\,a}}\put(3.5,3.3){\pp{}{v,\,\x}}
\put(3.5,0.4){\pp{t}{b}}\put(1.5,0.4){\pp{t}{b}}\put(0.4,1.5){\pp{r}{c}}
\put(1.45,2.55){\pp{br}{d}}\put(2.45,3.55){\pp{br}{e}}
\put(3.5,4.6){\pp{b}{f}}
\put(4.1,0.5){\pp{tl}{\b}}\put(4.1,4.5){\pp{bl}{\d}}
\multiput(2.5,1.5)(1,1){2}{\pp{}{\bullet}}
\put(4,2.5){\pp{}{\bullet}}
\multiput(1.72,0.5)(0.26,0){7}{\pp{}{.}}
\epic
\text{1.7}{5}{0.9}{2.5}{}{=}
\bpic(4.5,5)
\multiput(0.5,3.5)(2,-2){2}{\line(1,1){1}}
\put(0.5,3.5){\line(1,-1){3}}
\put(1.5,4.5){\line(1,-1){2}}
\put(1.5,2.5){\line(1,1){2}}
\multiput(3.5,0.5)(0,2){3}{\line(1,0){0.5}}
\multiput(4,0.5)(0,0.3){6}{\line(0,1){0.2}}
\put(4,2.3){\line(0,1){0.4}}\multiput(4,2.8)(0,0.3){6}{\line(0,1){0.2}}
\multiput(0.62,3.56)(1,-1){2}{\pp{l}{\downarrow}}
\put(1.5,3.5){\pp{}{u-v}}\put(2.56,2.5){\pp{}{\l-u-v}}
\put(3.5,1.7){\pp{}{r,\,a}}\put(3.5,1.3){\pp{}{v,\,\x}}
\put(3.5,3.7){\pp{}{r,\,a}}\put(3.5,3.3){\pp{}{u,\,\x}}
\put(3.5,0.4){\pp{t}{b}}
\put(2.45,1.45){\pp{tr}{c}}\put(1.45,2.45){\pp{tr}{d}}
\put(0.4,3.5){\pp{r}{e}}
\put(1.5,4.6){\pp{b}{f}}\put(3.5,4.6){\pp{b}{f}}
\put(4.1,0.5){\pp{tl}{\b}}\put(4.1,4.5){\pp{bl}{\d}}
\multiput(2.5,3.5)(1,-1){2}{\pp{}{\bullet}}
\put(4,2.5){\pp{}{\bullet}}
\multiput(1.72,4.5)(0.26,0){7}{\pp{}{.}}
\put(4.7,2.3){\p{}{,}}
\epic\ea\ee
for each $(b,c,d,e,f)\in\G^5$ with $\r^r_{ba}\,\r^r_{fa}>0$
and $\b\in\{1,\ldots,\r^r_{ba}\}$, $\d\in\{1,\ldots,\r^r_{fa}\}$.
This equation can be verified by expressing the bulk and boundary
weights in terms of bulk and boundary operators, using \er{WX} and
\er{B}, and applying the operator form of the boundary Yang-Baxter
equation, \er{TLBYBE}.  In doing this, the fusion matrix which is
formed on the interior of each side of the equation by the sum on $\g$
and \er{FMD} can be moved to the exterior by expressing the fusion
matrices in terms of fusion operators using \er{FM} and the boundary
operators in terms of fused row operators using \er{K} and applying
the push-through relations \er{PTR}.  Meanwhile, the orientation of
the central bulk weights on each side of \er{BYBE} can be changed to
that required for \er{TLBYBE} by using crossing symmetry \er{CS}, with
the eigenvector entries from \er{CS} canceling where needed with
those introduced explicitly through \er{B}.

We note that the boundary Yang-Baxter equation \er{BYBE} is still
satisfied after renormalization of the boundary weights,
\be\el{BNT}\B{ra}{b}{\b}{c}{d}{\d}{u,\x}\;\mapsto\;f^{ra}(u,\x)\;
\B{ra}{b}{\b}{c}{d}{\d}{u,\x}\,,\ee
where $f^{ra}$ are arbitrary functions.
It is also still satisfied
after gauge transformations of the boundary weights,
\be\el{BGT}\B{ra}{b}{\b}{c}{d}{\d}{u,\x}\;\mapsto\;
\sum_{\b'=1}^{\r^r_{ba}}\;\sum_{\d'=1}^{\r^r_{da}}\;
S^{ra}(b)_{\b\b'}\;S^{ra}(d)_{\d\d'}\;\B{ra}{b}{\b'}{c}{d}{\d'}{u,\x}\,,\ee
where $S^{ra}(e)$ are arbitrary orthonormal matrices,
\be\el{SON}
\sum_{\a''=1}^{\r^r_{ea}}\;S^{ra}(e)_{\a\a''}\;S^{ra}(e)_{\a'\a''}\;=\;
\d_{\a\a'}\,.\ee
Indeed, a transformation \er{UGT} of the fusion vectors simply induces a 
gauge transformation of the boundary weights with 
\be\el{SGT}S^{ra}(e)=R^r(e,a)\,.\ee

In addition to the boundary Yang-Baxter equation, some other important
local relations satisfied by the boundary weights are boundary reflection
symmetry,
\be\el{BRS}\B{ra}{b}{\b}{c}{d}{\d}{u,\x}\;=\;
\B{ra}{d}{\d}{c}{b}{\b}{u,\x}\,,\ee
boundary crossing symmetry,
\be\el{BCS}\ba{l}\ds\ru{4}\sum_{(b,e,d)\:\in\:\G^3_{bd}}
\W{c}{b}{e}{d}{2u\mi\l}\B{ra}{b}{\b}{e}{d}{\d}{u,\x}\\
\qquad\qquad\qquad\qquad\qquad\qquad\qquad\qquad=\;
s_0(2u)\;\B{ra}{b}{\b}{c}{d}{\d}{\l\mi u,\x}\,,\ea\ee
and the boundary inversion relation,
\be\el{BIR}\ba{l}\ds\ru{4}\sum_{\stackrel{\ru{0.7}\ss e\:\in\:\G}
{\sss(G_{ce}\r^r_{ea}>0)}}\sum_{\g=1}^{\r^r_{ea}}\;
\frac{\psi_b^\qt\:\psi_d^\qt\:\psi_e^\hf}{\psi_{c}}\;
\B{ra}{b}{\b}{c}{e}{\g}{u,\x}\B{ra}{e}{\g}{c}{d}{\d}{-\!u,\x}\\
\ds\qquad\qquad\qquad\qquad\qquad\qquad=\;
\frac{s_0(\x\mi u)\:s_0(\x\pl u)\:s_r(\x\mi u)\:s_r(\x\pl u)}{s_0(2\x)^2}\;
\d_{bd}\;\d_{\b\d}\,,\ea\ee
for each $(b,c,d)\in\G^3$ with $\r^r_{ba}\,\r^r_{da}>0$
and $\b\in\{1,\ldots,\r^r_{ba}\}$, $\d\in\{1,\ldots,\r^r_{da}\}$.  
Boundary reflection symmetry follows from \er{B}, boundary crossing
symmetry can be proved by expressing the boundary weights in \er{BCS}
in the form \er{CE} and repeatedly applying the Yang-Baxter equation
\er{YBE}, and the boundary inversion relation follows from the
operator form of the boundary inversion relation, \er{TLBIR}.

\sub{Boundary Edge Weights}
We now introduce boundary edge weights in terms of which the boundary weights
of the previous section can be expressed.

We begin by defining, for each boundary condition $(r,a)$, a 
set of boundary edges as
\be\el{BE}\ep^{ra}\;=\;\{(b,c)\in\G^2\mid\r^r_{ba}\,\r^{r+1}_{ca}>0\}\,.\ee

We note that the boundary edges are ordered pairs and that, in contrast to the
graph's set of edges for which
$(b,c)\in\G^2\Leftrightarrow(c,b)\in\G^2$, the appearance of $(b,c)$
in $\ep^{ra}$ need not imply the appearance of $(c,b)$ in $\ep^{ra}$.

The $(r,a)$ boundary edge weights are now given, for each
$(b,c)\in\ep^{ra}$, $\b\in\{1,\ldots,\r^r_{ba}\}$ and
$\g\in\{1,\ldots,\r^{r+1}_{ca}\}$, by
\be\el{BEW}\ba{l}\ru{2.5}E^{ra}(b,c)_{\b\g}\;=\\
\ds\qquad\frac{S_r^\hf\:\psi_c^\qt}{\psi_b^\qt}\!
\sum_{(b,d_1,\ldots,d_{r\mi2},a)\:\in\:\G^r_{ba}}
\!\!\!\!U^r(b,a)_{\b,(b,d_1,\ldots,d_{r\mi2},a)}\;
U^{r+1}(c,a)_{\g,(c,b,d_1,\ldots,d_{r\mi2},a)}\,.\ea\ee
We shall represent the boundary edge weights diagrammatically as
\setlength{\unitlength}{6mm}
\be\raisebox{-1\unitlength}[1\unitlength][1\unitlength]{
\text{2.9}{2}{1.4}{1}{}{E^{ra}(b,c)_{\b\g}}
\text{1.3}{2}{0.8}{1}{}{=}
\bpic(2.5,2)
\put(1.5,0.5){\line(1,0){0.5}}\put(0.5,1.5){\line(1,0){1.5}}
\put(0.5,1.5){\line(1,-1){1}}
\multiput(2,0.5)(0,0.9){2}{\line(0,1){0.1}}
\multiput(2,0.65)(0,0.25){3}{\line(0,1){0.2}}
\put(1.55,1){\pp{}{r,\,a}}
\put(1.45,0.45){\pp{tr}{b}}\put(0.4,1.52){\pp{br}{c}}
\put(2.03,0.42){\pp{tl}{\b}}\put(2.06,1.58){\pp{bl}{\g}}
\put(2.7,0.8){\p{}{.}}\epic}\ee

We see that 
\be\mbox{number of $(r,a)$ boundary edge weights}\;=\;
(\r^r\,G\,\r^{r+1})_{aa}\,.\ee

We note, as an example, that for the $(1,a)$ boundary condition,
\be\ep^{1a}\;=\;\{(a,c)\mid G_{ac}=1\}\,,\quad
E^{1a}(a,c)_{11}\;=\;\pm\,\psi_c^\qt/\psi_a^\qt\,.\ee

We now find, substituting \er{K1} into \er{B}
and using \er{FM}, \er{FMD} and \er{FMON}, a general expression for the
boundary weights in terms of boundary edge weights, 
\be\el{GBW}\ba{l}\ru{3.5}\ds\B{ra}{b}{\b}{c}{d}{\d}{u,\x}\;=\\
\qquad\ds
\frac{s_0(\x\mi u)\:s_r(\x\pl u)\;\psi_c^\hf}{s_0(2\x)\;\psi_{b}^\hf}\;
\d_{bd}\;\d_{\b\d}\;+\;\frac{s_0(2u)}{s_0(2\x)}\;
\sum_{\g=1}^{\r^{r+1}_{ca}}E^{ra}(b,c)_{\b\g}\:E^{ra}(d,c)_{\d\g}\,.\ea\ee

{}From this, we immediately see that at $u=\x$ the boundary weights 
are independent of $\x$ and can be decomposed as a sum of
products of boundary edge weights,
\setlength{\unitlength}{6mm}
\be\el{BP}\ba{r@{}c@{}l}\B{ra}{b}{\b}{c}{d}{\d}{\x,\x}&\;=\;&
\ds\sum_{\g=1}^{\r^{r+1}_{ca}}E^{ra}(b,c)_{\b\g}\:E^{ra}(d,c)_{\d\g}\\
\bpic(2.5,3.3)
\multiput(2,0.5)(0,1.9){2}{\line(0,1){0.1}}
\multiput(2,0.65)(0,0.25){7}{\line(0,1){0.2}}
\multiput(1.5,0.5)(0,2){2}{\line(1,0){0.5}}
\put(0.5,1.5){\line(1,-1){1}}\put(0.5,1.5){\line(1,1){1}}
\put(1.5,1.7){\pp{}{r,\,a}}\put(1.5,1.3){\pp{}{\x,\,\x}}
\put(1.45,0.47){\pp{tr}{b}}\put(0.4,1.5){\pp{r}{c}}
\put(1.45,2.53){\pp{br}{d}}
\put(2.1,0.5){\pp{tl}{\b}}\put(2.1,2.5){\pp{bl}{\d}}
\epic
&\text{0}{3}{0}{1.5}{}{=}&
\bpic(2.5,3)
\multiput(2,0.5)(0,1.9){2}{\line(0,1){0.1}}
\multiput(2,0.65)(0,0.25){7}{\line(0,1){0.2}}
\multiput(1.5,0.5)(0,2){2}{\line(1,0){0.5}}
\put(0.5,1.5){\line(1,-1){1}}\put(0.5,1.5){\line(1,1){1}}
\put(0.5,1.5){\line(1,0){1.5}}
\put(1.55,1.03){\pp{}{r,\,a}}
\put(1.55,1.87){\pp{}{r,\,a}}
\put(1.45,0.47){\pp{tr}{b}}\put(0.4,1.5){\pp{r}{c}}
\put(1.45,2.53){\pp{br}{d}}
\put(2.1,0.5){\pp{tl}{\b}}\put(2.1,2.5){\pp{bl}{\d}}
\put(2,1.5){\pp{}{\bullet}}
\put(2.7,1.3){\p{}{.}}\epic
\ea\ee
We note that the origin of this decomposition is the fact,
apparent from \er{K1}, that at $u=\x$ the boundary operators 
are proportional to fusion operators, so that the decomposition of
boundary weights is essentially equivalent to the eigenvector 
decomposition of projectors.

We shall refer to this point, $u=\x$, as the conformal point, since it
is here that certain lattice models are expected to exhibit conformal
behavior, with the set of $(r,a)$ boundary edge weights providing a
lattice realization of a particular conformal boundary condition.

We find, by using \er{K2} in \er{B}, that a decomposition similar
to \er{BP} also occurs at $u=-\x$,
\be\el{BM}\B{ra}{b}{\b}{c}{d}{\d}{-\x,\x}\;=\;
\ds\sum_{\stackrel{\ru{0.5}\ss\g=1}{\ss(r\ne1)}}^{\r^{r-1}_{ca}}
E^{r-1,a}(c,b)_{\g\b}\:E^{r-1,a}(c,d)_{\g\d}\,.\ee
We therefore see that, apart from an unimportant 
reversal of the order of the nodes in the boundary edges, the $(r,a)$
boundary condition at $u=-\x$ is equivalent to the $(r-1,a)$ boundary 
condition at $u=\x$.
In fact, decompositions of the form \er{BP} or \er{BM} also
occur at other points, for example at $u=-r\l\mi\x$, $u=r\l\pl\x$
or points related to these by trigonometric periodicity, but since
these points all involve the same boundary edge weights up to
reordering of the nodes in the boundary edges or relabeling of
$r$ as $r\plmi1$, we can, without loss of generality, 
restrict our attention to the single conformal point $u=\x$.
 
We also note that the sum in \er{BP} is empty if
$\r^{r+1}_{ca}=0$.  It is thus possible that for a particular boundary
condition, certain boundary weights are non-zero away from the conformal point
but vanish at the conformal point.  From \er{GBW}, we find that such
boundary weights are specifically those for which $b=d$, $\b=\d$,
$\r^r_{ba}\,G_{bc}>0$ and $\r^{r+1}_{ca}=0$.  

We see from \er{GBW} that the boundary weights satisfy the boundary
anisotropy property
\be\el{BCAP}\B{ra}{b}{\b}{c}{d}{\d}{0,\x}\;=\;
\frac{s_0(\x)\:s_r(\x)\;\psi_c^\hf}{s_0(2\x)\;\psi_{b}^\hf}\;
\d_{bd}\;\d_{\b\d}\,.\ee
The second term on the right side of \er{GBW} also
vanishes for $\x\rightarrow\pm i\infty$ so that, assuming $\l/\pi\not\in\Z$,
\be\el{BL}\B{ra}{b}{\b}{c}{d}{\d}{u,\pm i\infty}\;=\;
\pm\,\frac{i\:e^{\mp i r\l}\;\psi_c^\hf}{2\,\sin\!\l\;\psi_{b}^\hf}\;\,
\d_{bd}\;\d_{\b\d}\,.\ee
Comparing the right sides of \er{BCAP} or \er{BL} with the $(1,a)$
boundary weights \er{B1a}, we see that at $u=0$ or $\x\rightarrow\pm i\infty$
the nonzero $(r,a)$ boundary weights reduce, up to unimportant normalization,
to $(1,b)$ boundary weights, with each $(1,b)$ weight, for any $b$,
appearing $\r^r_{ab}$ (which may be zero) times.

Finally, we note that a gauge transformation of the boundary edge weights,
\be\el{EGT}E^{ra}(b,c)_{\b\g}\;\mapsto\;
\sum_{\b'=1}^{\r^r_{ba}}\;\sum_{\g'=1}^{\r^{r+1}_{ca}}\;
S^{ra}(b)_{\b\b'}\;S^{r+1,a}(c)_{\g\g'}\;E^{ra}(b,c)_{\b'\g'}\,,\ee
induces a gauge transformation \er{BGT} of the boundary weights,
for any $S^{ra}(e)$ satisfying \er{SON}.

\sub{Transfer Matrices and the Partition Function}\label{PFS}
We now proceed to a study of some aspects of the complete lattice.
We shall be considering a square lattice 
on a cylinder of width $N$ and circumference $2M$ lattice spacings,
with the left boundary
condition and boundary field given by $(r_1,a_1)$ and $\x_1$
and the right boundary condition and boundary field given by 
$(r_2,a_2)$ and $\x_2$.  In particular, we shall express the partition
function for the model using double-row transfer matrices,
these being defined in terms of the bulk and boundary weights of the
previous sections.  Double-row transfer matrices of this type were
introduced in~\cite{Skl88} and first used for interaction-round-a-face
models in~\cite{BehPeaObr96}.

An additional feature of the lattice to be considered here is that
alternate rows will be associated with spectral parameter values $u$
and $\l-u$.  Also, the left boundary will be associated with the
spectral parameter value $\l\mi u$ and the right boundary with value
$u$. These values are used since they result in the double-row
transfer matrices forming a commuting family.  The most physically
relevant point is the isotropic point $u=\l/2$, at which all rows and
both boundaries are associated with the same value of the spectral
parameter.  Combining this with the conformal point described in the
previous section, the case of most interest here is thus
$u=\xi_1=\xi_2=\l/2$.

We begin by defining a set of paths consistent with boundary conditions
$(r_1,a_1)$ and $(r_2,a_2)$ and lattice width $N$,
\be\el{GBC}\ba{l}\ru{2}\G^\N_{r_1\!a_1\!|r_2a_2}\;=\;
\Bigl\{(\b_1,b_0,\ldots,b_\N,\b_2)\,\Big|\,
(b_0,\ldots,b_\N)\in\G^{\N+1},\;\r^{r_1}_{a_1b_0}\r^{r_2}_{b_Na_2}>0,\\
\qquad\qquad\qquad\qquad\qquad\qquad\qquad
\b_1\in\{1,\ldots,\r^{r_1}_{a_1b_0}\},\;
\b_2\in\{1,\ldots,\r^{r_2}_{b_Na_2}\}\Bigr\}\,.\ea\ee

We see that
\be\el{SGBC}|\G^\N_{r_1\!a_1\!|r_2a_2}|\;=\;
\bigl(\r^{r_1}\,G^\N\r^{r_2}\bigr)_{a_1a_2}\,,\ee
which is invariant under interchange of $a_1$ and $a_2$ or of $r_1$
and $r_2$, since $\r^{r_1}$, $\r^{r_2}$ and $G$ are symmetric and
mutually commuting matrices.

We now introduce a double-row transfer matrix
$\vec{D}^\N_{\!r_1\!a_1\!|r_2a_2}\!(u,\x_1,\x_2)$  with rows 
and columns labeled by the paths of $\G^\N_{r_1\!a_1\!|r_2a_2}$
and entries defined by
\setlength{\unitlength}{9mm}
\be\el{DRTM}\ba{l}\ru{3.5}\vec{D}^\N_{\!r_1\!a_1\!|r_2a_2}\!
(u,\x_1,\x_2)_{(\b_1,b_0,\ldots,b_N,\b_2),(\d_1,d_0,\ldots,d_N,\d_2)}\;=\\
\ru{8.5}\ds\sum_{\stackrel{\ru{0.7}\ss(c_0,\ldots,c_N)\:\in\:
\G^{N\!+1}}{\ss(\sss\prod_{j=0}^N G_{b_jc_j}G_{c_jd_j}=1\ss)}}
\!\!\!\!\B{r_1a_1}{b_0}{\b_1}{c_0}{d_0}{\d_1}{\l\mi u,\x_1}\;\times\\
\qquad\ds\left[\prod_{j=0}^{\N-1}\W{b_j}{b_{j\pl1}}{c_{j\pl1}}{c_j}{u}
\W{c_j}{c_{j\pl1}}{d_{j\pl1}}{d_j}{\l\mi u}\right]\!
\B{r_2a_2}{b_\N}{\b_2}{c_\N}{d_\N}{\d_2}{u,\x_2}\\
\text{1.9}{3.4}{1.1}{1.5}{}{=}
\bpic(9,3)
\multiput(0.5,0.5)(0,2){2}{\line(1,0){0.5}}
\multiput(8,0.5)(0,2){2}{\line(1,0){0.5}}
\multiput(2,1.5)(6,-1){2}{\line(-1,1){1}}
\multiput(2,1.5)(6,1){2}{\line(-1,-1){1}}
\multiput(2,0.5)(0,1){3}{\line(1,0){5}}
\multiput(2,0.5)(1,0){2}{\line(0,1){2}}
\multiput(6,0.5)(1,0){2}{\line(0,1){2}}
\multiput(0.5,0.5)(0,0.3){7}{\line(0,1){0.2}}
\multiput(8.5,0.5)(0,0.3){7}{\line(0,1){0.2}}
\multiput(1.99,0.6)(0,1){2}{\pp{bl}{\searrow}}
\multiput(5.99,0.6)(0,1){2}{\pp{bl}{\searrow}}
\multiput(2.5,1)(4,0){2}{\pp{}{u}}
\multiput(2.5,2)(4,0){2}{\pp{}{\l-u}}
\put(1.1,1.65){\pp{}{r_1,\,a_1}}\put(1.15,1.35){\pp{}{\l-u,\,\x_1}}
\put(7.9,1.65){\pp{}{r_2,\,a_2}}\put(7.9,1.35){\pp{}{u,\,\x_2}}
\put(0.46,0.46){\pp{tr}{\b_1}}\put(0.46,2.54){\pp{br}{\d_1}}
\put(8.54,0.46){\pp{tl}{\b_2}}\put(8.54,2.54){\pp{bl}{\d_2}}
\multiput(1.1,0.39)(0.9,0){2}{\pp{t}{b_0}}\put(3,0.39){\pp{t}{b_1}}
\put(6.1,0.39){\pp{t}{b_{N\mi1}}}\multiput(7,0.39)(1,0){2}{\pp{t}{b_{N}}}
\multiput(1.1,2.61)(0.9,0){2}{\pp{b}{d_0}}\put(3,2.61){\pp{b}{d_1}}
\put(6.1,2.61){\pp{b}{d_{N\mi1}}}\multiput(7,2.61)(1,0){2}{\pp{b}{d_{N}}}
\multiput(2,1.5)(1,0){2}{\pp{}{\bullet}}
\multiput(6,1.5)(1,0){2}{\pp{}{\bullet}}
\multiput(1.125,0.5)(0.15,0){6}{\pp{}{.}}
\multiput(1.125,2.5)(0.15,0){6}{\pp{}{.}}
\multiput(7.125,0.5)(0.15,0){6}{\pp{}{.}}
\multiput(7.125,2.5)(0.15,0){6}{\pp{}{.}}
\put(8.9,1.4){\p{}{.}}\epic\ea\ee
We note that if $\G^\N_{r_1\!a_1\!|r_2a_2}$
is empty, then $\vec{D}^\N_{\!r_1\!a_1\!|r_2a_2}\!(u,\x_1,\x_2)$
is zero-dimensional with its value taken as $0$.

The partition function for the lattice model is now given by
\be\el{Z}\ba{rcl}
\vec{Z}^{\N\MM}_{r_1\!a_1\!|r_2a_2}\!(u,\x_1,\x_2)&=&
\mbox{tr}\,\bigl(\vec{D}^\N_{\!r_1\!a_1\!|r_2a_2}\!(u,\x_1,\x_2)\bigr)^\MM\\
&\text{0.4}{8.4}{0.2}{4.7}{}{=}&
\bpic(9.5,9)\put(-0.3,0.7){
\bpic(9,8)
\multiput(0.5,0.5)(0,2){2}{\line(1,0){0.5}}
\multiput(8,0.5)(0,2){2}{\line(1,0){0.5}}
\multiput(0.5,5.5)(0,2){2}{\line(1,0){0.5}}
\multiput(8,5.5)(0,2){2}{\line(1,0){0.5}}
\multiput(2,1.5)(6,-1){2}{\line(-1,1){1}}
\multiput(2,1.5)(6,1){2}{\line(-1,-1){1}}
\multiput(2,6.5)(6,-1){2}{\line(-1,1){1}}
\multiput(2,6.5)(6,1){2}{\line(-1,-1){1}}
\multiput(2,0.5)(0,1){3}{\line(1,0){5}}\multiput(2,5.5)(0,1){3}{\line(1,0){5}}
\multiput(2,0.5)(1,0){2}{\line(0,1){7}}\multiput(6,0.5)(1,0){2}{\line(0,1){7}}
\multiput(0.5,0.6)(0,0.3){23}{\line(0,1){0.2}}
\multiput(8.5,0.6)(0,0.3){23}{\line(0,1){0.2}}
\multiput(1.99,0.6)(0,1){2}{\pp{bl}{\searrow}}
\multiput(5.99,0.6)(0,1){2}{\pp{bl}{\searrow}}
\multiput(1.99,5.6)(0,1){2}{\pp{bl}{\searrow}}
\multiput(5.99,5.6)(0,1){2}{\pp{bl}{\searrow}}
\multiput(2.5,1)(4,0){2}{\pp{}{u}}\multiput(2.5,2)(4,0){2}{\pp{}{\l-u}}
\multiput(2.5,6)(4,0){2}{\pp{}{u}}\multiput(2.5,7)(4,0){2}{\pp{}{\l-u}}
\multiput(1.1,1.65)(0,5){2}{\pp{}{r_1,\,a_1}}
\multiput(1.15,1.35)(0,5){2}{\pp{}{\l-u,\,\x_1}}
\multiput(7.9,1.65)(0,5){2}{\pp{}{r_2,\,a_2}}
\multiput(7.9,1.35)(0,5){2}{\pp{}{u,\,\x_2}}
\multiput(2,0.5)(0,1){3}{\pp{}{\bullet}}
\multiput(3,0.5)(0,1){3}{\pp{}{\bullet}}
\multiput(6,0.5)(0,1){3}{\pp{}{\bullet}}
\multiput(7,0.5)(0,1){3}{\pp{}{\bullet}}
\multiput(2,5.5)(0,1){3}{\pp{}{\bullet}}
\multiput(3,5.5)(0,1){3}{\pp{}{\bullet}}
\multiput(6,5.5)(0,1){3}{\pp{}{\bullet}}
\multiput(7,5.5)(0,1){3}{\pp{}{\bullet}}
\multiput(0.5,0.5)(0,2){2}{\pp{}{\bullet}}
\multiput(8.5,0.5)(0,2){2}{\pp{}{\bullet}}
\multiput(0.5,5.5)(0,2){2}{\pp{}{\bullet}}
\multiput(8.5,5.5)(0,2){2}{\pp{}{\bullet}}
\multiput(1,0.5)(0,2){2}{\pp{}{\bullet}}
\multiput(8,0.5)(0,2){2}{\pp{}{\bullet}}
\multiput(1,5.5)(0,2){2}{\pp{}{\bullet}}
\multiput(8,5.5)(0,2){2}{\pp{}{\bullet}}
\multiput(1.125,0.5)(0.15,0){6}{\pp{}{.}}
\multiput(1.125,2.5)(0.15,0){6}{\pp{}{.}}
\multiput(7.125,0.5)(0.15,0){6}{\pp{}{.}}
\multiput(7.125,2.5)(0.15,0){6}{\pp{}{.}}
\multiput(1.125,5.5)(0.15,0){6}{\pp{}{.}}
\multiput(1.125,7.5)(0.15,0){6}{\pp{}{.}}
\multiput(7.125,5.5)(0.15,0){6}{\pp{}{.}}
\multiput(7.125,7.5)(0.15,0){6}{\pp{}{.}}
\multiput(4.3,0.4)(0,7.2){2}{\pp{}{.}}\multiput(4.36,0.26)(0,7.48){2}{\pp{}{.}}
\multiput(4.5,0.2)(0,7.6){2}{\pp{}{.}}\multiput(4.64,0.26)(0,7.48){2}{\pp{}{.}}
\multiput(4.7,0.4)(0,0.2){37}{\pp{}{.}}\epic}
\put(3.9,0.3){\vector(-1,0){2.2}}\put(4.5,0.3){\vector(1,0){2.2}}
\put(4.2,0.3){\pp{}{N}}
\put(9.2,4.4){\vector(0,-1){3.2}}\put(9.2,5){\vector(0,1){3.2}}
\put(9.2,4.7){\pp{}{2M}}\put(9.7,4.3){\p{}{.}}
\epic\ea\ee
We can see from the first equality of \er{Z} that the task of
evaluating the partition function is equivalent to that of evaluating
the eigenvalues $\Lambda^\N_{r_1\!a_1\!|r_2a_2}\!(u,\x_1,\x_2)_k$,
$k=1,\ldots,(\r^{r_1}G^\N\r^{r_2})_{a_1a_2}$, of
$\vec{D}^\N_{\!r_1\!a_1\!|r_2a_2}\!(u,\x_1,\x_2)$, with
\be\el{ZE}\vec{Z}^{\N\MM}_{r_1\!a_1\!|r_2a_2}\!(u,\x_1,\x_2)\;=\;\sum_k\;
\bigl(\Lambda^\N_{r_1\!a_1\!|r_2a_2}\!(u,\x_1,\x_2)_k\bigr)^\MM\,.\ee

\sub{Transfer Matrix Properties}\label{TMP}
We shall now show that the double-row transfer matrices satisfy a
variety of important properties.  In particular, we shall find that
applying certain transformations to the parameters of the model
results only in similarity transformations of the transfer matrix.
This implies that these parameter transformations are symmetries of
the model, since the transfer matrix eigenvalues and partition
function are invariant under any such similarity transformation.

\subsub{Commutation}\label{SDC}
It follows from the Yang-Baxter equation \er{YBE}, inversion relation
\er{IR} and boundary Yang-Baxter equation \er{BYBE} that the
double-row transfer matrices commute for any two values, $u$ and $v$,
of the spectral parameter,
\be\el{DC}\Bigl[\vec{D}^\N_{\!r_1\!a_1\!|r_2a_2}\!(u,\x_1,\x_2)\:,\:
\vec{D}^\N_{\!r_1\!a_1\!|r_2a_2}\!(v,\x_1,\x_2)\Bigr]\;=\;0\,.\ee
This can be proved diagrammatically, as done in Section 3.4 of
\cite{BehPeaObr96}.

\subsub{Transposition}
It follows straightforwardly from \er{CS} and \er{BRS} that each double-row 
transfer matrix is similar to a symmetric matrix.  More specifically,
defining
\be\el{A}\vec{A}^\N_{r_1\!a_1\!|r_2a_2\!\raisebox{-0.3ex}{
$\ss(\b_1,b_0,\ldots,b_N,\b_2),(\d_1,d_0,\ldots,d_N,\d_2)$}}\,\;=\;\,
\frac{\psi_{b_N}^\qt}{\psi_{b_0}^\qt}\;\,\delta_{(\b_1,b_0,\ldots,b_N,\b_2),
(\d_1,d_0,\ldots,d_N,\d_2)}\ee
and
\be\tilde{\vec{D}}^\N_{\!r_1\!a_1\!|r_2a_2}\!(u,\x_1,\x_2)\;=\;
\vec{A}^\N_{r_1\!a_1\!|r_2a_2}\;
\vec{D}^\N_{\!r_1\!a_1\!|r_2a_2}\!(u,\x_1,\x_2)\;
(\vec{A}^\N_{r_1\!a_1\!|r_2a_2})^{-1}\,,\ee
we have
\be\el{DT}\ru{2.5}
\Bigl(\tilde{\vec{D}}^\N_{\!r_1\!a_1\!|r_2a_2}\!(u,\x_1,\x_2)\Bigr)^T\;=\;
\tilde{\vec{D}}^\N_{\!r_1\!a_1\!|r_2a_2}\!(u,\x_1,\x_2)\,.\ee

This implies that if
$\tilde{\vec{D}}^\N_{\!r_1\!a_1\!|r_2a_2}\!(u,\x_1,\x_2)$ is real, as
for example occurs if $\psi$ is the Perron-Frobenius eigenvector and
$u$, $\x_1$ and $\x_2$ are appropriately chosen, then the eigenvalues
of $\vec{D}^\N_{\!r_1\!a_1\!|r_2a_2}\!(u,\x_1,\x_2)$ are all real.

\subsub{Gauge Invariance}
We see that the model is invariant under any gauge transformation \er{BGT}
of the boundary weights since this results only in a similarity transformation 
of the double-row transfer matrix,
\be\el{DGT}\vec{D}^\N_{\!r_1\!a_1\!|r_2a_2}\!(u,\x_1,\x_2)\;\mapsto\;
\vec{S}^\N_{\!r_1\!a_1\!|r_2a_2}\;
\vec{D}^\N_{\!r_1\!a_1\!|r_2a_2}\!(u,\x_1,\x_2)\;
(\vec{S}^\N_{\!r_1\!a_1\!|r_2a_2})^{-1}\,,\ee
where
\be\ba{l}\ru{2.5}\el{GIS}\vec{S}^\N_{\!r_1\!a_1\!|r_2a_2\!\raisebox{-0.3ex}{
$\ss(\b_1,b_0,\ldots,b_N,\b_2),(\d_1,d_0,\ldots,d_N,\d_2)$}}\;=\\
\qquad\qquad\qquad\qquad\qquad\qquad
S^{r_1a_1}(b_0)_{\b_1\d_1}\;S^{r_2a_2}(b_N)_{\b_2\d_2}\;
\delta_{(b_0,\ldots,b_N\!),(d_0,\ldots,d_N\!)}\,.\ea\ee

\subsub{Simplification at Completely Anisotropic Points}
It follows from the anisotropy property \er{CAP}, boundary anisotropy 
property \er{BCAP}, crossing symmetry \er{CS} and 
boundary crossing symmetry \er{BCS} that at the
completely anisotropic points, $u=0$ and $u=\l$, the  
double-row transfer matrices are proportional to the identity,
\be\el{DCAP}\vec{D}^\N_{\!r_1\!a_1\!|r_2a_2}\!(0,\x_1,\x_2)\:=\:
\vec{D}^\N_{\!r_1\!a_1\!|r_2a_2}\!(\l,\x_1,\x_2)\:=\:
\frac{S_2\,s_0(\x_1)\,s_0(\x_2)\,s_{r_1}(\x_1)\,s_{r_2}(\x_2)}
{s_0(2\x_1)\,s_0(2\x_2)}\:\vec{I}\,.\ee

\subsub{Crossing Symmetry}
It follows from the Yang-Baxter equation \er{YBE}, inversion relation \er{IR}
and boundary crossing symmetry \er{BCS} that the double-row transfer matrices
satisfy crossing symmetry,
\be\el{DCS}\vec{D}^\N_{\!r_1\!a_1\!|r_2a_2}\!(\l\mi u,\x_1,\x_2)\;=\;
\vec{D}^\N_{\!r_1\!a_1\!|r_2a_2}\!(u,\x_1,\x_2)\,.\ee
This, like \er{DC}, can be proved diagrammatically, as done in Section 3.3 of
\cite{BehPeaObr96}.

\subsub{Left-Right Symmetry}\label{LR}
It follows from reflection symmetry \er{RS} and boundary reflection
symmetry \er{BRS} that on interchanging the left and right
boundary conditions and boundary fields, we have
\be\el{DLRS}\vec{D}^\N_{\!r_2a_2|r_1\!a_1}\!(u,\x_2,\x_1)\;
\vec{R}^\N_{r_1\!a_1\!|r_2a_2}\;=\;
\vec{R}^\N_{r_1\!a_1\!|r_2a_2}\;
\Bigl(\vec{D}^\N_{\!r_1\!a_1\!|r_2a_2}\!(\l\mi u,\x_1,\x_2)\Bigr)^T\,,\ee
where $\vec{R}^\N_{r_1\!a_1\!|r_2a_2}$ is a square matrix
with rows labeled by the paths of $\G^\N_{r_2a_2|r_1\!a_1}$,
columns labeled by the paths of $\G^\N_{r_1\!a_1\!|r_2a_2}$
and entries given by
\be\vec{R}^\N_{r_1\!a_1\!|r_2a_2\!\raisebox{-0.3ex}{
$\ss(\b_1,b_0,\ldots,b_N,\b_2),(\d_1,d_0,\ldots,d_N,\d_2)$}}\;=\;
\delta_{(\b_2,b_N,\ldots,b_0,\b_1),(\d_1,d_0,\ldots,d_N,\d_2)}\,.\ee
Combining \er{DLRS} and the invertibility of $\vec{R}^\N_{r_1\!a_1\!|r_2a_2}$
with \er{DT} and \er{DCS}, we see that 
$\vec{D}^\N_{\!r_1\!a_1\!|r_2a_2}\!(u,\x_1,\x_2)$ and
$\vec{D}^\N_{\!r_2a_2|r_1\!a_1}\!(u,\x_2,\x_1)$ are related by
a similarity transformation, so that this complete interchanging of the
left and right boundaries is a symmetry of the model.

\subsub{Symmetry under Interchange of $r_1$, $\x_1$ and 
$r_2$, $\x_2$}\label{IS}
It can also be shown that under interchange of $r_1$, $\x_1$ 
and $r_2$, $\x_2$, we have
\be\el{DPIS}\vec{D}^\N_{\!r_2a_1\!|r_1\!a_2}\!(u,\x_2,\x_1)\;
\vec{C}^\N_{\!r_1\!a_1\!|r_2a_2}\!(\x_1,\x_2)\;=\;
\vec{C}^\N_{\!r_1\!a_1\!|r_2a_2}\!(\x_1,\x_2)\;
\vec{D}^\N_{\!r_1\!a_1\!|r_2a_2}\!(u,\x_1,\x_2)\,,\ee
where $\vec{C}^\N_{\!r_1\!a_1\!|r_2a_2}\!(\x_1,\x_2)$ is a square matrix
with rows labeled by the paths of $\G^\N_{r_2a_1\!|r_1\!a_2}$,
columns labeled by the paths of $\G^\N_{r_1\!a_1\!|r_2a_2}$
and entries given by
\be\el{C}\ba{l}\ru{4}\ds\vec{C}^\N_{\!r_1\!a_1\!|r_2a_2}\!(\x_1,\x_2)_
{(\b_1,b_0,\ldots,b_N,\b_2),(\d_1,d_0,\ldots,d_N,\d_2)}\;=\;
\frac{\psi_{b_0}^\qt\:\psi_{b_N}^\qt\:\psi_{d_0}^\qt\:\psi_{d_N}^\qt}
{\psi_{a_1}^\hf\:\psi_{a_2}^\hf}\;\times\\
\ru{8.5}\ds\sum_{\stackrel{\ru{0.7}\ss(c_0,\ldots,c_{N-1})\:\in\:\G^{N}}
{\ss(\sss G_{c_{N\mi1}a_2}\prod_{j=0}^{N\mi1}\!
\r^{r_1}_{\!b_jc_j}\r^{r_2}_{\!c_jd_j}>0\ss)}}\!\!
\sum_{\a_0=1}^{\r^{r_1}_{\!b_0c_0}}\ldots\!\!\!\!\!
\sum_{\a_{N\mi1}=1}^{\r^{r_1}_{\!b_{N\mi1}c_{N\mi1}}}\;
\sum_{\g_0=1}^{\r^{r_2}_{\!c_0d_0}}\ldots\!\!\!\!\!
\sum_{\g_{N\mi1}=1}^{\r^{r_2}_{\!c_{N\mi1}d_{N\mi1}}}
\Wf{r_2r_1}{\Wi{a_1}{\b_1}{b_0}{\a_0}{c_0}{\g_0}{d_0}{\d_1}}{\!\ba{c}
\x_1\mi\x_2\\ \;\;+\l\ea\!}\\
\ru{6}\ds\qquad\qquad\times\,\left[\prod_{j=0}^{\N-1}
\Wf{2r_1}{\Wi{b_j}{\;1\;}{b_{j\pl1}}{\a_{j\pl1}}{c_{j\pl1}}{1}{c_j}{\a_j}}
{\x_1\pl2\l}\:
\Wf{2r_2}{\Wi{c_j}{\;1\;}{c_{j\pl1}}{\g_{j\pl1}}{d_{j\pl1}}{1}{d_j}{\g_j}}
{\x_2\pl\l}\right]\,,\ea\ee
in which we take $c_\N=a_2$, $\a_\N=\b_2$ and $\g_\N=\d_2$.

The weights which appear in \er{C} are fused bulk weights which
are defined, 
for each $r,s\in\{1,\ldots,g\mi1\}$, as fused $r\mi1$ by $s\mi1$ 
blocks of bulk weights,
\setlength{\unitlength}{10mm}
\be\ba{l}\Wf{rs}{\Wi{a}{\:\a\:}{b}{\b}{c}{\g}{d}{\d}}{u}\;=\\
\bpic(12.2,5.6)\put(3.8,0.3){
\bpic(5,5)
\multiput(0.5,0.5)(1,0){2}{\line(0,1){4}}
\multiput(0.5,0.5)(0,1){2}{\line(1,0){4}}
\multiput(3.5,0.5)(1,0){2}{\line(0,1){4}}
\multiput(0.5,3.5)(0,1){2}{\line(1,0){4}}
\multiput(0.48,0.58)(3,0){2}{\pp{bl}{\searrow}}
\multiput(0.48,3.58)(3,0){2}{\pp{bl}{\searrow}}
\put(4,1){\pp{c}{u}}
\put(0.6,1.17){\pp{l}{u-}}\put(1.45,0.9){\pp{r}{(r-2)\l}}
\put(3.6,4.17){\pp{l}{u+}}\put(4.45,3.9){\pp{r}{(s-2)\l}}
\put(0.6,4.17){\pp{l}{u+}}\put(1.45,3.9){\pp{r}{(s-r)\l}}
\put(0.48,0.48){\pp{tr}{a}}\put(4.52,0.48){\pp{tl}{b}}
\put(0.48,4.52){\pp{br}{d}}\put(4.52,4.52){\pp{bl}{c}}
\put(1.5,0.39){\pp{t}{e_1}}\put(3.6,0.39){\pp{t}{e_{r-2}}}
\put(4.61,1.5){\pp{l}{f_1}}\put(4.61,3.5){\pp{l}{f_{\!s-2}}}
\put(1.5,4.64){\pp{b}{g_1}}\put(3.6,4.64){\pp{b}{g_{r-2}}}
\put(0.39,1.5){\pp{r}{h_1}}\put(0.39,3.5){\pp{r}{h_{s-2}}}
\multiput(0.5,1.5)(1,0){2}{\pp{}{\bullet}}
\multiput(0.5,3.5)(1,0){2}{\pp{}{\bullet}}
\multiput(3.5,1.5)(1,0){2}{\pp{}{\bullet}}
\multiput(3.5,3.5)(1,0){2}{\pp{}{\bullet}}
\multiput(1.5,0.5)(2,0){2}{\pp{}{\bullet}}
\multiput(1.5,4.5)(2,0){2}{\pp{}{\bullet}}\epic}
\put(6.5,0.05){\p{b}{U^r(a,b)_{\a,(a,e_1,\ldots,e_{r\mi2},b)}}}
\put(12.2,2.8){\p{r}{U^s(b,c)_{\b,(b,f_1,\ldots,f_{\!s\mi2},c)}}}
\put(6.5,5.65){\p{t}{U^r(d,c)_{\g,(d,g_1,\ldots,g_{r\mi2},c)}}}
\put(0.5,2.8){\p{l}{U^s(a,d)_{\d,(a,h_1,\ldots,h_{s\mi2},d)}}}
\put(12.35,2.6){\p{}{.}}\epic\ea\ee
These fused weights can be regarded as lattice generalizations of the
$1$ by $r\mi1$ and $r\mi1$ by $1$ fused blocks considered in \er{Y}.
They satisfy a generalized Yang-Baxter equation, as
given in (3.39) of \cite{ZhoPea94a}, and it is by using this equation,
and expressing the boundary weights in the double row transfer
matrices in the form \er{CE}, that \er{DPIS} can be proved.  In doing
this, it is the $(\x_1\pl2\l)$-dependent $1$ by $r_1\mi1$ fused 
weights in \er{C} which, together with the generalized Yang-Baxter
equation, allow the $\x_1$-dependent $r_1\mi1$ by $2$ fused block of
bulk weights within the left boundary weight to be propagated 
to the right.  Similarly, it is the $(\x_2\pl\l)$-dependent $1$ by
$r_2\mi1$ fused weights which allow the $\x_2$-dependent
$r_2\mi1$ by $2$ fused block within the right boundary
weight to be propagated to the left, and it is the
$(\x_1\mi\xi_2\pl\l)$-dependent $r_2\mi1$ by $r_1\mi1$ fused 
weight which allows the two blocks of bulk weights from the left and
right boundary weights to be interchanged.  

We also note that the generalized Yang-Baxter equation allows the
fused weights within $\vec{C}^\N_{\!r_1\!a_1\!|r_2a_2}\!(\x_1,\x_2)$
to be rearranged.  In particular, the $(\x_1\mi\xi_2\pl\l)$-dependent
fused weight, which in \er{C} is on the far left of the lattice row,
can be propagated to an arbitrary position further to the right, while
reversing the order of the $(\x_1\pl2\l)$- and $(\x_2\pl\l)$-dependent
fused weights to the left of its final position.  This essentially
corresponds to the fact that, in proving \er{DPIS}, the order with
which the blocks from the left and right boundaries are interchanged
is arbitrary.

By using a generalized inversion relation, it can also be shown that
$\vec{C}^\N_{\!r_1\!a_1\!|r_2a_2}\!(\x_1,\x_2)$ is invertible (except
at isolated values of $\x_1$ or $\x_2$ which, by continuity, are
unimportant), its inverse being proportional to
$\vec{C}^\N_{\!r_2a_1\!|r_1\!a_2}\!(\x_2,\x_1)$.  It thus follows that
transfer matrices with $r_1$, $\x_1$ and $r_2$, $\x_2$ interchanged
are related by a similarity transformation, so that this partial
interchanging of the left and right boundaries is a symmetry of the
model.

Finally, we note that a simple generalization of the results of this
section is that it is also possible, up to similarity transformation,
to propagate the fused block of bulk weights within either boundary
weight to an arbitrary position within the interior of the transfer
matrix.  The importance of this observation is that it is consistent
with the viewpoint in conformal field theory of the boundary
conditions corresponding to local operators.

\subsub{Properties Arising from $\x\rightarrow i\infty$}
Various properties follow from the $\x\rightarrow i\infty$
(or equivalently $\x\rightarrow-i\infty$) form \er{BL} of the 
boundary weights. 

For example, applying this 
limit to both the left and right boundary fields we find that 
$\vec{D}^\N_{\!r_1\!a_1\!|r_2a_2}\!(u,i\infty,i\infty)$ 
is proportional to a direct sum, over ${a_1\!}',{a_2\!}'\in\G$,
of $\vec{D}^\N_{\!1{a_1\!}'|1{a_2}'}(u,\x_1,\x_2)$, with this term 
being repeated
$\r^{r_1}_{a_1{a_1\!}'}\r^{r_2}_{a_2{a_2}'}$ times in the sum.

As a further example, we can attach an additional $r_1\mi1$ by $2$
fused block on the left and an additional $r_2\mi1$ by $2$ fused block
on the right of $\vec{D}^\N_{\!{r_1\!}'a_1\!|{r_2}'a_2}\!(u,\x_1,\x_2)$,
where each of these blocks is of the same form as
that in \er{CE} and we take $\x\rightarrow i\infty$ in each.  
We then find, from \er{BL}, that the resulting matrix is proportional
to a corresponding direct sum and, using methods similar to those
of Section~\ref{IS},
that the two attached blocks can be interchanged up to similarity 
transformation.  This therefore gives the result
\be\el{DII}\bigoplus_{{a_1\!}',{a_2\!}'\in\G}
\!\!\!\!\!{}^{\r^{r_1}_{a_1{a_1\!}'}\r^{r_2}_{a_2{a_2\!}'}}\;
\vec{D}^\N_{\!{r_1\!}'{a_1\!}'|{r_2}'{a_2}'}(u,\x_1,\x_2)\;\,\approx\;
\bigoplus_{{a_1\!}',{a_2\!}'\in\G}
\!\!\!\!\!{}^{\r^{r_2}_{a_1{a_1\!}'}\r^{r_1}_{a_2{a_2\!}'}}\;
\vec{D}^\N_{\!{r_1\!}'{a_1\!}'|{r_2}'{a_2}'}(u,\x_1,\x_2)\,,\ee
where $\approx$ indicates equality up to similarity transformation and
the superscripts on $\bigoplus$ indicate the number of times
that the corresponding terms appear in the direct sum.  This relation
implies that
\be\el{ZII}\ba{l}\ds\ru{2.5}\sum_{{a_1\!}',{a_2\!}'\in\G}
\r^{r_1}_{a_1{a_1\!}'}\;\r^{r_2}_{a_2{a_2\!}'}\;
\vec{Z}^{\N\MM}_{{r_1\!}'{a_1\!}'|{r_2}'{a_2}'}(u,\x_1,\x_2)\;=\\
\qquad\qquad\qquad\qquad\qquad\qquad\ds
\sum_{{a_1\!}',{a_2\!}'\in\G}
\r^{r_2}_{a_1{a_1\!}'}\;\r^{r_1}_{a_2{a_2\!}'}\;
\vec{Z}^{\N\MM}_{{r_1\!}'{a_1\!}'|{r_2}'{a_2}'}(u,\x_1,\x_2)\,.\ea\ee

\subsub{\protect Decomposition into Single-Row Transfer Matrices at the\\
Conformal Point}
It follows from \er{BP} that if both boundaries are at
their conformal point, that is if $u=\l\mi\x_1=\x_2\equiv\x$, then
the double-row transfer matrix decomposes into a product
of two single-row transfer matrices. We denote these, 
generally not square, matrices as 
$\stackrel{\sss\vee}{\vec{T}}{}\!\!^\N_{\!r_1\!a_1\!|r_2a_2}\!(\x)$,
with rows labeled by the paths of $\G^\N_{r_1\!a_1\!|r_2a_2}$ 
and columns labeled by the paths of $\G^\N_{r_1\pl1,a_1\!|r_2+\!1,a_2}$,
and $\stackrel{\sss\wedge}{\vec{T}}{}\!\!^\N_{\!r_1\!a_1\!|r_2a_2}\!(\x)$,
with rows labeled by the paths of $\G^\N_{r_1\pl1,a_1\!|r_2+\!1,a_2}$
and columns labeled by the paths of $\G^\N_{r_1\!a_1\!|r_2a_2}$.
The entries of these matrices are given by
\setlength{\unitlength}{9mm}
\be\el{LSRTM}\ba{l}\ru{3.5}
\stackrel{\sss\vee}{\vec{T}}{}\!\!^\N_{\!r_1\!a_1\!|r_2a_2}\!
(\x)_{(\b_1,b_0,\ldots,b_N,\b_2),(\g_1,c_0,\ldots,c_N,\g_2)}\;=\\
\qquad\left\{\ba{l}\ru{5}\ds E^{r_1a_1}(b_0,c_0)_{\b_1\g_1}
\prod_{j=0}^{\N-1}\W{b_j}{b_{j\pl1}}{c_{j\pl1}}{c_j}{\x}\,
E^{r_2a_2}(b_\N,c_\N)_{\b_2\g_2},\;\prod_{j=0}^{\N}G_{b_jc_j}=1\\
\ds\;0\,,\quad\prod_{j=0}^{\N}G_{b_jc_j}=0\ea\right.\\
\text{1.9}{2.3}{1.1}{1}{}{=}
\bpic(9,2)
\multiput(0.5,0.5)(7.5,0){2}{\line(1,0){0.5}}
\put(2,1.5){\line(-1,-1){1}}\put(7,1.5){\line(1,-1){1}}
\put(2,0.5){\line(1,0){5}}\put(0.5,1.5){\line(1,0){8}}
\multiput(2,0.5)(1,0){2}{\line(0,1){1}}
\multiput(6,0.5)(1,0){2}{\line(0,1){1}}
\multiput(0.5,0.5)(0,0.9){2}{\line(0,1){0.1}}
\multiput(0.5,0.65)(0,0.25){3}{\line(0,1){0.2}}
\multiput(8.5,0.5)(0,0.9){2}{\line(0,1){0.1}}
\multiput(8.5,0.65)(0,0.25){3}{\line(0,1){0.2}}
\multiput(1.99,0.6)(4,0){2}{\pp{bl}{\searrow}}
\multiput(2.5,1)(4,0){2}{\pp{}{\x}}
\put(0.57,1){\pp{l}{r_1,\,a_1}}
\put(8.43,1){\pp{r}{r_2,\,a_2}}
\put(0.46,0.46){\pp{tr}{\b_1}}\put(0.46,1.54){\pp{br}{\g_1}}
\put(8.54,0.46){\pp{tl}{\b_2}}\put(8.54,1.54){\pp{bl}{\g_2}}
\multiput(1.1,0.39)(0.9,0){2}{\pp{t}{b_0}}\put(3,0.39){\pp{t}{b_1}}
\put(6.1,0.39){\pp{t}{b_{N\mi1}}}\multiput(7,0.39)(1,0){2}{\pp{t}{b_{N}}}
\put(2,1.61){\pp{b}{c_0}}\put(3,1.61){\pp{b}{c_1}}
\put(6.1,1.61){\pp{b}{c_{N\mi1}}}\put(7,1.61){\pp{b}{c_{N}}}
\multiput(1.125,0.5)(0.15,0){6}{\pp{}{.}}
\multiput(7.125,0.5)(0.15,0){6}{\pp{}{.}}
\epic\ea\ee
and
\be\el{USRTM}\ba{l}\ru{3.5}
\stackrel{\sss\wedge}{\vec{T}}{}\!\!^\N_{\!r_1\!a_1\!|r_2a_2}\!
(\x)_{(\g_1,c_0,\ldots,c_N,\g_2),(\b_1,b_0,\ldots,b_N,\b_2)}\;=\\
\quad\left\{\ba{l}\ru{5}\ds E^{r_1a_1}(b_0,c_0)_{\b_1\g_1}
\prod_{j=0}^{\N-1}\W{c_j}{c_{j\pl1}}{b_{j\pl1}}{b_j}{\!\l\mi\x}\,
E^{r_2a_2}(b_\N,c_\N)_{\b_2\g_2},\;\prod_{j=0}^{\N}G_{c_jb_j}=1\\
\ds\;0\,,\quad\prod_{j=0}^{\N}G_{c_jb_j}=0\ea\right.\\
\text{1.9}{2.3}{1.1}{1}{}{=}
\bpic(9,2)
\multiput(0.5,1.5)(7.5,0){2}{\line(1,0){0.5}}
\put(1,1.5){\line(1,-1){1}}\put(7,0.5){\line(1,1){1}}
\put(2,1.5){\line(1,0){5}}\put(0.5,0.5){\line(1,0){8}}
\multiput(2,0.5)(1,0){2}{\line(0,1){1}}
\multiput(6,0.5)(1,0){2}{\line(0,1){1}}
\multiput(0.5,0.5)(0,0.9){2}{\line(0,1){0.1}}
\multiput(0.5,0.65)(0,0.25){3}{\line(0,1){0.2}}
\multiput(8.5,0.5)(0,0.9){2}{\line(0,1){0.1}}
\multiput(8.5,0.65)(0,0.25){3}{\line(0,1){0.2}}
\multiput(1.99,0.6)(4,0){2}{\pp{bl}{\searrow}}
\multiput(2.5,1)(4,0){2}{\pp{}{\l-\x}}
\put(0.57,1){\pp{l}{r_1,\,a_1}}
\put(8.45,1){\pp{r}{r_2,\,a_2}}
\put(0.46,0.46){\pp{tr}{\g_1}}\put(0.46,1.54){\pp{br}{\b_1}}
\put(8.54,0.46){\pp{tl}{\g_2}}\put(8.54,1.54){\pp{bl}{\b_2}}
\put(2,0.39){\pp{t}{c_0}}\put(3,0.39){\pp{t}{c_1}}
\put(6.1,0.39){\pp{t}{c_{N\mi1}}}\put(7,0.39){\pp{t}{c_{N}}}
\multiput(1.1,1.61)(0.9,0){2}{\pp{b}{b_0}}\put(3,1.61){\pp{b}{b_1}}
\put(6.1,1.61){\pp{b}{b_{N\mi1}}}\multiput(7,1.61)(1,0){2}{\pp{b}{b_{N}}}
\multiput(1.125,1.5)(0.15,0){6}{\pp{}{.}}
\multiput(7.125,1.5)(0.15,0){6}{\pp{}{.}}
\put(9.1,0.8){\p{}{.}}\epic\ea\ee

We now see, using \er{BP}, that the decomposition 
of the double-row transfer matrix at the conformal point is
\be\el{DD}\vec{D}^\N_{\!r_1\!a_1\!|r_2a_2}\!(\x,\l\mi\x,\x)\;=\;\:
\stackrel{\sss\vee}{\vec{T}}{}\!\!^\N_{\!r_1\!a_1\!|r_2a_2}\!(\x)\;
\stackrel{\sss\wedge}{\vec{T}}{}\!\!^\N_{\!r_1\!a_1\!|r_2a_2}\!(\x)\,.\ee
We emphasise that the orientations of the boundary edge weights 
in the two single-row transfer matrices in \er{DD}
are opposite with respect to a fixed direction along the boundary.
Thus, even at the isotropic point $\x=\l/2$, at which
adjacent rows of the lattice become indistinguishable in the bulk,
the alternating orientations of the boundary edge weights will
still distinguish adjacent rows at the boundaries.

We see from \er{CS} that $\ru{2.5}
\!\!\stackrel{\sss\wedge}{\vec{T}}{}\!\!^\N_{\!r_1\!a_1\!|r_2a_2}\!(\x)
=(\vec{A}^\N_{r_1\pl1,a_1\!|r_2+\!1,a_2})^{-2}
(\stackrel{\sss\vee}{\vec{T}}{}\!\!^\N_{\!r_1\!a_1\!|r_2a_2}\!(\x))^T\!
(\vec{A}^\N_{r_1\!a_1\!|r_2a_2})^{2}$,
where $\vec{A}^\N_{r_1\!a_1\!|r_2a_2}$ is given by \er{A},
and therefore that
\be\vec{D}^\N_{\!r_1\!a_1\!|r_2a_2}\!(\x,\l\mi\x,\x)\;=\;
(\vec{A}^\N_{r_1\!a_1\!|r_2a_2})^{-1}\;
\tilde{\vec{T}}^\N_{\!r_1\!a_1\!|r_2a_2}\!(\x)\;
\Bigl(\tilde{\vec{T}}^\N_{\!r_1\!a_1\!|r_2a_2}\!(\x)\Bigr)^T\;
\vec{A}^\N_{r_1\!a_1\!|r_2a_2}\,,\ee
where $\ru{2.5}\tilde{\vec{T}}^\N_{\!r_1\!a_1\!|r_2a_2}\!(\x)=
\vec{A}^\N_{r_1\!a_1\!|r_2a_2}
\stackrel{\sss\vee}{\vec{T}}{}\!\!^\N_{\!r_1\!a_1\!|r_2a_2}\!(\x)
(\vec{A}^\N_{r_1\!a_1\!|r_2a_2})^{-1}$.  

This immediately implies that if
$\tilde{\vec{T}}^\N_{\!r_1\!a_1\!|r_2a_2}\!(\x)$ is real, then
$\vec{D}^\N_{\!r_1\!a_1\!|r_2a_2}\!(\x,\l\mi\x,\x)$ has all
nonnegative eigenvalues.

\subsub{Properties Arising from Graph Bicolorability}\label{TMBC}
We now consider some properties which arise if the graph $\G$ is
bicolorable.  Although we have not assumed until now that $\G$
is bicolorable, all of the specific $A$, $D$ and $E$ cases to
be considered in the next section have this property.

Bicolorability of $\G$ means that a parity $\pi_a\in\{-1,1\}$ can
be assigned to each node $a\in\G$ so that adjacent nodes always
have opposite parity; that is,
\be\el{BC}G_{ab}=1\;\Longrightarrow\;\pi_a\,\pi_b=-1\,.\ee

We now note that it follows from \er{RR} that each $\r^r$ with $r$
even/odd is an odd/even polynomial in $G$, which leads to a
generalization of \er{BC} to a selection rule on the fused adjacency
matrices,
\be\el{BCSR}\r^r_{ab}>0\;\Longrightarrow\;\pi_a\,\pi_b=(-1)^{r+1}\,.\ee

Proceeding to the effect of graph bicolorability on the lattice
model, a square lattice can be naturally divided into two
interpenetrating sublattices, with the nearest neighbors of each site
on one sublattice being sites on the other.  Thus, if $\G$ is
bicolorable then the condition that the spin states on neighboring
lattice sites be adjacent nodes on $\G$ implies that, in each spin
assignment, the spin states on one sublattice all have parity $1$,
while those on the other all have parity $-1$.

Using \er{BCSR}, there is also a consistency condition between the
left and right boundary conditions,
\be\el{CC}\pi_{a_1}\,\pi_{a_2}\;=\;(-1)^{r_1+r_2+N}\,,\ee 
since otherwise $\G^\N_{r_1\!a_1\!|r_2a_2}$ is empty and
$\vec{D}^\N_{\!r_1\!a_1\!|r_2a_2}\!(u,\x_1,\x_2)$ and
$\vec{Z}^{\N\MM}_{r_1\!a_1\!|r_2a_2}\!(u,\x_1,\x_2)$ are zero.  If
this condition is satisfied, then the sublattices are fixed by the
boundary conditions; that is, the sublattice which contains all parity
$1$, or all parity $-1$, spin states is the same for all possible 
assignments.

Finally, we note that \er{BCSR} also implies that for each boundary
edge $(b,c)\in\ep^{ra}$, the parities of $b$ and $c$ are respectively
opposite to and the same as $\pi_a\,(-1)^r$, and that therefore any
pair of nodes can appear in only one order in a particular set of
boundary edges.

\newpage
\sect{Critical Unitary $A$--$D$--$E$ Models} 
In this section, we specialize to the critical unitary $A$--$D$--$E$
models, for which $\G$ is an $A$, $D$ or $E$ Dynkin diagram with
Coxeter number $g$ and $\psi$ is the Perron-Frobenius eigenvector of
the adjacency matrix.  In these cases, $\l=\pi/g$ and the regime of
interest here is $0<u<\l$.  This class of models was first identified
and studied in \cite{Pas87a}.  The $A$ and $D$ models are the critical
limits of models introduced in \cite{AndBaxFor84} and \cite{Pas87b}
respectively, but the $E$ models do not have off-critical
counterparts.  We also note that if $\psi$ is instead taken as a
submaximal adjacency matrix eigenvector, then the critical nonunitary
$A$--$D$--$E$ models are obtained.

Fusion of the $A$ models was introduced in
\cite{DatJimMiwOka86,DatJimMiwOka87a,DatJimKunMiwOka88}, while fusion
of the $D$ and $E$ models was first studied in \cite{ZhoPea94a}.  We
shall also consider intertwiner relations among these models, these
having been studied in detail in \cite{Dif92,DifZub90a,Roc90,PeaZho93}.
In particular, we shall find that certain symmetries in the fusion and
intertwiner properties of these models lead to various additional
properties of the boundary conditions, transfer matrices and partition
functions.

We shall also study the explicit forms of the boundary weights and
boundary edge weights for these models. For the $A$ models, various
methods have previously been used to obtain sets of diagonal boundary
weights in \cite{BehPea96,AhnKoo96a,BehPeaObr96,BehPea97} and sets
containing non-diagonal weights in
\cite{BehPea96,FanHouShi95,AhnKoo96a,BehPea97,AhnYou98}, and all of
the $A$ boundary weights found here represent certain cases of these
previously-known weights.  For the $D$ and $E$ models, sets of
diagonal boundary weights were found by direct solution of the
boundary Yang-Baxter equation in \cite{BehPea97}, but most of the sets
obtained here contain non-diagonal weights and were not previously
known.

Finally, we shall consider the connection between the lattice model
boundary conditions at the conformal and isotropic points and the
conformal boundary conditions of the corresponding unitary minimal
theories.

\sub{Graphs and Adjacency Matrices} 
The $A$, $D$ and $E$ Dynkin diagrams with Coxeter number $g$ are
explicitly given by
\setlength{\unitlength}{9mm}
\be\raisebox{-0.5\unitlength}[0.5\unitlength][0.5\unitlength]{
\text{2.3}{1}{0}{0.5}{l}{A_{g-1}\;\;=}
\bpic(6,1)\multiput(0,0.5)(3.5,0){2}{\line(1,0){2.5}}
\multiput(2.7,0.5)(0.4,0){2}{\line(1,0){0.2}}
\multiput(0,0.5)(1,0){2}{\pp{}{\bullet}}
\multiput(5,0.5)(1,0){2}{\pp{}{\bullet}}
\put(0,0.35){\pp{t}{1}}\put(1,0.35){\pp{t}{2}}
\put(5,0.35){\pp{t}{g-2}}\put(6,0.35){\pp{t}{g-1}}\epic
\text{4.35}{1}{4.35}{0.5}{r}{,\qquad g\,=\,2,3,\ldots\,,}}\ee
\be\raisebox{-1\unitlength}[1\unitlength][1\unitlength]{
\text{2.35}{2}{0}{1}{l}{D_{\gg+1}\;\;=}
\bpic(7,2)\multiput(0,1)(3.5,0){2}{\line(1,0){2.5}}
\multiput(2.7,1)(0.4,0){2}{\line(1,0){0.2}}
\put(6,1){\line(2,1){1}}\put(6,1){\line(2,-1){1}}
\multiput(0,1)(1,0){2}{\pp{}{\bullet}}
\multiput(5,1)(1,0){2}{\pp{}{\bullet}}
\multiput(7,0.5)(0,1){2}{\pp{}{\bullet}}
\put(0,0.85){\pp{t}{1}}\put(1,0.85){\pp{t}{2}}
\put(4.9,0.85){\pp{t}{\gg-2}}\put(5.92,0.85){\pp{t}{\gg-1}}
\put(7,1.7){\pp{b}{\gg^{\!-}}}\put(7,0.37){\pp{t}{\gg^{\!+}}}\epic
\text{4.35}{2}{4.35}{1}{r}{,\qquad g\,=\,6,8,\ldots\,,}}\ee
\be\raisebox{-0.5\unitlength}[1.5\unitlength][0.5\unitlength]{
\text{1.9}{2}{0}{0.5}{l}{E_6\;\;=} \bpic(4,2)
\put(0,0.5){\line(1,0){4}}\put(2,0.5){\line(0,1){1}}
\multiput(0,0.5)(1,0){5}{\pp{}{\bullet}}\put(2,1.5){\pp{}{\bullet}}
\put(0,0.35){\pp{t}{1}}\put(1,0.35){\pp{t}{2}}
\put(2,0.35){\pp{t}{3}}\put(3,0.35){\pp{t}{4}}
\put(4,0.35){\pp{t}{5}}\put(2,1.65){\pp{b}{6}}\epic
\text{3.1}{2}{3.1}{0.5}{r}{,\qquad g\,=\,12\,,}}\ee
\be\raisebox{-0.5\unitlength}[1.5\unitlength][0.5\unitlength]{
\text{1.9}{2}{0}{0.5}{l}{E_7\;\;=} \bpic(5,2)
\put(0,0.5){\line(1,0){5}}\put(3,0.5){\line(0,1){1}}
\multiput(0,0.5)(1,0){6}{\pp{}{\bullet}}\put(3,1.5){\pp{}{\bullet}}
\put(0,0.35){\pp{t}{1}}\put(1,0.35){\pp{t}{2}}
\put(2,0.35){\pp{t}{3}}\put(3,0.35){\pp{t}{4}}
\put(4,0.35){\pp{t}{5}}\put(5,0.35){\pp{t}{6}}
\put(3,1.65){\pp{b}{7}}\epic \text{3.1}{2}{3.1}{0.5}{r}{,\qquad
g\,=\,18\,,}}\ee and
\be\raisebox{-0.5\unitlength}[1.5\unitlength][0.5\unitlength]{
\text{1.9}{2}{0}{0.5}{l}{E_8\;\;=} \bpic(6,2)
\put(0,0.5){\line(1,0){6}}\put(4,0.5){\line(0,1){1}}
\multiput(0,0.5)(1,0){7}{\pp{}{\bullet}}\put(4,1.5){\pp{}{\bullet}}
\put(0,0.35){\pp{t}{1}}\put(1,0.35){\pp{t}{2}}
\put(2,0.35){\pp{t}{3}}\put(3,0.35){\pp{t}{4}}
\put(4,0.35){\pp{t}{5}}\put(5,0.35){\pp{t}{6}}
\put(6,0.35){\pp{t}{7}}\put(4,1.65){\pp{b}{8}}\epic
\text{3.1}{2}{3.1}{0.5}{r}{,\qquad g\,=\,30\,.}}\ee We note that when
referring to the nodes $\gg^-$ and $\gg^+$ of $D_{\gg+1}$, we shall
use the convention that if the superscript is not specified then the
choice is immaterial.  Furthermore, if a numerical value for $\gg^-$
or $\gg^+$ is required in any equation, it is to be taken as $\gg$.

The eigenvalues of the adjacency matrices of
these graphs are $2\cos(k_j\,\pi/g)$, $j=1,\ldots,|\G|$, where
$k_j$ are the Coxeter exponents, as given explicitly by
\be\ba{r@{\,:\qquad}l}
\ru{2}A_{g-1}&k_j\;=\;j\,,\quad j=1,\ldots,g\mi1\\
\ru{4}D_{\gg+1}&k_j=\left\{\ba{l}
\ru{1.5}2j-1\,,\quad j=1,\ldots,\gg\\
\gg\,,\quad j=\gg+1\end{array}\right.\\
\ru{2}E_6&(k_1,\ldots,k_6)\;=\;(1,4,5,7,8,11)\\
\ru{2}E_7&(k_1,\ldots,k_7)\;=\;(1,5,7,9,11,13,17)\\
E_8&(k_1,\ldots,k_8)\;=\;(1,7,11,13,17,19,23,29)\,.\ea\ee

It can also be shown that for an integer $k$, and any $A$--$D$--$E$ graph
with Coxeter number $g$, 
\be\el{NU}\ba{l}\ru{1.2}\mbox{$\exists$ an adjacency matrix eigenvector 
with all non-}\\
\mbox{zero entries and associated eigenvalue $2\cos(k\pi/g)$}\ea\;
\;\Longleftrightarrow\;\:\mbox{$k$ is coprime to $g$}\,.\ee
We note, as already mentioned in Section~\ref{MFL}, that \er{NU} implies
that, for any $A$--$D$--$E$ case, the maximum fusion level and Coxeter
number are equal, and that they can thus both be denoted by $g$.

Throughout the rest of Section~4, we shall restrict our attention to
the unitary $A$--$D$--$E$ models, which are associated with the
Perron-Frobenius eigenvector and thus with the Coxeter exponent
$k_1=1$.  Therefore, from this point on, we shall take 
\be\l=\pi/g\ee 
and $\psi$ to be the Perron-Frobenius eigenvector, which can be
explicitly given by
\be\el{ADEPF}\psi_a\;=\;\left\{\begin{array}{@{}ll}
\ru{2.5}\;\;\:S_a\,,\;\;\,a=1,\ldots,g\mi1\;;&A_{g-1}\\
\ru{4.5}\left\{\begin{array}{@{}ll}
\ru{1.5}S_a\,,&a=1,\ldots,\gg-1\\
1\,/\,(2\sin\!\l)\,,&a=\gg\end{array}\right.\;;&D_{\gg+1}\\
\left\{\begin{array}{@{}ll}
\ru{1.5}S_a\,,&a=1,\ldots,l\mi3\\
\ru{1.5}\sin((l\mi1)\l)\,/\sin(2\l)\,,&a=l\mi2\\
\ru{1.5}2\cos((l\mi2)\l)\,,&a=l\mi1\\
\sin((l\mi3)\l)\,/\sin(2\l)\,,&a=l\end{array}\right.\;;&E_l\,,\;l=6,7,8\,,
\end{array}\right.\ee
where, from \er{SD}, $S_a=\sin(a\l)\,/\sin\!\l$.

It is also important here to consider a particular involution
$a\mapsto\ab$ of the nodes of each graph, this being given by the
graph's $\Z_2$ symmetry transformation for the $A$,
$D_{\mbox{\scriptsize odd}}$ and $E_6$ cases and by the identity for
the $D_{\mbox{\scriptsize even}}$, $E_7$ and $E_8$ cases.  More
explicitly, we have
\be\el{ADEI}\ab\;=\;\left\{\ba{@{\:}ll}\ru{2.5}g\mi a\,,\quad&a\in A_{g-1}\\
\ru{2.5}\gg^\mp\,,&a=\gg^\pm\in D_{\gg+1}\,,\quad\gg\pl1\mbox{ odd}\\
\ru{2}6\mi a\,,&a\in E_6\backslash\{6\}\\
\ru{0}a\,,&\mbox{otherwise}\,.\ea\right.\ee
We see that the eigenvector entries \er{ADEPF} are invariant under this
involution,
\be\el{PFI}\psi_{\ab}=\psi_a\,.\ee

We also observe that each $A$, $D$ and $E$ graph is bicolorable and that
we may set the parities as
\be\el{ADEP}\pi_a\;=\;(-1)^a\,.\ee

We now consider the $A$, $D$ and $E$ fused adjacency matrices.  A more
comprehensive treatment of these can be found in Appendix B of
\cite{BehPeaPetZub00}.

For $A_{g-1}$, we have explicitly
\be\el{AFAM}\r^r_{ab}\;=\;
\left\{\begin{array}{@{\:}ll}
\ru{2}1\,;&\mbox{$a\pl b\pl r$ odd, \ $|a-b|\le r\mi1$ \
and \ $r\pl1\le a+b\le 2g\mi r\mi1$}\\
0\,;&\mbox{otherwise}\,.\ea\right.\ee
We note that $\r^r_{ab}$ for $A_{g-1}$, with $r\ne g$, is actually
symmetric in all three indices.

For $D_{\gg+1}$, we have explicitly
\be\el{DFAM}\r^r_{ab}\;=\;
\left\{\begin{array}{@{\;}ll}
2\,;&\ru{1.5}\mbox{$a,b\ne\gg$, \ $a\pl b\pl r$ odd, \ 
$|a-b|\le\gg\mi|\gg\mi r|\mi1$}\\
&\ru{2.5}\quad\mbox{and \ $a\pl b\mi\gg\ge|\gg\mi r|\pl1$}\\
1\,;&\ru{1.5}\mbox{$a,b\ne\gg$, \ $a\pl b\pl r$ odd, \ 
$|a-b|\le\gg\mi|\gg\mi r|\mi1$}\\
&\ru{2.5}\quad\mbox{and \ 
$|a\pl b\mi\gg|\le|\gg\mi r|\mi1$}\\
1\,;&\ru{2.5}\mbox{$a\ne\gg$, \ $b=\gg$, \ $a\pl\gg\pl r$ odd \ and \
$a\ge|\gg\mi r|\pl1$}\\ 
1\,;&\ru{2.5}\mbox{$a=\gg$, \ $b\ne\gg$, \ $b\pl\gg\pl r$ odd \ and \
$b\ge|\gg\mi r|\pl1$}\\ 
1\,;&\ru{2.5}\mbox{$a=\gg^\pm$, \ $b=\gg^\pm$ \ and \ 
$r\equiv1$ mod $4$}\\ 
1\,;&\ru{2.5}\mbox{$a=\gg^\pm$, \ $b=\gg^\mp$ \ and \ 
$r\equiv3$ mod $4$}\\ 
0\,;&\mbox{otherwise}\,.\ea\right.\ee

The fused adjacency matrices for $E_6$ are given explicitly in
Section~\ref{E6}.  We shall not give the $E_7$ and $E_8$ fused
adjacency explicitly, since they can be obtained straightforwardly
using a computer from the recursive definition \er{RR}, but we note
that for $E_7$ each entry is in $\{0,\ldots,4\}$ and that for $E_8$
each entry is in $\{0,\ldots,6\}$.

We now list some properties which apply to all of the $A$, $D$ and $E$
fused adjacency matrices.  These properties can be proved by decomposing 
these matrices in terms of their eigenvalues and eigenvectors as given
in \cite{BehPeaPetZub00}.

For $r=g$ we have
\be\el{FAMg}\r^g\;=\;0\,.\ee
Meanwhile, for $r\in\{1,\ldots,g-1\}$, $\r^r$ form the basis of a
commutative matrix algebra, which is a representation of the Verlinde
fusion algebra.  These $g\mi1$ matrices are therefore often referred
to as Verlinde matrices and denoted $V_r$.

These matrices can also be used to define related intertwiner 
matrices $I^a$, for
each $a\in\G$, with rows labeled by $1,\ldots,g\mi1$, columns
labeled by the nodes of $\G$ and entries given by
\be\el{IM}I^a_{rb}=\r^r_{ab}\,.\ee 
Each $I^a$ then intertwines the
fused adjacency matrices of $\G$ and those of $A_{g-1}$,
\be\el{ITR}I^a\,\r(\G)^r\;=\;\r(A_{g-1})^r\:I^a\,.\ee
For $\G=A_{g-1}$, this property simply amounts to the commutation of the
$A_{g-1}$ fused adjacency matrices, but if $\G$ is a $D$ or $E$ graph
it forms the basis of various relationships between $\G$ and the $A$
graph with the same Coxeter number.

Finally, a property of the $A$, $D$ and $E$ fused adjacency matrices of
particular relevance here is that, for $r\in\{1,\ldots,g\mi1\}$, 
\be\el{FAMI}\r^{g-r}_{ab}\;=\;\r^r_{\ab b}\;=\;\r^r_{a\bb}\,.\ee
An important special case of this is $r=1$, for which
\be\el{FAMIg}\r^{g-1}_{ab}\;=\;\delta_{a\bb}\,.\ee

\sub{Bulk Weights, Fusion Matrices and Fusion Vectors}\label{BWFMFV}
We now consider the bulk weights, fusion matrices and fusion vectors
of the critical unitary $A$, $D$ and $E$ models.

We see from \er{W} and \er{ADEPF} that the $A_{g-1}$ bulk weights are
\be\ba{rcl}
\ru{3.5}
\W{a}{a\mipl1}{a}{a\plmi1}{u}&=&\ds s_1(-u)\\
\el{WA}\ru{3.5}\W{a\mipl1}{a}{a\plmi1}{a}{u}&=&\ds
\frac{(S_{a-1}\:S_{a+1})^\hf\:s_0(u)}{S_a}\\
\W{a\plmi1}{a}{a\plmi1}{a}{u}&=&\ds\frac{s_a(\pm u)}{S_a}\,.\ea\ee

It can also be shown, using \er{PE}, \er{WX} and \er{WA} together
with certain results on the fusion of $A_{g-1}$ bulk weights from
\cite{DatJimKunMiwOka88}, that the $A_{g-1}$ fusion matrices are
given explicitly by
\be\el{AFM}\ba{l}\ru{3}P^r(a,b)_{(a,c_1,\ldots,c_{r\mi2},b),
(a,d_1,\ldots,d_{r\mi2},b)}\;=\\
\quad\ds\Biggl(\,\prod_{m=2}^{\frac{r+a-b-1}{2}}\!\!\!S_m\Biggr)
\Biggl(\,\prod_{m=2}^{\frac{r-a+b-1}{2}}\!\!\!S_m\Biggr)
\Biggl(\,\prod_{m=\frac{a+b-r+1}{2}}^{\frac{a+b+r-1}{2}}\!S_m\Biggr)
\Biggl(\,\prod_{m=0}^{r-1}\frac{\e_{c_m}\:
\e_{d_m}}{(S_{c_m}\,S_{d_m})^\hf}\Biggr)
\bigg/\Biggl(\,\prod_{m=2}^{r-1}S_m\!\Biggr),\ea\ee
where in the fourth product we set $c_0=d_0=a$ and $c_{r-1}=d_{r-1}=b$, 
and where
\be\e_a\;=\;\left\{\ba{r@{}l}1&,\quad a\equiv0\mbox{ or }1\mbox{ mod }4\\
-1&,\quad a\equiv2\mbox{ or }3\mbox{ mod }4\,.\ea\right.\ee
In fact, the only properties of the sign factors 
required here are $\e_a\in\{-1,1\}$
and $\e_{a\mi1}\,\e_{a\pl1}=-1$, so any of the three other assignments
which satisfy these could be used instead.

We can see from \er{AFM} that, in keeping with \er{r} and \er{AFAM},
each nonzero $A_{g-1}$ fusion matrix has rank $1$ and that the 
corresponding fusion
vectors, which are thus uniquely defined up to sign, are given, 
up to this sign, by
\be\el{UA}\ba{l}\ru{2.5}U^r(a,b)_{1,(a,c_1,\ldots,c_{r\mi2},b)}\;=\\
\qquad\ds
\Biggl[\Biggl(\,\prod_{m=2}^{\frac{r+a-b-1}{2}}\!\!\!S_m\Biggr)
\Biggl(\,\prod_{m=2}^{\frac{r-a+b-1}{2}}\!\!\!S_m\Biggr)
\Biggl(\,\prod_{m=\frac{a+b-r+1}{2}}^{\frac{a+b+r-1}{2}}\!S_m\Biggr)
\bigg/\Biggl(\,\prod_{m=2}^{r-1}S_m\Biggr)\Biggr]^{\!\hf}
\Biggl(\,\prod_{m=0}^{r-1}\frac{\e_{c_m}}{S^\hf_{c_m}}\Biggr),\ea\ee
where in the last product we set $c_0=a$ and $c_{r-1}=b$.

Proceeding to $D_{\gg+1}$, it is possible to write expressions,
similarly explicit to those for $A_{g-1}$, for the bulk weights and
entries of the fusion matrices and fusion vectors, for a certain
natural choice of these vectors.  Since each of these expressions
involves many different cases, analogous to those of \er{DFAM}, we do
give them here.  However, we note that, as with $A_{g-1}$, each
$D_{\gg+1}$ bulk weight and fusion vector entry can be expressed as a
single product of terms.

We also note that each $D_{\gg+1}$ bulk weight and fusion matrix entry
can be expressed as a linear combination of $A_{g-1}$ bulk weights and
fusion matrix entries, with the coefficients being products of entries
of so-called intertwiner cells.  These intertwiner cells, whose exact
properties are outlined in \cite{Dif92,DifZub90a,Roc90,PeaZho93},
serve a similar role at the level of the bulk weights to that served
at the level of the adjacency matrices by the intertwiner matrices
\er{IM}.  While the expressions provided by these intertwiner cells
may not be particularly compact when applied to specific cases, they
are still useful for deriving certain general properties using known
properties of the intertwiner cells and of the $A_{g-1}$ bulk weights
and fusion matrices.

With regard to the $E$ graphs, the number of different cases is
particularly large so that the numerical values of fusion matrix and
fusion vector entries are probably best evaluated using a computer.
While this may result in a somewhat unnatural choice of the fusion
vectors, this is largely immaterial since the lattice properties of
interest are independent of this choice.  We also note that, exactly
as with $D_{\gg+1}$, each $E$ bulk weight and fusion matrix entry can
be expressed as a linear combination of $A_{g-1}$ bulk weights or
fusion matrix entries using intertwiner cells and that these
expressions can be used to obtain general properties of the $E$ bulk
weights, fusion matrices and fusion vectors.

We now consider some properties which apply to all of the $A$, $D$ and $E$
graphs. We first note, using \er{PFI}, that the bulk weights 
are invariant under the involution \er{ADEI}, 
\be\el{ADEWI}\W{a}{b}{c}{d}{u}\;=\;\W{\ab}{\bb}{\bar{c}}{\bar{d}}{u}\,.\ee

We also see similarly that for the fusion matrices,
\be\el{ADEFMI}
P^r(a,b)_{(a,c_1,\ldots,c_{r\mi2},b),(a,d_1,\ldots,d_{r\mi2},b)}\;=\;
P^r(\ab,\bb)_{(\ab,\bar{c}_1,\ldots,\bar{c}_{r\mi2},\bb),
(\ab,\bar{d}_1,\ldots,\bar{d}_{r\mi2},\bb)}\,.\ee
It follows from this that $U^r(a,b)_{\a,(a,c_1,\ldots,c_{r\mi2},b)}$ and 
$U^r(\ab,\bb)_{\a,(\ab,\bar{c}_1,\ldots,\bar{c}_{r\mi2},\bb)}$
correspond to two orthonormal decompositions of both 
$P^r(a,b)$ and $P^r(\ab,\bb)$, and therefore that these fusion
vectors are related by a transformation \er{UGT},
\be\el{ADEUI}U^r(\ab,\bb)_{\a,(\ab,\bar{c}_1,\ldots,\bar{c}_{r\mi2},\bb)}\;=\;
\sum_{\a'=1}^{\r^r_{ab}}\;R^r(a,b)_{\a\a'}\;
U^r(a,b)_{\a',(a,c_1,\ldots,c_{r\mi2},b)}\,.\ee

Finally, we note that the fusion matrices
$P^{g-1}(a,\ab)$, which from \er{FAMIg} all have rank~1, can be used to 
generate the fusion matrices of lower fusion level according to
\be\el{ADEFMP}\ba{l}\ds\ru{3}
\!\!\!\!\sum_{(b,e_1,\ldots,e_{g\mi r\mi2},\ab)\:\in\:\G^{g-r}_{b\ab}}
\!\!\!P^{g-1}(a,\ab)_{(a,c_1,\ldots,c_{r\mi2},b,e_1,\ldots,e_{g\mi r\mi2},\ab),
(a,d_1,\ldots,d_{r\mi2},b,e_1,\ldots,e_{g\mi r\mi2},\ab)}\\
\ds\qquad\qquad\qquad\qquad\qquad\qquad\qquad\quad
=\;\frac{\psi_b}{S_r\,\psi_a}\;P^r(a,b)_{(a,c_1,\ldots,c_{r\mi2},b),
(a,d_1,\ldots,d_{r\mi2},b)}\,.\ea\ee
This can be proved by first obtaining the result for the $A$ graphs
using \er{AFM} and then proceeding to the $D$ and $E$ graphs using
intertwiner cells.

\sub{\protect Additional Boundary Condition and Transfer Matrix\\Properties}
We now show that in addition to satisfying all of the properties
outlined in Section~\ref{TMP}, including those of Section~\ref{TMBC}
arising from bicolorability with parity \er{ADEP}, the critical
unitary $A$--$D$--$E$ models possess further important symmetries
associated with the involutions \er{ADEI} and the intertwiner
relations \er{IM}.

\subsub{Symmetry under $a_1\mapsto\ab_1$ and $a_2\mapsto\ab_2$}
We first consider the relationship between the $(r,a)$ and $(r,\ab)$ 
boundary conditions. It follows straightforwardly from \er{FAMI} and \er{BE} 
that the sets of boundary edges for these boundary conditions are related by
\be\el{BPS1}\ep^{r\ab}\;=\;
\bigl\{\bigl(\bb,\bar{c}\bigr)\,\big|\,(b,c)\in\ep^{ra}\bigr\}\,.\ee

Proceeding to the corresponding boundary edge weights and boundary weights, 
we find, using \er{ADEUI}, that
\be\el{EWS1}E^{r\ab}(\bb,\bar{c})_{\b\g}\;=\;
\sum_{\b'=1}^{\r^r_{ba}}\;\sum_{\g'=1}^{\r^{r+1}_{ca}}\;
S^{ra}(b)_{\b\b'}\;S^{r+1,a}(c)_{\g\g'}\;E^{ra}(b,c)_{\b'\g'}\ee
and 
\be\el{ADEBWI}\B{r\ab}{\bb}{\b}{\bar{c}}{\bar{d}}{\d}{u,\x}\;=\;
\sum_{\b'=1}^{\r^r_{ba}}\;\sum_{\d'=1}^{\r^r_{da}}\;
S^{ra}(b)_{\b\b'}\;S^{ra}(d)_{\d\d'}\;\B{ra}{b}{\b'}{c}{d}{\d'}{u,\x},\ee
where the orthonormal transformation matrices in \er{ADEUI} and
those in \er{EWS1} and \er{ADEBWI} are related by \er{SGT}.

It now follows from \er{ADEWI} and \er{ADEBWI} that 
$\vec{D}^\N_{\!r_1\!\ab_1\!|r_2\ab_2}\!(u,\x_1,\x_2)$
and $\vec{D}^\N_{\!r_1\!a_1\!|r_2a_2}\!(u,\x_1,\x_2)$
are related by a similarity transformation,
\be\el{DIS1}\ba{l}\ru{2}
\vec{D}^\N_{\!r_1\!\ab_1\!|r_2\ab_2}\!(u,\x_1,\x_2)\;
\vec{H}^\N_{r_1\!a_1\!|r_2a_2}\;
\vec{S}^\N_{r_1\!a_1\!|r_2a_2}\;=\\
\qquad\qquad\qquad\qquad\qquad\qquad\qquad\qquad
\vec{H}^\N_{r_1\!a_1\!|r_2a_2}\;
\vec{S}^\N_{r_1\!a_1\!|r_2a_2}\;
\vec{D}^\N_{\!r_1\!a_1\!|r_2a_2}\!(u,\x_1,\x_2)\,,\ea\ee
where $\vec{S}^\N_{r_1\!a_1\!|r_2a_2}$ is given by
\er{GIS}, using the same $S^{ra}(e)$ as in \er{ADEBWI},
and $\vec{H}^\N_{r_1\!a_1\!|r_2a_2}$ is a square matrix
with rows labeled by the paths of $\G^\N_{r_1\!\ab_1\!|r_2\ab_2}$,
columns labeled by the paths of $\G^\N_{r_1\!a_1\!|r_2a_2}$
and entries given by
\be\vec{H}^\N_{r_1\!a_1\!|r_2a_2\!\raisebox{-0.3ex}{
$\ss(\b_1,b_0,\ldots,b_N,\b_2),(\d_1,d_0,\ldots,d_N,\d_2)$}}\;=\;
\delta_{(\b_1,\bb_0,\ldots,\bb_N,\b_1),(\d_1,d_0,\ldots,d_N,\d_2)}\,.\ee

\subsub{\protect Equivalence of $(r,a)$ and $(g\mi r\mi1,\ab)$ Boundary 
Conditions at the\\Conformal Point}
We now show that, at the conformal point and for
$r\in\{1,\ldots,g\mi2\}$, the $(r,a)$ and $(g\mi r\mi1,\ab)$ boundary
conditions are equivalent.  This equivalence takes the form of the
boundary edge weights for each of these boundary conditions being the
same, except for a gauge transformation and a reversal of the order of
the nodes in each boundary pair, neither of which affects any
properties of interest. We shall denote such an equivalence of
boundary conditions by $\leftrightarrow$.

In terms of the set of boundary edges \er{BE}, it follows
from \er{FAMI} that, for $r\in\{1,\ldots,g\mi2\}$,
the sets $\ep^{ra}$ and $\ep^{g-r-1,\ab}$ contain the same pairs of
nodes but with opposite ordering; that is, 
\be\el{BPS2}
\ep^{g-r-1,\ab}\;=\;\bigl\{(c,b)\,\big|\,(b,c)\in\ep^{ra}\bigr\}\,.\ee
We also see, from \er{FAMg}, that 
\be\ep^{g-1,a}\;=\;\emptyset\,,\ee
so that at the conformal point there are no $(g\mi1,a)$ boundary conditions.

For the boundary edge weights, the equivalence of
the $(r,a)$ and $(g\mi r\mi1,\ab)$ boundary conditions is
\be\el{EWS2}E^{g-r-1,\ab}(c,b)_{\g\b}\;=\;
\sum_{\b'=1}^{\r^r_{ba}}\;\sum_{\g'=1}^{\r^{r+1}_{ca}}\;
S^{ra}(b)_{\b\b'}\;S^{r+1,a}(c)_{\g\g'}\;E^{ra}(b,c)_{\b'\g'}\,,\ee
where $S^{ra}(e)$ are orthonormal matrices satisfying \er{SON}
which are defined by
\be\ba{l}\ds\ru{2.5}\el{S2}S^{ra}(e)_{\a\a'}\;=\;
(S_r\,\psi_a\,/\psi_e)^\hf\;\times\\
\ds\ru{4.5}
\sum_{(e,b_1,\ldots,b_{g\mi r\mi2},\ab)\:\in\:\G^{g-r}_{e\ab}}\;
\sum_{(e,c_1,\ldots,c_{r\mi2},a)\:\in\:\G^r_{ea}}\!\!
U^{g-1}(\ab,a)_{1,(\ab,b_{g\mi r\mi2},\ldots,b_1,e,c_1,\ldots,c_{r\mi2},a)}
\;\times\\
\qquad\qquad\qquad\qquad\qquad\qquad
U^{g-r}(e,\ab)_{\a,(e,b_1,\ldots,b_{g\mi r\mi2},\ab)}\;
U^{r}(e,a)_{\a',(e,c_1,\ldots,c_{r\mi2},a)}\,.\ea\ee
We note that $S^{ra}(e)$ also satisfy
\be\el{ST}S^{ra}(e)^T\;=\;S^{g-r,\ab}(e)\,.\ee
These relations, $\er{EWS2}$ and $\er{ST}$, and the orthonormality
$\er{SON}$ can all be proved using the general properties of the
fusion matrices and fusion vectors, \er{FMP}, \er{FMT}, \er{FMD},
\er{FMON} and those which follow from \er{PPG}, together with 
\er{ADEFMP}.

We now note that, as labels, $(r,a)$ and $(g\mi r\mi1,\ab)$ are always
distinct since for an $A_{\mbox{\scriptsize odd}}$, $D$ or $E$ graph
$g$ is even so that $r$ and $g\mi r\mi1$ are different, while for an
$A_{\mbox{\scriptsize even}}$ graph $g$ is odd so that $a$ and
$\ab=g\mi a$ are different.  Thus, due to the equivalence \er{EWS2},
there are at most $(g\mi2)\,|\G|/2$ distinct boundary conditions.
Furthermore, by examining the sets of boundary edges \er{BE} 
for each $A$, $D$ and $E$ case,
we find that there are no further equivalences between these sets,
either direct or through reversing the order of nodes in each edge.  We
therefore conclude that
\be\mbox{number of boundary conditions at the conformal
point}\;=\;(g \mi2)\,|\G|/2\,.\ee

Due to the consistency condition \er{CC},
the implementation of a given left and right boundary condition 
at their conformal points on a lattice of fixed width can be achieved using 
only two of the four possibilities which arise from the two
versions of each boundary condition.  If $(r_1,a_1)$ and $(r_2,a_2)$
is one of these possibilities, then the other is 
$(g\mi r_1\mi1,\ab_1)$ and $(g\mi r_2\mi1,\ab_2)$ and
the transfer matrices for the two are related by
\be\ba{l}\ru{2}
\vec{D}^\N_{\!g-r_1\mi1,\ab_1\!|g-r_2-\!1,\ab_2}\!(\x,\l\mi\x,\x)\;=\\
\qquad\qquad\vec{S}^\N_{\!r_1\pl1,a_1\!|r_2+\!1,a_2}\;
\stackrel{\sss\wedge}{\vec{T}}{}\!\!^\N_{\!r_1\!a_1\!|r_2a_2}\!(\l\mi\x)\;
\stackrel{\sss\vee}{\vec{T}}{}\!\!^\N_{\!r_1\!a_1\!|r_2a_2}\!(\l\mi\x)\;
(\vec{S}^\N_{\!r_1\pl1,a_1\!|r_2+\!1,a_2})^{-1}\,,\ea\ee
where $\vec{S}^\N_{\!r_1\pl1,a_1\!|r_2+\!1,a_2}$ is given by \er{GIS}
and \er{S2}.  Comparing this with \er{DD} we see that the ordering in
the products of single row transfer matrices is different in each,
which does not affect the nonzero eigenvalues (although unimportant
differences in the number of zero eigenvalues will arise due to the
different dimensions of the oppositely-ordered products). We thus see
that the partition functions are related by
\be\el{ZS2}\vec{Z}^{\N\MM}_{g-r_1\mi1,\ab_1\!|g-r_2-\!1,\ab_2}\!
(\x,\l\mi\x,\x)\;=\;
\vec{Z}^{\N\MM}_{r_1\!a_1\!|r_2a_2}\!(\l\mi\x,\x,\l\mi\x)\,.\ee

Finally, we find, using \er{K2} in \er{B}, that 
the boundary edge weight relation
\er{EWS2} implies the boundary weight relation
\be\el{BWS2}\B{g-r,\ab}{b}{\b}{c}{d}{\d}{u,\x}\;=\;
\sum_{\b'=1}^{\r^r_{ba}}\;\sum_{\d'=1}^{\r^r_{da}}\;
S^{ra}(b)_{\b\b'}\;S^{ra}(d)_{\d\d'}\;\B{ra}{b}{\b'}{c}{d}{\d'}{u,-\x},\ee
with $S^{ra}(e)$ again given by \er{S2}. This then implies that 
\be\vec{Z}^{\N\MM}_{r_1\!a_1\!|r_2a_2}\!(u,\x_1,\x_2)\;=\;
\vec{Z}^{\N\MM}_{g-r_1,\ab_1\!|r_2a_2}\!(u,-\x_1,\x_2)\;=\; 
\vec{Z}^{\N\MM}_{r_1\!a_1\!|g-r_2,\ab_2}\!(u,\x_1,-\x_2)\,.\ee
This relation taken at $u=\l\mi\x_1=\x_2=\l\mi\x$ is consistent
with \er{ZS2} through the equivalence, which
follows from \er{BM}, of the $(r,a)$ boundary condition at $u=-\x$ 
and the $(r\mi1,a)$ boundary condition at $u=\x$. 

\subsub{Intertwiner Symmetry}
We now consider the relationship between the double-row transfer
matrices and partition functions of the model based on $\G$, with
Coxeter number $g$, and those of the model based on $A_{g-1}$.  In
particular, we shall find that any critical unitary $A$--$D$--$E$
partition function can be expressed as a sum of certain $A$ partition
functions.

It can be shown using intertwiner cells that, 
for each $r_1,r_2,s'\in\{1,\ldots,g\mi1\}$ and $a_1,a_2\in\G$, we have
\be\el{DITR}\ba{rcl}\ru{3.5}
\ds\bigoplus_{a\in\G}{}^{\r^{s'}_{a_1\!a}}\;
\vec{D}^\N_{\!r_1\!a|r_2a_2}(u,\x_1,\x_2)&\approx&\ds
\bigoplus_{s=1}^{g-1}{}^{\r^{s}_{a_1\!a_2}}\;
\vec{D}^{\N\!,\,A_{g-1}}_{\!r_1\!s'|r_2s}(u,\x_1,\x_2)\\
\ds\bigoplus_{a\in\G}{}^{\r^{s'}_{a_2a}}\;
\vec{D}^\N_{\!r_1\!a_1\!|r_2a}(u,\x_1,\x_2)&\approx&\ds
\bigoplus_{s=1}^{g-1}{}^{\r^{s}_{a_1\!a_2}}\;
\vec{D}^{\N\!,\,A_{g-1}}_{\!r_1\!s|r_2s'}(u,\x_1,\x_2)\,,\ea\ee
where we are using $\approx$ and the superscripts on $\bigoplus$ in the
same ways as in \er{DII}, and where the fused adjacency matrices on
both sides and the transfer matrices on the left sides refer to $\G$,
while the transfer matrices on the right sides refer, as indicated, to
$A_{g-1}$.  

Since the two forms of this relation can be proved similarly, and are
related through the symmetries of Sections \ref{LR} and \ref{IS}, we
shall consider, from now on, only the first form.

We first discuss the details of the proof.  The similarity
transformation in the first line of \er{DITR} can be implemented by an
invertible matrix $\vec{J}^{s'\N}_{r_1\!a_1\!|r_2a_2}$ which
premultiplies the left side and postmultiplies the right side.  The
rows of $\vec{J}^{s'\N}_{r_1\!a_1\!|r_2a_2}$ and the rows and columns
of $\ds\bigoplus_{s=1}^{g-1}{}^{\r^{s}_{a_1\!a_2}}
\vec{D}^{\N\!,\,A_{g-1}}_{\!r_1\!s'|r_2s}(u,\x_1,\x_2)$
are labeled by the paths of
\[\Bigl\{(t_0,\ldots,t_\N,t,\a_2)\,\Big|\,
(1,t_0,\ldots,t_\N,1)\!\in\!(A_{g-1})^\N_{r_1\!s'|r_2t},\,
\r^t_{a_1a_2}\!>\!0,\,\a_2\!\in\!\{1,\ldots,\r^t_{a_1a_2}\}\Bigr\}\,,\]
while the columns of $\vec{J}^{s'\N}_{r_1\!a_1\!|r_2a_2}$ and the
rows and columns of $\ds\bigoplus_{a\in\G}{}^{\r^{s'}_{a_1\!a}}
\vec{D}^\N_{\!r_1\!a|r_2a_2}(u,\x_1,\x_2)$
are labeled by the paths of 
\[\Bigl\{(\a_1,b,\b_1,b_0,\ldots,b_\N,\b_2)\,\Big|\,
(\b_1,b_0,\ldots,b_\N,\b_2)\!\in\!\G^\N_{r_1\!b|r_2a_2},\,
\r^{s'}_{a_1b}\!>\!0,\,\a_1\!\in\!\{1,\ldots,\r^{s'}_{a_1b}\}\Bigr\}\,.\]
The entries of these matrices are given by
\be\el{J}\ba{l}\ru{3}\ds\vec{J}^{s'\N}_{r_1\!a_1\!|r_2a_2\!
\raisebox{-0.3ex}{
$\ss(t_0,\ldots,t_N,t,\a_2),(\a_1,b,\b_1,b_0,\ldots,b_N,\b_2)$}}\,\;=\\
\qquad\ds\sum_{\g_0=1}^{\r^{t_0}_{\!a_1b_0}}\ldots\!
\sum_{\g_N=1}^{\r^{t_N}_{\!a_1b_N}}
\Qf{r_1a_1\!}{s'}{t_0}{\g_0}{b_0}{\b_1}{b}{\a_1}
\left[\prod_{j=0}^{\N-1}
\Q{a_1\!}{t_j}{t_{j+\!1}}{\g_{j+\!1}}{b_{j+\!1}}{b_j}{\g_j}\right]
\Qf{r_2a_1\!}{t_\N}{t}{\a_2}{a_2}{\b_2}{b_\N}{\g_\N}\,,\ea\ee
\be\ba{l}\ru{4.5}\ds\Biggl[\,\bigoplus_{a\in\G}{}^{\r^{s'}_{a_1\!a}}\;
\vec{D}^\N_{\!r_1\!a|r_2a_2}(u,\x_1,\x_2)
\Biggr]_{(\a_1,b,\b_1,b_0,\ldots,b_\N,\b_2),
(\a_1',d,\d_1,d_0,\ldots,d_\N,\d_2)}\;=\\
\qquad\qquad\qquad\qquad\qquad\d_{\a_1{\a_1}'}\;\d_{bd}\;
\vec{D}^\N_{\!r_1\!b|r_2a_2}(u,\x_1,\x_2)_{(\b_1,b_0,\ldots,b_\N,\b_2),
(\d_1,d_0,\ldots,d_\N,\d_2)}\ea\ee
and similarly for $\ru{3}\ds\bigoplus_{s=1}^{g-1}{}^{\r^{s}_{a_1\!a_2}}
\vec{D}^{\N\!,\,A_{g-1}}_{\!r_1\!s'|r_2s}(u,\x_1,\x_2)$.  In \er{J},
$Q^{a_1}$ are intertwiner cells associated with the intertwiner
matrix $I^{a_1}$ of \er{IM}, and $Q^{r_1a_1}$ and $Q^{r_2a_1}$ are 
fused blocks of such cells, of widths $r_1\mi1$ and $r_2\mi1$ respectively.
Thus, $\vec{J}^{s'\N}_{r_1\!a_1\!|r_2a_2}$ can be viewed as a row of
intertwiner cells in which the spins on the lower left and upper right
corners are fixed to $s'$ and $a_2$ respectively, the lower row
of spins between $s'$ and $a_2$ on $A_{g-1}$ label the rows of the
matrix and the upper row of spins between $s'$ and $a_2$ on $\G$
label the columns of the matrix. We also see that all of the
matrices here are square with dimension
$(I^{a_1}\r^{r_1}G^\N\r^{r_2})_{s'a_2}$, it being possible using
\er{ITR} to propagate $I^{a_1}$ to the right of this expression while
replacing the adjacency matrices of $\G$ with those of $A_{g-1}$.
The intertwiner cells are assumed to
satisfy an intertwiner relation, as given in
(4.6a) of \cite{DifZub90a}, as well as two inversion relations, as
given in (4.6b) and (4.6c) of \cite{DifZub90a}.  It follows
immediately from the inversion relations that
$\vec{J}^{s'\N}_{r_1\!a_1\!|r_2a_2}$
is invertible, with its inverse being given, up to gauge
transformations on the intertwiner cells, by its transpose.  Meanwhile,
the equation corresponding to the first line of \er{DITR} can be
obtained by using one of
the inversion relations to insert a pair of cells 
between $\vec{J}^{s'\N}_{r_1\!a_1\!|r_2a_2}$ and
$\ru{3}\ds\bigoplus_{a\in\G}{}^{\r^{s'}_{a_1\!a}}
\vec{D}^\N_{\!r_1\!a|r_2a_2}(u,\x_1,\x_2)$, using the intertwiner
relation and the form \er{CE} of the boundary weights
to propagate these cells around a single loop, and then using
an inversion relation again to remove the inserted cells, thus giving
$\ru{3}\ds\bigoplus_{s=1}^{g-1}{}^{\r^{s}_{a_1\!a_2}}
\vec{D}^{\N\!,\,A_{g-1}}_{\!r_1\!s'|r_2s}(u,\x_1,\x_2)\,
\vec{J}^{s'\N}_{r_1\!a_1\!|r_2a_2}$ and completing the proof.
We note that this last process is probably best understood
diagrammatically, with the lower part of the loop involving the
conversion in the lower row of the transfer matrix from $\G$ weights
to $A_{g-1}$ weights and the propagation of the row of intertwiner
cells to a position between the rows of the transfer matrix, and the
upper part of the loop involving the conversion of weights in the
upper row of the transfer matrix and the propagation of the row of
cells to its final position.

We note that \er{DITR} is still nontrivial for $\G=A_{g-1}$ and that
it then corresponds to certain cases of \er{DII}.  In fact,
the $A_{g-1}$ intertwiner cells $Q^a$ can be obtained by taking a
$u\rightarrow i\infty$ limit on fused $a\mi1$ by $1$ blocks of
$A_{g-1}$ bulk weights.

We also note that the intertwiner cells $Q^a$ for $\G$ a $D$ or $E$ 
graph have only been found explicitly, in 
\cite{Dif92,DifZub90a,Roc90,PeaZho93}, for $a=1$.  However, for
various reasons, the existence of these cells for other values of
$a$ seems guaranteed.

Finally, we observe that a particularly important case of 
\er{DITR} is $s'=1$, for which
\be\el{DITR1}
\vec{D}^\N_{\!r_1\!a_1\!|r_2a_2}(u,\x_1,\x_2)\;\approx\;
\bigoplus_{s=1}^{g-1}{}^{\r^{s}_{a_1\!a_2}}\;
\vec{D}^{\N\!,\,A_{g-1}}_{\!r_1\!1|r_2s}(u,\x_1,\x_2)\,,\ee
which in turn implies that
\be\el{ZITR}
\vec{Z}^{\N\MM}_{r_1\!a_1\!|r_2a_2}\!(u,\x_1,\x_2)\;=\;
\sum_{s=1}^{g-1}\:\r^{s}_{a_1a_2}\;
\vec{Z}^{\N\MM\!,\,A_{g-1}}_{r_1\!1|r_2s}(u,\x_1,\x_2)\,.\ee
We thus see that the task of evaluating the partition functions of all
of the $A$--$D$--$E$ models with boundary conditions $(r_1,a_1)$ and
$(r_2,a_2)$ has been reduced to that of evaluating the partition
functions of just the $A$ models with boundary conditions $(r_1,1)$
and $(r_2,s)$.

\sub{Boundary Weights}
We now consider the explicit forms of the boundary weights and
boundary edge weights for the critical unitary $A$--$D$--$E$ models.

\subsub{$A$ Graphs} 
For $A_{g-1}$, we find, using \er{UA} in \er{BEW} and then applying a
simple gauge transformation to remove a factor $\e_c$ which arises,
that the boundary edge weights are given explicitly by
\be\el{ABEW}E^{ra}(c\plmi1,c)_{11}\;=\;
\frac{(S_{(r\mp c+a)\!/2}\:S_{(c\pm a\mp r)\!/2})^\hf}
{(S_{c\pm1}\:S_c)^\qt}\,.\ee 

We see that each of these weights is
positive.  We also see that, for these weights, the relations \er{EWS1}
and \er{EWS2} are 
\be E^{r,g-a}(g\mi b,g\mi c)_{11}\;=\;E^{ra}(b,c)_{11}\,,\quad
E^{g-r-1,g-a}(c,b)_{11}\;=\;E^{ra}(b,c)_{11}\,.\ee 

Substituting
\er{ABEW} into \er{GBW} we now find that the $A_{g-1}$ boundary weights
are
\be\el{ABW}\ba{l}\ru{3.5}\B{ra}{c\plmi1}{1}{c}{c\plmi1}{1}{u,\xi}\;=\\
\ds\ru{5}\;\frac{S_{(r\mp c+a)\!/\!2}\,S_{(c\pm a\mp r)\!/\!2}\,
s_0(\x\pl u)\,s_r(\x\mi u)+ S_{(r\pm c+a)\!/\!2}\,S_{(c\mp a\pm
r)\!/\!2}\, s_0(\x\mi u)\,s_r(\x\pl u)}
{S_r\,(S_c\,S_{c\pm1})^\hf\,s_0(2\x)}\\
\ds\B{ra}{c\plmi1}{1}{c}{c\mipl1}{1}{u,\xi}\;=\;
\frac{(S_{(r-c+a)\!/2}\,S_{(r+c-a)\!/2}\,S_{(c+a-r)\!/2}\,
S_{(c+a+r)\!/2})^\hf\:s_0(2u)}{(S_{c-1}\,S_{c+1})^\qt\:S_c^\hf\:s_0(2\x)}
\,.\ea\ee 
All of the boundary weights which were found as solutions of the
boundary Yang-Baxter equation for the $A_{g-1}$ models and their
off-critical extensions in
\cite{BehPea96,FanHouShi95,AhnKoo96a,BehPeaObr96,BehPea97,AhnYou98}
can be related to those of \er{ABW} by using appropriate values for
the various parameters involved.  In particular, we note that more
general boundary weights which depend on two boundary field parameters
are known, as for example given in (3.14--3.15) of \cite{BehPea96},
and that these reduce to the weights \er{ABW} when one of these
parameters is set to $i\infty$.

Some important special cases here are the
$(r,1)$ and $(g\mi r\mi1,g\mi1)$ boundary conditions for which
\be\ba{c}\ds\ru{2.5}\ep^{r1}\;=\;\{(r,r\pl1)\}\,,\qquad
\ep^{g-r-1,g-1}\;=\;\{(r\pl1,r)\}\,,\\
E^{r1}(r,r\pl1)\,=\,E^{g-r-1,g-1}(r\pl1,r)\,=\,(S_r\:S_{r+1})^\qt\,.\ea\ee
The corresponding boundary weights are all diagonal and given by
\be\ba{l}\ru{3}\B{r1}{r}{1}{r\plmi1}{r}{1}{u,\x}\;=\;
\B{g-r,g-1}{r}{1}{r\plmi1}{r}{1}{u,-\x}\;=\\
\qquad\qquad\qquad\qquad\qquad\qquad\qquad\qquad\qquad\qquad\qquad\ds
\frac{S_{r\pm1}^\hf\:s_0(\x\plmi u)\:s_r(\x\mipl u)}{S_r^\hf\:s_0(2\x)}\,,
\ea\ee
these weights matching those found in \cite{BehPeaObr96}.  We see that
for $r\in\{2,\ldots,g\mi2\}$ these cases provide an example in which
boundary weights which are nonzero away from the conformal point
vanish at the conformal point.  More specifically, away from the
conformal point, these cases represent semi-fixed boundary conditions
in which the state of every alternate boundary spin is fixed to be $r$
and that of each spin between these can be $r-1$ or $r+1$, while
at the conformal point they represent completely fixed boundary
conditions in which only a single boundary spin configuration 
$\ldots r,r+1,r,r+1\ldots$ contributes to the partition function.

Other important cases here are the $(1,a)\leftrightarrow(g\mi2,g\mi a)$ 
boundary conditions, with $a\in\{2,\ldots,g\mi2\}$, for which
\be\ba{c}\ds\ru{2.5}\ep^{1a}\;=\;\{(a,a\mi1),(a,a\pl1)\}\,,\quad
\ep^{g-2,g-a}\;=\;\{(a\mi1,a),(a\pl1,a)\}\,,\\
\ds E^{1a}(a,a\plmi1)\;=\;E^{g-2,g-a}(a\plmi1,a)\;=\;
(S_{a\pm1}/S_a)^\qt\,.\ea\ee
As already seen for the general case in \er{B1a}, 
these are semi-fixed boundary conditions in which
the state of every alternate boundary spin is fixed to be $a$.

We now observe that the only cases in which every edge of $A_{g-1}$ 
appears, in one order, in the set of boundary edges are
\be\ba{rl}\ru{5.5}g\mbox{ odd:}&
\left\{\ba{l}\ru{2}\ep^{\frac{g-1}{2},\frac{g-1}{2}}\;=\;
\{(1,2),(3,2),\ldots,(g\mi2,g\mi3),(g\mi2,g\mi1)\}\\
\ru{0}\ep^{\frac{g-1}{2},\frac{g+1}{2}}\;=\;
\{(2,1),(2,3),\ldots,(g\mi3,g\mi2),(g\mi1,g\mi2)\}\ea
\right.\\
g\mbox{ even:}&\left\{\ba{l}\ru{2}\ep^{\gg-1,\gg}\;=\;
\{(2,1),(2,3),\ldots,(g\mi2,g\mi3),(g\mi2,g\mi1)\}\\
\ru{0}\ep^{\gg,\gg}\;=\;\{(1,2),(3,2),\ldots,(g\mi3,g\mi2),(g\mi1,g\mi2)\}
\,.\ea\right.\ea\ee
These cases therefore represent boundary conditions in which each
configuration of boundary spins consistent with fixed even-spin and
odd-spin sublattices contributes a nonzero weight to the partition
function at the conformal point.  If these weights are all equal,
which in fact only occurs for $A_3$, we refer to the boundary
condition as free, while if the weights are not all equal we refer to it as
quasi-free.

Finally, we note that all other boundary conditions not of the fixed,
semi-fixed, quasi-free or free type can be regarded as intermediate between
these types.

\subsub{Example: $A_3$}
We now consider, as an example, the case $A_3$, this being the
simplest nontrivial $A$ model.  For any spin assignment in the $A_3$
model, the spin states on one sublattice are all $2$, while each spin
state on the other is either $1$ or $3$.  The former sublattice, being
frozen in a single configuration, can therefore be discarded and the
model viewed as a two-state model on the other sublattice.  It can be
shown that the bulk weights of this model are those of the critical
Ising model, with the horizontal and vertical coupling constants
suitably parameterized in terms of the spectral parameter.  It is
therefore natural to relabel the nodes of $A_3$ as a frozen state $0$
and Ising states $+$ and $-$; that is,
\setlength{\unitlength}{9mm}
\be\raisebox{-0.5\unitlength}[0.5\unitlength][0.5\unitlength]{
\text{1.9}{1}{0}{0.5}{l}{A_3\;\;=}
\bpic(2,1)\put(0,0.5){\line(1,0){2}}
\multiput(0,0.5)(1,0){3}{\pp{}{\bullet}}
\put(0,0.35){\pp{t}{+}}\put(1,0.35){\pp{t}{0}}
\put(2,0.35){\pp{t}{-}}\put(2.6,0.3){\p{}{.}}
\epic}\ee

The $A_3$ fused adjacency matrices are, from \er{AFAM},
\be\ba{c}
\r^1=\left(\begin{array}{@{\;}c@{\;\:}c@{\;\:}c@{\;}}
1&0&0\\0&1&0\\0&0&1\ea\right)\quad
\r^2=\left(\begin{array}{@{\;}c@{\;\:}c@{\;\:}c@{\;}}
0&1&0\\1&0&1\\0&1&0\ea\right)\quad
\r^3=\left(\begin{array}{@{\;}c@{\;\:}c@{\;\:}c@{\;}}
0&0&1\\0&1&0\\1&0&0\ea\right)\quad
\r^4=\left(\begin{array}{@{\;}c@{\;\:}c@{\;\:}c@{\;}}
0&0&0\\0&0&0\\0&0&0\ea\right).\ea\ee
Using these matrices to determine each set of 
boundary edges and then \er{ABEW} to give the corresponding weights,
we find that these weights are, up to normalization, as given in
Table 1.
\setlength{\unitlength}{1.2mm}\stepcounter{tab}
\bc\bpic(74,30)
\put(0,0){\line(1,0){74}}
\multiput(0,6)(0,8){4}{\line(1,0){74}}
\put(0,0){\line(0,1){30}}
\multiput(6,0)(34,0){3}{\line(0,1){30}}
\put(0,0){\line(1,1){6}}
\put(23,3){\p{}{1}}\put(57,3){\p{}{2}}
\put(3,10){\p{}{+}}\put(3,18){\p{}{0}}\put(3,26){\p{}{-}}
\put(5.4,0.9){\p{br}{r}}\put(0.8,5.2){\p{tl}{a}}
\put(23,10){\pp{}{E(+,0)_{11}\;=\;1}}
\put(57,10){\pp{}{E(0,-)_{11}\;=\;1}}
\put(23,18){\pp{}{E(0,+)_{11}\;=\;E(0,-)_{11}\;=\;1}}
\put(57,18){\pp{}{E(+,0)_{11}\;=\;E(-,0)_{11}\;=\;1}}
\put(23,26){\pp{}{E(-,0)_{11}\;=\;1}}
\put(57,26){\pp{}{E(0,+)_{11}\;=\;1}}
\epic\\
Table \thetab: $A_3$ boundary edge weights\ec
We see that the three $A_3$ boundary conditions at the conformal
point are:
\nc{\hsp}{\;\;}\nc{\vsp}{\ru{1.8}}
\[\ba{@{\bullet\hsp\;}c@{\hsp\leftrightarrow\hsp}c@{\hsp\leftrightarrow\hsp}l}
(1,+)&(2,-)&\mbox{$+$ fixed}\vsp\\
(1,-)&(2,+)&\mbox{$-$ fixed}\vsp\\
(1,0)&(2,0)&\mbox{free.}\ea\]
We note that the last of these boundary conditions is invariant under the
model's $\Z_2$ symmetry, while the first two are not.

\subsub{Example: $A_4$}
We now consider, as another example, the case $A_4$.  It is known that
this model can related to the tricritical hard square and tricritical
Ising models.

The $A_4$ fused adjacency matrices are
\be\ba{c}\ru{7}
\r^1=\left(\begin{array}{@{\;}c@{\;\:}c@{\;\:}c@{\;\:}c@{\;}}
1&0&0&0\\0&1&0&0\\0&0&1&0\\0&0&0&1\ea\right)\quad
\r^2=\left(\begin{array}{@{\;}c@{\;\:}c@{\;\:}c@{\;\:}c@{\;}}
0&1&0&0\\1&0&1&0\\0&1&0&1\\0&0&1&0\ea\right)\quad
\r^3=\left(\begin{array}{@{\;}c@{\;\:}c@{\;\:}c@{\;\:}c@{\;}}
0&0&1&0\\0&1&0&1\\1&0&1&0\\0&1&0&0\ea\right)\\
\r^4=\left(\begin{array}{@{\;}c@{\;\:}c@{\;\:}c@{\;\:}c@{\;}}
0&0&0&1\\0&0&1&0\\0&1&0&0\\1&0&0&0\ea\right)\quad
\r^5=\left(\begin{array}{@{\;}c@{\;\:}c@{\;\:}c@{\;\:}c@{\;}}
0&0&0&0\\0&0&0&0\\0&0&0&0\\0&0&0&0\ea\right).\ea\ee
Using these and \er{ABEW}, we find that the $A_4$ boundary edge weights
are, up to normalization, as given in Table 2.
\setlength{\unitlength}{1.2mm}\stepcounter{tab}
\bc\bpic(104,54)
\multiput(0,0)(0,54){2}{\line(1,0){104}}
\multiput(0,6)(0,8){2}{\line(1,0){104}}
\multiput(0,30)(0,16){2}{\line(1,0){104}}
\multiput(0,0)(104,0){2}{\line(0,1){54}}
\multiput(6,0)(32,0){2}{\line(0,1){54}}
\put(72,0){\line(0,1){54}}
\put(0,0){\line(1,1){6}}
\put(22,3){\p{}{1}}\put(55,3){\p{}{2}}\put(88,3){\p{}{3}}
\put(3,10){\p{}{1}}\put(3,22){\p{}{2}}\put(3,38){\p{}{3}}
\put(3,50){\p{}{4}}
\put(5.4,0.9){\p{br}{r}}\put(0.8,5.2){\p{tl}{a}}
\put(22,10){\pp{}{E(1,2)_{11}\;=\;1}}
\put(55,10){\pp{}{E(2,3)_{11}\;=\;1}}
\put(88,10){\pp{}{E(3,4)_{11}\;=\;1}}
\put(22,20){\pp{}{E(2,3)_{11}\;=\;(\sqrt{5}+1)^\et}}
\put(22,24){\pp{}{E(2,1)_{11}\;=\;(\sqrt{5}-1)^\et}}
\put(42.5,26.3){\pp{l}{E(1,2)_{11}\;=\;E(3,4)_{11}}}
\put(48,22.3){\pp{l}{=\;(\sqrt{5}+1)^\et}}
\put(42.5,17.7){\pp{l}{E(3,2)_{11}\;=\;(5\sqrt{5}-11)^\et}}
\put(88,20){\pp{}{E(4,3)_{11}\;=\;(\sqrt{5}-1)^\et}}
\put(88,24){\pp{}{E(2,3)_{11}\;=\;(\sqrt{5}+1)^\et}}
\put(22,36){\pp{}{E(3,4)_{11}\;=\;(\sqrt{5}-1)^\et}}
\put(22,40){\pp{}{E(3,2)_{11}\;=\;(\sqrt{5}+1)^\et}}
\put(42.5,42.3){\pp{l}{E(2,1)_{11}\;=\;E(4,3)_{11}}}
\put(48,38.3){\pp{l}{=\;(\sqrt{5}+1)^\et}}
\put(42.5,33.7){\pp{l}{E(2,3)_{11}\;=\;(5\sqrt{5}-11)^\et}}
\put(88,36){\pp{}{E(3,2)_{11}\;=\;(\sqrt{5}+1)^\et}}
\put(88,40){\pp{}{E(1,2)_{11}\;=\;(\sqrt{5}-1)^\et}}
\put(22,50){\pp{}{E(4,3)_{11}\;=\;1}}
\put(55,50){\pp{}{E(3,2)_{11}\;=\;1}}
\put(88,50){\pp{}{E(2,1)_{11}\;=\;1}}
\epic\\
Table \thetab: $A_4$ boundary edge weights\ec
We see that the six $A_4$ boundary conditions at the conformal point are:
\[\ba{@{\bullet\hsp}c@{\hsp\leftrightarrow\hsp}c@{\hsp\leftrightarrow\hsp}l}
(1,1)&(3,4)&\ldots1,2,1,2 \ldots\mbox{ fixed}\vsp\\
(1,4)&(3,1)&\ldots3,4,3,4\ldots\mbox{ fixed}\vsp\\
(2,1)&(2,4)&\ldots2,3,2,3\ldots\mbox{ fixed}\vsp\\
(1,2)&(3,3)&\ldots2,1/3,2,1/3\ldots\mbox{ semi-fixed}\vsp\\
(1,3)&(3,2)&\ldots3,2/4,3,2/4\ldots\mbox{ semi-fixed}\vsp\\
(2,2)&(2,3)&\mbox{quasi-free.}\ea\]

\subsub{$D$ Graphs}
For $D_{\gg+1}$, we find, by substituting the fusion vector entries described
in Section~\ref{BWFMFV} into \er{BEW}
and then applying certain simple gauge transformations, that
the boundary edge weights can be taken explicitly as
\be\el{DBEW}\ba{r@{\;}c@{\;}l}
\ru{8.5}E^{ra}(b,c)_{11}&=&\left\{\ba{l}
\ru{2.5}2^{-\qt}|_{b,c=\gg}\;\:\E^{ra}(b,c)_{11}
\:;\;\;a\ne\gg\,,\,\;r\le\gg\mi1\\
\ru{2.5}2^{-\qt}|_{b,c=\gg}\;\:\E^{r,g-a}(b,c)_{11}
\:;\;\;a\ne\gg\,,\,\;r\ge\gg\\
\ru{0}2^\qt|_{b,c=\gg}\;\:\E^{r,\gg}(b,c)_{11}\:;\;\;a=\gg
\ea\right.\\
\ru{8.5}E^{ra}(b,c)_{12}&=&\left\{\ba{l}
\ru{2.5}\,0\:;\;\;b\ne\gg\\
\ru{2.5}
\!\pm\!2^{-\qt}\;\:\E^{r,g-a}(\gg,\gg\mi1)_{11}
\:;\;\;b=\gg^\pm,\,\;r\le\gg\mi1\\
\ru{0}\!\pm\!2^{-\qt}\;\:\E^{ra}(\gg,\gg\mi1)_{11}
\:;\;\;b=\gg^\pm,\,\;r\ge\gg\ea\right.\\
\ru{8.5}E^{ra}(b,c)_{21}&=&\left\{\ba{l}
\ru{2.5}\,0\:;\;\;c\ne\gg\\
\ru{2.5}
\!\pm\!2^{-\qt}\;\:\E^{r,g-a}(\gg\mi1,\gg)_{11}
\:;\;\;c=\gg^\pm,\,\;r\le\gg\mi1\\
\ru{0}\!\pm\!2^{-\qt}\;\:\E^{ra}(\gg\mi1,\gg)_{11}\:;\;\;c=\gg^\pm,\,\;r\ge\gg
\ea\right.\\
E^{ra}(b,c)_{22}&=&\left\{\ba{l}
\ru{2.5}\E^{r,g-a}(b,c)_{11}\:;\;\;r\le\gg\mi1\\
\E^{ra}(b,c)_{11}\:;\;\;r\ge\gg\,,
\ea\right.\ea\ee
where by $X|_{b,c=\gg}$ we mean that $X$ is only to be included if
$b=\gg$ or $c=\gg$, and where $\E^{ra}(b,c)_{11}$ are the
$A_{g-1}$ boundary edge weights as given by \er{ABEW}.
The fact that the $D_{\gg+1}$ boundary edge weights can be expressed in
terms of $A_{g-1}$ boundary edge weights follows from the 
intertwiner relations between the corresponding models.

We see that for any boundary condition, and in this gauge, there is at
most one negative boundary edge weight.  We also see that, for these
weights, the relations \er{EWS1} and \er{EWS2} are
\be\el{DEWS}E^{r\ab}(\bb,\bar{c})_{\b\g}\;=\;
\sigma_{\b\g}\:E^{ra}(b,c)_{\b\g}\,,
\qquad E^{g-r-1,\ab}(c,b)_{\g\b}\;=\;E^{ra}(b,c)_{\b\g}\,,\ee
where 
\be\sigma_{\b\g}\;=\;\left\{\ba{r@{}l}\ru{1.5}
-1&,\quad D_{\mbox{\scriptsize odd}}\mbox{ with }\b\ne\g\\
1&,\quad\mbox{otherwise.}\ea\right.\ee
In fact, for $D_{\mbox{\scriptsize even}}$ the first relation
of \er{DEWS} is trivial since the involution $a\mapsto\ab$
is the identity, but we note that the relation still holds 
if the involution is instead taken to be the graph's $\Z_2$ 
symmetry transformation and if $\sigma_{12}=\sigma_{21}$ are
replaced by $-1$.

Boundary weights for $D_{\gg+1}$ can be obtained by substituting the
edge weights \er{DBEW} into \er{GBW}.  For some of the boundary
conditions, all of the boundary weights are diagonal and these weights
can be related to those previously found in \cite{BehPea97}.  However,
non-diagonal boundary weights for the $D_{\gg+1}$ models, apart from
one case of $D_4$ considered in \cite{BehPeaZub98}, were not previously
known.  We also note that since some simple, but $(r,a)$-dependent,
gauge transformations have been included in the boundary edge weights
of \er{DBEW}, some corresponding gauge factors need to be included in
equations which relate boundary weights at different values of
$(r,a)$, in particular \er{BM} and \er{BWS2}.

Finally, we note that, as with $A_{g-1}$, certain of the $D_{\gg+1}$ boundary
conditions can be identified as being of fixed, semi-fixed, free or
quasi-free type.

\subsub{Example: $D_4$}
We now consider, as an example, the case $D_4$.  For any spin
assignment in the $D_4$ model, the spin states on one sublattice are
all $2$, while each spin state on the other is $1$, $3$ or $4$, so
that the model can be regarded as a three-state model on the latter
sublattice.  The bulk weights of this model can be shown to be those
of the critical three-state Potts model. These bulk weights are also
invariant under any ${\cal S}_3$ permutation of the Potts spin states.
We shall use the more natural labeling of the nodes of
$D_4$, $1\mapsto A$, $2\mapsto 0$, $3^+\mapsto B$ and $3^-\mapsto C$;
that is,
\setlength{\unitlength}{9mm}
\be\raisebox{-1\unitlength}[1\unitlength][1\unitlength]{
\text{1.9}{2}{0}{1}{l}{D_4\;\;=}\bpic(2,2)
\put(0,1){\line(1,0){1}}\put(1,1){\line(3,5){0.6}}
\put(1,1){\line(3,-5){0.6}}
\multiput(0,1)(1,0){2}{\pp{}{\bullet}}
\multiput(1.6,0)(0,2){2}{\pp{}{\bullet}}
\put(0,0.8){\pp{t}{A}}\put(0.9,0.88){\pp{t}{0}}
\put(1.8,0){\pp{l}{B}}\put(1.8,2){\pp{l}{C}}
\put(2.5,0.8){\p{}{.}}
\epic}\ee

The $D_4$ fused adjacency matrices are, from \er{DFAM},
\be\ba{c}\ru{7}
\r^1=\r^5=\left(\begin{array}{@{\;}c@{\;\:}c@{\;\:}c@{\;\:}c@{\;}}
1&0&0&0\\0&1&0&0\\0&0&1&0\\0&0&0&1\ea\right)\quad
\r^2=\r^4=\left(\begin{array}{@{\;}c@{\;\:}c@{\;\:}c@{\;\:}c@{\;}}
0&1&0&0\\1&0&1&1\\0&1&0&0\\0&1&0&0\ea\right)\\
\r^3=\left(\begin{array}{@{\;}c@{\;\:}c@{\;\:}c@{\;\:}c@{\;}}
0&0&1&1\\0&2&0&0\\1&0&0&1\\1&0&1&0\ea\right)\qquad
\r^5=\left(\begin{array}{@{\;}c@{\;\:}c@{\;\:}c@{\;\:}c@{\;}}
0&0&0&0\\0&0&0&0\\0&0&0&0\\0&0&0&0\ea\right),\ea\ee
where the rows and columns are ordered $A$, $0$, $B$, $C$.
Using these and \er{DBEW}, we find that the $D_4$ boundary edge weights
are, up to normalization and a simple gauge transformation on
the $(2,0)$ and $(3,0)$ boundary conditions which makes their
${\cal S}_3$ symmetry properties more apparent, as given in Table 3.
\setlength{\unitlength}{1.2mm}\stepcounter{tab}
\bc\bpic(106,70)
\multiput(0,0)(0,6){2}{\line(1,0){106}}
\put(0,18){\line(1,0){106}}
\multiput(0,46)(0,12){3}{\line(1,0){106}}
\multiput(0,0)(6,0){2}{\line(0,1){70}}
\multiput(28,0)(28,0){3}{\line(0,1){70}}
\put(106,0){\line(0,1){70}}
\put(0,0){\line(1,1){6}}
\put(17,3){\p{}{1}}\put(42,3){\p{}{2}}\put(70,3){\p{}{3}}
\put(95,3){\p{}{4}}
\put(3,12){\p{}{A}}\put(3,32){\p{}{0}}\put(3,52){\p{}{B}}
\put(3,64){\p{}{C}}
\put(5.4,0.9){\p{br}{r}}\put(0.8,5.2){\p{tl}{a}}
\put(17,12){\pp{}{E(A,0)_{11}\;=\;1}}
\put(35,10){\pp{l}{E(0,C)_{11}\;=\;1}}
\put(35,14){\pp{l}{E(0,B)_{11}\;=}}
\put(63,10){\pp{l}{E(C,0)_{11}\;=\;1}}
\put(63,14){\pp{l}{E(B,0)_{11}\;=}}
\put(95,12){\pp{}{E(0,A)_{11}\;=\;1}}
\put(10,28){\pp{l}{E(0,C)_{11}\;=\;1}}
\put(10,32){\pp{l}{E(0,B)_{11}\;=}}
\put(10,36){\pp{l}{E(0,A)_{11}\;=}}
\put(32,22){\pp{l}{E(C,0)_{12}\;=\;-\sqrt{3}/2}}
\put(32,26){\pp{l}{E(C,0)_{11}\;=\;-1/2}}
\put(32,30){\pp{l}{E(B,0)_{12}\;=\;\sqrt{3}/2}}
\put(32,34){\pp{l}{E(B,0)_{11}\;=\;-1/2}}
\put(32,38){\pp{l}{E(A,0)_{12}\;=\;0}}
\put(32,42){\pp{l}{E(A,0)_{11}\;=\;1}}
\put(60,22){\pp{l}{E(0,C)_{21}\;=\;-\sqrt{3}/2}}
\put(60,26){\pp{l}{E(0,C)_{11}\;=\;-1/2}}
\put(60,30){\pp{l}{E(0,B)_{21}\;=\;\sqrt{3}/2}}
\put(60,34){\pp{l}{E(0,B)_{11}\;=\;-1/2}}
\put(60,38){\pp{l}{E(0,A)_{21}\;=\;0}}
\put(60,42){\pp{l}{E(0,A)_{11}\;=\;1}}
\put(88,28){\pp{l}{E(C,0)_{11}\;=\;1}}
\put(88,32){\pp{l}{E(B,0)_{11}\;=}}
\put(88,36){\pp{l}{E(A,0)_{11}\;=}}
\put(17,52){\pp{}{E(B,0)_{11}\;=\;1}}
\put(35,50){\pp{l}{E(0,C)_{11}\;=\;1}}
\put(35,54){\pp{l}{E(0,A)_{11}\;=}}
\put(63,50){\pp{l}{E(C,0)_{11}\;=\;1}}
\put(63,54){\pp{l}{E(A,0)_{11}\;=}}
\put(95,52){\pp{}{E(0,B)_{11}\;=\;1}}
\put(17,64){\pp{}{E(C,0)_{11}\;=\;1}}
\put(35,62){\pp{l}{E(0,B)_{11}\;=\;1}}
\put(35,66){\pp{l}{E(0,A)_{11}\;=}}
\put(63,62){\pp{l}{E(B,0)_{11}\;=\;1}}
\put(63,66){\pp{l}{E(A,0)_{11}\;=}}
\put(95,64){\pp{}{E(0,C)_{11}\;=\;1}}
\epic\\
Table \thetab: $D_4$ boundary edge weights\ec
We see that the eight $D_4$ boundary conditions at the conformal point are:
\[\ba{c@{}c@{}c@{}c@{}c@{}l}
\bullet\hsp&(1,A)&\hsp\leftrightarrow\hsp&(4,A)&\hsp\leftrightarrow\hsp&
A\mbox{ fixed}\vsp\\
\bullet\hsp&(1,B)&\hsp\leftrightarrow\hsp&(4,B)&\hsp\leftrightarrow\hsp&
B\mbox{ fixed}\vsp\\
\bullet\hsp&(1,C)&\hsp\leftrightarrow\hsp&(4,C)&\hsp\leftrightarrow\hsp&
C\mbox{ fixed}\vsp\\
\bullet\hsp&(2,A)&\hsp\leftrightarrow\hsp&(3,A)&\hsp\leftrightarrow\hsp&
\mbox{$B$ and $C$ mixed with equal weight}\vsp\\
\bullet\hsp&(2,B)&\hsp\leftrightarrow\hsp&(3,B)&\hsp\leftrightarrow\hsp&
\mbox{$A$ and $C$ mixed with equal weight}\vsp\\
\bullet\hsp&(2,C)&\hsp\leftrightarrow\hsp&(3,C)&\hsp\leftrightarrow\hsp&
\mbox{$A$ and $B$ mixed with equal weight}\vsp\\
\bullet\hsp&(1,0)&\hsp\leftrightarrow\hsp&(4,0)&\hsp\leftrightarrow\hsp&
\mbox{free}\vsp\\
\bullet\hsp&(2,0)&\hsp\leftrightarrow\hsp&(3,0)&\hsp\leftrightarrow\hsp&
\mbox{quasi-free in which same-spin pairs have weight $1$}\ru{1.4}\\
&&&&&\mbox{and different-spin pairs have weight $-1/2$}\ea\]
The nature of the last boundary condition is best seen by
considering the $(2,0)$ boundary weights at the conformal point, these
being, up to normalization, 
\be\el{B20}\B{2,0}{b}{1}{0}{d}{1}{\x,\x}\;=\;
\left\{\ba{l}\ru{1.5}\;1\,;\;\:b=d\\-1/2\,;\;\:b\ne d\,,\qquad
b,d\in\{A,B,C\}\,.\ea\right.\ee
This is therefore a boundary condition on nearest-neighbor
pairs of Potts spins, in which like and unlike neighbors are associated 
with weights $1$ and $-1/2$ respectively.

We see that the last two $D_4$ boundary conditions are ${\cal S}_3$
symmetric, while the first six are not.  In fact, the $(2,0)$ boundary
weights \er{B20} represent the only possibility, other than
reproducing the $(1,0)\leftrightarrow(4,0)$ weights, which is
${\cal S}_3$ symmetric and consistent with a decomposition \er{BP} in which
$\g$ is summed over two values.

\subsub{$E$ Graphs}\label{E6} 
For the $E$ graphs, there are a large number of boundary
conditions at the conformal point (specifically, $30$ for $E_6$, $56$ for $E_7$
and $112$ for $E_8$) and many of these are in turn associated with a
large number of boundary edge weights. Therefore, since it is
straightforward and more practical to obtain the numerical values for
these weights using a computer,
we do not list them here.  However, as an example, we do give the sets
of $E_6$ boundary edges.

The $E_6$ fused adjacency matrices are
\be\ba{c}
\r^1=\left(\begin{array}{@{\;}c@{\;\:}c@{\;\:}c@{\;\:}c@{\;\:}c@{\;\:}c@{\;}}
1&0&0&0&0&0\\0&1&0&0&0&0\\0&0&1&0&0&0\\
0&0&0&1&0&0\\0&0&0&0&1&0\\0&0&0&0&0&1\ea\right)\;
\r^2=\left(\begin{array}{@{\;}c@{\;\:}c@{\;\:}c@{\;\:}c@{\;\:}c@{\;\:}c@{\;}}
0&1&0&0&0&0\\1&0&1&0&0&0\\0&1&0&1&0&1\\
0&0&1&0&1&0\\0&0&0&1&0&0\\0&0&1&0&0&0\ea\right)\;
\r^3=\left(\begin{array}{@{\;}c@{\;\:}c@{\;\:}c@{\;\:}c@{\;\:}c@{\;\:}c@{\;}}
0&0&1&0&0&0\\0&1&0&1&0&1\\1&0&2&0&1&0\\
0&1&0&1&0&1\\0&0&1&0&0&0\\0&1&0&1&0&0\ea\right)\\
\\
\r^4=\left(\begin{array}{@{\;}c@{\;\:}c@{\;\:}c@{\;\:}c@{\;\:}c@{\;\:}c@{\;}}
0&0&0&1&0&1\\0&0&2&0&1&0\\0&2&0&2&0&1\\
1&0&2&0&0&0\\0&1&0&0&0&1\\1&0&1&0&1&0\ea\right)\;
\r^5=\left(\begin{array}{@{\;}c@{\;\:}c@{\;\:}c@{\;\:}c@{\;\:}c@{\;\:}c@{\;}}
0&0&1&0&1&0\\0&1&0&2&0&1\\1&0&3&0&1&0\\
0&2&0&1&0&1\\1&0&1&0&0&0\\0&1&0&1&0&1\ea\right)\;
\r^6=\left(\begin{array}{@{\;}c@{\;\:}c@{\;\:}c@{\;\:}c@{\;\:}c@{\;\:}c@{\;}}
0&1&0&1&0&0\\1&0&2&0&1&0\\0&2&0&2&0&2\\
1&0&2&0&1&0\\0&1&0&1&0&0\\0&0&2&0&0&0\ea\right)\\
\\
\r^7=\left(\begin{array}{@{\;}c@{\;\:}c@{\;\:}c@{\;\:}c@{\;\:}c@{\;\:}c@{\;}}
1&0&1&0&0&0\\0&2&0&1&0&1\\1&0&3&0&1&0\\
0&1&0&2&0&1\\0&0&1&0&1&0\\0&1&0&1&0&1\ea\right)\;
\r^8=\left(\begin{array}{@{\;}c@{\;\:}c@{\;\:}c@{\;\:}c@{\;\:}c@{\;\:}c@{\;}}
0&1&0&0&0&1\\1&0&2&0&0&0\\0&2&0&2&0&1\\
0&0&2&0&1&0\\0&0&0&1&0&1\\1&0&1&0&1&0\ea\right)\;
\r^9=\left(\begin{array}{@{\;}c@{\;\:}c@{\;\:}c@{\;\:}c@{\;\:}c@{\;\:}c@{\;}}
0&0&1&0&0&0\\0&1&0&1&0&1\\1&0&2&0&1&0\\
0&1&0&1&0&1\\0&0&1&0&0&0\\0&1&0&1&0&0\ea\right)\\
\\
\r^{10}=
\left(\begin{array}{@{\;}c@{\;\:}c@{\;\:}c@{\;\:}c@{\;\:}c@{\;\:}c@{\;}}
0&0&0&1&0&0\\0&0&1&0&1&0\\0&1&0&1&0&1\\
1&0&1&0&0&0\\0&1&0&0&0&0\\0&0&1&0&0&0\ea\right)\;
\r^{11}=
\left(\begin{array}{@{\;}c@{\;\:}c@{\;\:}c@{\;\:}c@{\;\:}c@{\;\:}c@{\;}}
0&0&0&0&1&0\\0&0&0&1&0&0\\0&0&1&0&0&0\\
0&1&0&0&0&0\\1&0&0&0&0&0\\0&0&0&0&0&1\ea\right)\;
\r^{12}=
\left(\begin{array}{@{\;}c@{\;\:}c@{\;\:}c@{\;\:}c@{\;\:}c@{\;\:}c@{\;}}
0&0&0&0&0&0\\0&0&0&0&0&0\\0&0&0&0&0&0\\
0&0&0&0&0&0\\0&0&0&0&0&0\\0&0&0&0&0&0\ea\right)\!.\ea\ee
Using these, we find that the $E_6$ boundary edges 
are as given in Table 4.  In this table we also give, for each
$(b,c)\in\ep^{ra}$, the values of
$F^r_{ba}$ and $F^{r+1}_{ca}$ as successive superscripts. 
\setlength{\unitlength}{1.19mm}\stepcounter{tab}
\bc\bpic(126,122)
\put(0,0){\line(0,1){122}}
\multiput(6,0)(12,0){11}{\line(0,1){122}}
\put(0,0){\line(1,0){126}}
\multiput(0,6)(0,116){2}{\line(1,0){126}}
\multiput(0,20)(0,22){4}{\line(1,0){126}}
\put(0,100){\line(1,0){126}}
\put(0,0){\line(1,1){6}}
\put(12,3){\p{}{1}}\put(24,3){\p{}{2}}\put(36,3){\p{}{3}}
\put(48,3){\p{}{4}}\put(60,3){\p{}{5}}\put(72,3){\p{}{6}}
\put(84,3){\p{}{7}}\put(96,3){\p{}{8}}\put(108,3){\p{}{9}}
\put(120,3){\p{}{10}}
\put(3,13){\p{}{1}}\put(3,31){\p{}{2}}\put(3,53){\p{}{3}}
\put(3,75){\p{}{4}}\put(3,93){\p{}{5}}\put(3,111){\p{}{6}}
\put(5.4,1.1){\p{br}{r}}\put(1,5){\p{tl}{a}}
\put(12,13){\pp{}{(1,2)^{11}}}
\put(24,13){\pp{}{(2,3)^{11}}}
\put(36,15){\pp{}{(3,4)^{11}}}
\put(36,11){\pp{}{(3,6)^{11}}}
\put(48,17){\pp{}{(4,3)^{11}}}
\put(48,13){\pp{}{(4,5)^{11}}}
\put(48,9){\pp{}{(6,3)^{11}}}
\put(60,17){\pp{}{(3,2)^{11}}}
\put(60,13){\pp{}{(3,4)^{11}}}
\put(60,9){\pp{}{(5,4)^{11}}}
\put(72,17){\pp{}{(2,1)^{11}}}
\put(72,13){\pp{}{(2,3)^{11}}}
\put(72,9){\pp{}{(4,3)^{11}}}
\put(84,17){\pp{}{(1,2)^{11}}}
\put(84,13){\pp{}{(3,2)^{11}}}
\put(84,9){\pp{}{(3,6)^{11}}}
\put(96,15){\pp{}{(2,3)^{11}}}
\put(96,11){\pp{}{(6,3)^{11}}}
\put(108,13){\pp{}{(3,4)^{11}}}
\put(120,13){\pp{}{(4,5)^{11}}}
\put(12,33){\pp{}{(2,1)^{11}}}
\put(12,29){\pp{}{(2,3)^{11}}}
\put(24,37){\pp{}{(1,2)^{11}}}
\put(24,33){\pp{}{(3,2)^{11}}}
\put(24,29){\pp{}{(3,4)^{11}}}
\put(24,25){\pp{}{(3,6)^{11}}}
\put(36,37){\pp{}{(2,3)^{12}}}
\put(36,33){\pp{}{(4,3)^{12}}}
\put(36,29){\pp{}{(4,5)^{11}}}
\put(36,25){\pp{}{(6,3)^{12}}}
\put(48,37){\pp{}{(3,2)^{21}}}
\put(48,33){\pp{}{(3,4)^{22}}}
\put(48,29){\pp{}{(3,6)^{21}}}
\put(48,25){\pp{}{(5,4)^{12}}}
\put(60,39){\pp{}{(2,1)^{11}}}
\put(60,35){\pp{}{(2,3)^{12}}}
\put(60,31){\pp{}{(4,3)^{22}}}
\put(60,27){\pp{}{(4,5)^{21}}}
\put(60,23){\pp{}{(6,3)^{12}}}
\put(72,39){\pp{}{(1,2)^{12}}}
\put(72,35){\pp{}{(3,2)^{22}}}
\put(72,31){\pp{}{(3,4)^{21}}}
\put(72,27){\pp{}{(3,6)^{21}}}
\put(72,23){\pp{}{(5,4)^{11}}}
\put(84,37){\pp{}{(2,1)^{21}}}
\put(84,33){\pp{}{(2,3)^{22}}}
\put(84,29){\pp{}{(4,3)^{12}}}
\put(84,25){\pp{}{(6,3)^{12}}}
\put(96,37){\pp{}{(1,2)^{11}}}
\put(96,33){\pp{}{(3,2)^{21}}}
\put(96,29){\pp{}{(3,4)^{21}}}
\put(96,25){\pp{}{(3,6)^{21}}}
\put(108,37){\pp{}{(2,3)^{11}}}
\put(108,33){\pp{}{(4,3)^{11}}}
\put(108,29){\pp{}{(4,5)^{11}}}
\put(108,25){\pp{}{(6,3)^{11}}}
\put(120,33){\pp{}{(3,4)^{11}}}
\put(120,29){\pp{}{(5,4)^{11}}}
\put(12,57){\pp{}{(3,2)^{11}}}
\put(12,53){\pp{}{(3,4)^{11}}}
\put(12,49){\pp{}{(3,6)^{11}}}
\put(24,61){\pp{}{(2,1)^{11}}}
\put(24,57){\pp{}{(2,3)^{12}}}
\put(24,53){\pp{}{(4,3)^{12}}}
\put(24,49){\pp{}{(4,5)^{11}}}
\put(24,45){\pp{}{(6,3)^{12}}}
\put(36,61){\pp{}{(1,2)^{12}}}
\put(36,57){\pp{}{(3,2)^{22}}}
\put(36,53){\pp{}{(3,4)^{22}}}
\put(36,49){\pp{}{(3,6)^{21}}}
\put(36,45){\pp{}{(5,4)^{12}}}
\put(48,61){\pp{}{(2,1)^{21}}}
\put(48,57){\pp{}{(2,3)^{23}}}
\put(48,53){\pp{}{(4,3)^{23}}}
\put(48,49){\pp{}{(4,5)^{21}}}
\put(48,45){\pp{}{(6,3)^{13}}}
\put(60,61){\pp{}{(1,2)^{12}}}
\put(60,57){\pp{}{(3,2)^{32}}}
\put(60,53){\pp{}{(3,4)^{32}}}
\put(60,49){\pp{}{(3,6)^{32}}}
\put(60,45){\pp{}{(5,4)^{12}}}
\put(72,61){\pp{}{(2,1)^{21}}}
\put(72,57){\pp{}{(2,3)^{23}}}
\put(72,53){\pp{}{(4,3)^{23}}}
\put(72,49){\pp{}{(4,5)^{21}}}
\put(72,45){\pp{}{(6,3)^{23}}}
\put(84,61){\pp{}{(1,2)^{12}}}
\put(84,57){\pp{}{(3,2)^{32}}}
\put(84,53){\pp{}{(3,4)^{32}}}
\put(84,49){\pp{}{(3,6)^{31}}}
\put(84,45){\pp{}{(5,4)^{12}}}
\put(96,61){\pp{}{(2,1)^{21}}}
\put(96,57){\pp{}{(2,3)^{22}}}
\put(96,53){\pp{}{(4,3)^{22}}}
\put(96,49){\pp{}{(4,5)^{21}}}
\put(96,45){\pp{}{(6,3)^{12}}}
\put(108,61){\pp{}{(1,2)^{11}}}
\put(108,57){\pp{}{(3,2)^{21}}}
\put(108,53){\pp{}{(3,4)^{21}}}
\put(108,49){\pp{}{(3,6)^{21}}}
\put(108,45){\pp{}{(5,4)^{11}}}
\put(120,57){\pp{}{(2,3)^{11}}}
\put(120,53){\pp{}{(4,3)^{11}}}
\put(120,49){\pp{}{(6,3)^{11}}}
\put(12,77){\pp{}{(4,3)^{11}}}
\put(12,73){\pp{}{(4,5)^{11}}}
\put(24,81){\pp{}{(3,2)^{11}}}
\put(24,77){\pp{}{(3,4)^{11}}}
\put(24,73){\pp{}{(3,6)^{11}}}
\put(24,69){\pp{}{(5,4)^{11}}}
\put(36,81){\pp{}{(2,1)^{11}}}
\put(36,77){\pp{}{(2,3)^{12}}}
\put(36,73){\pp{}{(4,3)^{12}}}
\put(36,69){\pp{}{(6,3)^{12}}}
\put(48,81){\pp{}{(1,2)^{12}}}
\put(48,77){\pp{}{(3,2)^{22}}}
\put(48,73){\pp{}{(3,4)^{21}}}
\put(48,69){\pp{}{(3,6)^{21}}}
\put(60,83){\pp{}{(2,1)^{21}}}
\put(60,79){\pp{}{(2,3)^{22}}}
\put(60,75){\pp{}{(4,3)^{12}}}
\put(60,71){\pp{}{(4,5)^{11}}}
\put(60,67){\pp{}{(6,3)^{12}}}
\put(72,83){\pp{}{(1,2)^{11}}}
\put(72,79){\pp{}{(3,2)^{21}}}
\put(72,75){\pp{}{(3,4)^{22}}}
\put(72,71){\pp{}{(3,6)^{21}}}
\put(72,67){\pp{}{(5,4)^{12}}}
\put(84,81){\pp{}{(2,3)^{12}}}
\put(84,77){\pp{}{(4,3)^{22}}}
\put(84,73){\pp{}{(4,5)^{21}}}
\put(84,69){\pp{}{(6,3)^{12}}}
\put(96,81){\pp{}{(3,2)^{21}}}
\put(96,77){\pp{}{(3,4)^{21}}}
\put(96,73){\pp{}{(3,6)^{21}}}
\put(96,69){\pp{}{(5,4)^{11}}}
\put(108,81){\pp{}{(2,1)^{11}}}
\put(108,77){\pp{}{(2,3)^{11}}}
\put(108,73){\pp{}{(4,3)^{11}}}
\put(108,69){\pp{}{(6,3)^{11}}}
\put(120,77){\pp{}{(1,2)^{11}}}
\put(120,73){\pp{}{(3,2)^{11}}}
\put(12,93){\pp{}{(5,4)^{11}}}
\put(24,93){\pp{}{(4,3)^{11}}}
\put(36,95){\pp{}{(3,2)^{11}}}
\put(36,91){\pp{}{(3,6)^{11}}}
\put(48,97){\pp{}{(2,1)^{11}}}
\put(48,93){\pp{}{(2,3)^{11}}}
\put(48,89){\pp{}{(6,3)^{11}}}
\put(60,97){\pp{}{(1,2)^{11}}}
\put(60,93){\pp{}{(3,2)^{11}}}
\put(60,89){\pp{}{(3,4)^{11}}}
\put(72,97){\pp{}{(2,3)^{11}}}
\put(72,93){\pp{}{(4,3)^{11}}}
\put(72,89){\pp{}{(4,5)^{11}}}
\put(84,97){\pp{}{(3,4)^{11}}}
\put(84,93){\pp{}{(3,6)^{11}}}
\put(84,89){\pp{}{(5,4)^{11}}}
\put(96,95){\pp{}{(4,3)^{11}}}
\put(96,91){\pp{}{(6,3)^{11}}}
\put(108,93){\pp{}{(3,2)^{11}}}
\put(120,93){\pp{}{(2,1)^{11}}}
\put(12,111){\pp{}{(6,3)^{11}}}
\put(24,113){\pp{}{(3,2)^{11}}}
\put(24,109){\pp{}{(3,4)^{11}}}
\put(36,117){\pp{}{(2,1)^{11}}}
\put(36,113){\pp{}{(2,3)^{11}}}
\put(36,109){\pp{}{(4,3)^{11}}}
\put(36,105){\pp{}{(4,5)^{11}}}
\put(48,119){\pp{}{(1,2)^{11}}}
\put(48,115){\pp{}{(3,2)^{11}}}
\put(48,111){\pp{}{(3,4)^{11}}}
\put(48,107){\pp{}{(3,6)^{11}}}
\put(48,103){\pp{}{(5,4)^{11}}}
\put(60,115){\pp{}{(2,3)^{12}}}
\put(60,111){\pp{}{(4,3)^{12}}}
\put(60,107){\pp{}{(6,3)^{12}}}
\put(72,115){\pp{}{(3,2)^{21}}}
\put(72,111){\pp{}{(3,4)^{21}}}
\put(72,107){\pp{}{(3,6)^{21}}}
\put(84,119){\pp{}{(2,1)^{11}}}
\put(84,115){\pp{}{(2,3)^{11}}}
\put(84,111){\pp{}{(4,3)^{11}}}
\put(84,107){\pp{}{(4,5)^{11}}}
\put(84,103){\pp{}{(6,3)^{11}}}
\put(96,117){\pp{}{(1,2)^{11}}}
\put(96,113){\pp{}{(3,2)^{11}}}
\put(96,109){\pp{}{(3,4)^{11}}}
\put(96,105){\pp{}{(5,4)^{11}}}
\put(108,113){\pp{}{(2,3)^{11}}}
\put(108,109){\pp{}{(4,3)^{11}}}
\put(120,111){\pp{}{(3,6)^{11}}}
\epic\\
Table \thetab: $E_6$ boundary edges\ec
From this table, the properties \er{BPS1} and \er{BPS2} are
immediately apparent, and we can also gain some understanding of the
nature of each of the $30$ boundary conditions at the conformal point.

\sub{Realization of Conformal Boundary Conditions}
We now consider the relationship between the integrable boundary
conditions of the critical unitary $A$--$D$--$E$ lattice models and
the conformal boundary conditions of the critical series of $\slt$
unitary minimal conformal field theories.

Each conformal field theory of this type on a torus or cylinder is
associated with two graphs, the $A$ graph $A_{g-2}$ and an $A$, $D$ or
$E$ graph $\G$ with Coxeter number $g$, this theory being denoted
$\M(A_{g-2},\G)$.  As shown in \cite{Hus84,Pas87d,Pas87e}, the lattice
model based on $\G$, with $\psi$ the Perron-Frobenius eigenvector and
$0<u<\l$, can be associated with the field theory $\M(A_{g-2},\G)$.

In \cite{BehPeaPetZub00}, it was found that the complete set of
conformal boundary conditions of $\M(A_{g-2},\G)$ on a cylinder is
labeled by the set of pairs $(r,a)$, with $r\in\{1,\ldots,g-2\}$,
$a\in\G$ and $(r,a)$ and $(g\mi r\mi1,\ab)$ being considered as
equivalent.  We immediately see that this classification is identical
to that of the boundary conditions of the corresponding lattice model
at the conformal point.

It was also shown in \cite{BehPeaPetZub00}
that the partition function of $\M(A_{g-2},\G)$ on a cylinder with conformal 
boundary conditions $(r_1,a_1)$ and $(r_2,a_2)$ is given by
\be\el{ZC}Z_{r_1\!a_1\!|r_2a_2}(q)\;=\;\sum_{r=1}^{g-2}\;\sum_{s=1}^{g-1}\;
\r(A_{g-2})^r_{r_1r_2}\;\r(\G)^s_{a_1a_2}\;\chi_{(r,s)}(q)\,,\ee
where $q$ is the modular parameter, 
$\r(A_{g-2})^r$ and $\r(\G)^s$ denote the fused adjacency matrices of
$A_{g-2}$ and $\G$, and $\chi_{(r,s)}$ is the
character of the irreducible representation  with highest weight 
\be\el{CW1}
\Delta_{(r,s)}\;=\;\frac{(r\,g\mi s\,(g\mi1))^2-1}{4\,g\,(g\mi1)}\ee
of the Virasoro algebra with central charge
\be\el{CW2}c\;=\;1-\frac{6}{g\,(g\mi1)}\,.\ee

The equivalence of the $(r,a)$ and $(g\mi r\mi1,\ab)$ conformal boundary
conditions is apparent by using \er{FAMI} and the relation
$\Delta_{(g-r-\!1,g-s)}=\Delta_{(r,s)}$ 
to observe that the partition function
\er{ZC} is unchanged by applying this transformation to either
of the boundary condition labels.

We now assert that, in the continuum scaling limit, the $(r,a)$ boundary
condition in the lattice model at the conformal and isotropic point 
provides a realization of the $(r,a)$ conformal boundary condition of
the corresponding field theory.
In particular, we expect that, as $N\rightarrow\infty$, the eigenvalues of
$\vec{D}^\N_{\!r_1\!a_1\!|r_2a_2}\!(\l/2,\l/2,\l/2)$ can be arranged
in towers, with each tower labeled by a pair $(r,s)$
and the multiplicity of tower $(r,s)$ given by
$\r(A_{g-2})^r_{r_1r_2}\;\r(\G)^s_{a_1a_2}$.  We further expect that
the $j$'th largest eigenvalue in this tower has the form
\be\el{TS}\Lambda^{\!(r,s)j}_{r_1\!a_1\!|r_2a_2}\:=\;
\exp\Bigl[-2N\F\,-\,2f_{r_1\!|r_2}\,+\,2\pi(c/24-\Delta_{(r,s)}-
k_{(r,s)j})/N\,+\,\mbox{o}(1/N)\Bigr],\ee
where $\F$ is the bulk free energy per lattice face, which
depends only on $g$,
$f_{r_1\!|r_2}$ is the boundary free energy per lattice row, which
depends only on $g$, $r_1$ and $r_2$,
$c$ and $\Delta_{(r,s)}$ are as given in \er{CW1} and \er{CW2}, and 
$k_{(r,s)j}$ are nonnegative integers given through the expansion of the
Virasoro characters by
\be\chi_{(r,s)}(q)\;=\;q^{-c/24+\Delta_{(r,s)}}\;\sum_{j=1}^\infty
\:q^{k_{(r,s)j}}\,,\quad k_{(r,s)j}\le k_{(r,s)j+\!1}\,.\ee

With the eigenvalues appearing in this tower structure, 
it follows using \er{ZE}
that, as $N\rightarrow\infty$, the lattice model and conformal
partition functions are related by 
\be\el{ZZ}
\ba{l}\ru{2}\vec{Z}^{\N\MM}_{r_1\!a_1\!|r_2a_2}\!(\l/2,\l/2,\l/2)\;\sim\\
\qquad\qquad\quad
\exp(-2MN\F\mi2Mf_{r_1\!|r_2})\;Z_{r_1\!a_1\!|r_2a_2}(q)\,,\quad
q=\exp(\!-2\pi M/N)\,.\ea\ee
We note that we also expect related conformal behavior in a lattice model
which is at the conformal point and has $0<u<\l$, but which is no
longer at the isotropic point $u=\l/2$.  

The expectation that the lattice model boundary conditions correspond
in this way to the conformal boundary conditions is supported by the
results of numerical studies we have performed, the matching of the
identifications made here of the nature of the lattice realizations of
certain conformal boundary conditions with those made in other
studies, the consistency of all of the symmetry properties of the
lattice model partition function with those of the conformal partition
function, and analytic confirmation in several cases.

In our numerical studies, we evaluated the eigenvalues of
$\vec{D}^\N_{\!r_1\!1|r_2s}(\l/2,\l/2,\l/2)$ for certain $A$ graphs,
selected values of $r_1$, $r_2$ and $s$, and several successive values
of $N$.  We then extrapolated these results to large $N$ and verified
consistency with \er{TS} for these cases.  This numerical data, used
with \er{DITR1}, also implied consistency with \er{TS} for all of the
related $A$, $D$ and $E$ cases.

Regarding the identification of the nature of the lattice realizations
of particular conformal boundary conditions, this was done for all
$A_3$ cases and all $D_4$ cases except $(2,0)$ in \cite{Car89,Car86b},
for all $(1,a)$ and $(r,1)$ cases of $A_{g-1}$ and $D_{\gg+1}$ in
\cite{SalBau89}, for all $A_4$ cases in \cite{Chi96}, and for the
$(2,0)$ case of $D_4$ in \cite{AffOshSal98}.  In all of these studies,
the lattice model boundary conditions were shown to have exactly the
same basic features as those found here.

Proceeding to the consistency of symmetry properties, it follows
straightforwardly from \er{ZC} and the properties of the fused
adjacency matrices that the $\M(A_{g-2},\G)$ partition function
satisfies 
\be\el{ZCS}\ba{c}\ru{2}
Z_{r_1\!a_1\!|r_2a_2}(q)\;=\;Z_{r_2a_2|r_1\!a_1}(q)
\;=\;Z_{r_2a_1\!|r_1\!a_2}(q)\\
\ru{3.5}\qquad\qquad\qquad=\;Z_{r_1\!\ab_1\!|r_2\ab_2}(q)\;=\;
Z_{g-r_1\mi1,\ab_1\!|g-r_2-\!1,\ab_2}(q)\\
\ds\sum_{{a_1\!}',{a_2\!}'\in\G}
\r(\G)^{r_1}_{a_1{a_1\!}'}\;\r(\G)^{r_2}_{a_2{a_2\!}'}\;
Z_{{r_1\!}'{a_1\!}'|{r_2\!}'{a_2\!}'}(q)\;=\qquad\qquad\qquad\\
\ru{4}\qquad\qquad\qquad\qquad\qquad
\ds\sum_{{a_1\!}',{a_2\!}'\in\G}
\r(\G)^{r_2}_{a_1{a_1\!}'}\;\r(\G)^{r_1}_{a_2{a_2\!}'}\;
Z_{{r_1\!}'{a_1\!}'|{r_2\!}'{a_2\!}'}(q)\\
\ds\sum_{a\in\G}\r(\G)^{s'}_{a_1a}\;Z_{r_1\!a|r_2a_2}(q)\;=\;
\sum_{s=1}^{g-1}\r(\G)^{s}_{a_1a_2}\;Z_{r_1\!s'|r_2s}(q)\,.\ea\ee 
We immediately see that these equalities are 
consistent with the lattice relations \er{DLRS}, \er{DPIS},
\er{DIS1}, \er{ZS2}, \er{ZII} and \er{DITR} respectively.  We also note that
certain cases of the last equality of \er{ZCS} and its lattice version
\er{DITR}, mostly with $s'=r_1=r_2=1$, were considered numerically and
analytically in \cite{Car89,SalBau89,DifZub90b,PasSal90,Dif92,Dor93}, 
while in \cite{Bat98} free combinations
of $(1,a)$ lattice boundary conditions in $A_{g-1}$ were studied
analytically leading to \er{ZITR} with $r_1\!=\!r_2\!=\!1$ and a sum on
$a_1$ and $a_2$.  However, in all of these studies, the orientation of
the lattice differed from that used here by a rotation of 45 degrees.

Finally, the partition function relation \er{ZZ} has been proved
analytically using techniques based on the Yang-Baxter and boundary
Yang-Baxter equations for a lattice with the same orientation as that
used here for all $A_3$ cases in \cite{ObrPeaWar96} and for all $A_4$
cases which lead to a single character on the right side of \er{ZC} in
\cite{ObrPeaWar97}.

\sect{Discussion} 
We have obtained various results on integrable boundary conditions for
general graph-based and, in particular, critical unitary $A$--$D$--$E$
lattice models.  More specifically, we have systematically constructed
boundary weights, derived general symmetry properties and, for the
$A$--$D$--$E$ cases, studied the relationship with conformal boundary
conditions.

The general formalism presented here, or certain natural extensions of
it, can be applied to various other integrable lattice models which
are associated with rational conformal field theories of interest.  We
expect that the corresponding integrable boundary conditions provide
realizations of the conformal boundary conditions of these theories,
although we acknowledge that for many of these models only the bulk
weights are currently known and that explicitly obtaining boundary
weights may involve certain technical challenges.  Nevertheless, in
conclusion, we list these other cases and indicate their connections
with those studied here: \itm{If, using \er{NU}, we choose an
eigenvector of an $A$--$D$--$E$ adjacency matrix corresponding to a
Coxeter exponent $k$ with $1<k<g\mi1$ and $k$ coprime to $g$, then a
nonunitary $A$--$D$--$E$ model is obtained which corresponds to the
nonunitary minimal theory $\M(A_{g-k-1},\G)$.  This enables the
consideration of all of the $\M(A_{h-1},\G)$ minimal theories with
$h<g$. The $\M(A_{h-1},\G)$ theories with $h>g$ and $\G$ a $D$ or $E$
graph can not be obtained in this way, but all of the
$\M(A_{h-1},A_{g-1})$ theories are accessible since in this case $h$
and $g$ are interchangeable.}  \itm{By taking, in Section~3, $\G$ as
an $A^{(1)}$, $D^{(1)}$ or $E^{(1)}$ Dynkin diagram, and $\psi$ as the
Perron-Frobenius eigenvector of its adjacency matrix, lattice models
are obtained which correspond to certain conformal field theories with
central charge $c=1$.  As noted in Section~\ref{MFL}, for these models
$s$ is given by the second case of \er{sD} and there are infinitely
many fusion levels.}  \itm{By replacing the relations of \er{TLA} with
those of the Hecke algebra, certain lattice models and conformal field
theories associated with $\hat{s\ell}(n)$, for $n>2$, can be obtained.
In this case, although the lattice models become significantly
different, we expect that most of the results of at least Section~2
remain unchanged} \itm{By using the dilute Temperley-Lieb algebra
instead of the Temperley-Lieb algebra, lattice models which correspond
to the so-called tricritical series of unitary minimal theories,
$\M(A_g,\G)$, can be obtained.  However, we note that the dilute
Temperley-Lieb algebra contains considerably more generators than the
Temperley-Lieb algebra, so that the formalism would be more
complicated from the outset.}  \itm{It is also apparent that lattice
models whose bulk weights are given by fused square blocks of
$A$--$D$--$E$ bulk weights could be considered by applying some
relatively straightforward extensions to the formalism. The field
theories associated with these models include certain superconformal
theories.  The cases of the $A$ models involving only diagonal
boundary weights were studied in~\cite{BehPeaObr96}, where it was
shown that the double-row transfer matrices of the standard and fused
model together satisfy a hierarchy of functional equations.  We expect
that equations of a similar form can be derived, using the boundary
inversion relation~\er{BIR} and other local properties, for the
remaining $A$--$D$--$E$ cases, such equations being important in the
analytic determination of transfer matrix eigenvalues.}  \itm{Finally,
we mention the off-critical $A$ and $D$ models, which can be
associated with perturbed conformal field theories.  The bulk weights
in these cases can no longer be expressed in terms of the
Temperley-Lieb algebra, but a fusion procedure still exists and
boundary weights constructed from fused blocks of bulk weights
attached to diagonal boundary weights still satisfy the boundary
Yang-Baxter equation. Some integrable boundary weights are known for
these cases, as listed in \cite{BehPea97}, and it is expected that
each of the $A$ and $D$ boundary conditions found here corresponds to
a critical limit of an off-critical integrable boundary condition.}

\section*{Acknowledgements}
PAP is supported by the Australian Research Council.
We are thankful to Jean-Bernard Zuber for useful discussions and
for his hospitality at CEA-Saclay, where some of this work was done.

\end{document}